\documentclass[10pt, twocolumn,article]{IEEEtran}

\usepackage[utf8]{inputenc}

\usepackage[noadjust]{cite}
\usepackage{filecontents}

\usepackage{setspace}
\usepackage{epsf}\epsfverbosetrue
\usepackage{graphics,epsfig}
\usepackage{graphicx,epsfig}
\usepackage{epsfig}
\usepackage{multirow}
\usepackage{alltt}
\usepackage{subfigure}
\usepackage{tcolorbox}
\usepackage{float}
\usepackage{url}
\usepackage{graphicx}
\usepackage{verbatim} %for giving multiple line comments by mubashir
\usepackage{footnote} % by mubashir
\usepackage{latexsym} % by mubashir
\usepackage{sidecap} %by mubashir
\usepackage{wrapfig}
\usepackage{eso-pic}
\usepackage{fix-cm}
\usepackage{algorithm}
\usepackage{algorithmic}
\usepackage{color}
\usepackage{wrapfig}
\usepackage{multicol}

\usepackage{pifont}
\newcommand{\tick}{\ding{52}}%
\newcommand{\cross}{\ding{55}}%
\newcommand{\mult}{\ding{93}}%
\newcommand\tikzmark[1]{\tikz[remember picture] \node (#1) {};}

\newcommand{\foo}{\hspace{-2.3pt}$\bullet$ \hspace{5pt}}

\usepackage{layout}

\usepackage{amsmath}
\usepackage[
  locale = DE % comma as decimal mark
]{siunitx}

\usepackage{textcomp}
\usepackage{multirow}
\usepackage{mathtools}
\usepackage{soul} 
\usepackage{amsmath} 
\usepackage{verbatim}
\usepackage{csvsimple} 
\usepackage{array}
\usepackage{subfig}
\usepackage[normalem]{ulem}
\usepackage{booktabs}% http://ctan.org/pkg/booktabs
\newcolumntype{P}[1]{>{\RaggedRight\arraybackslash}p{#1}}
\newcommand{\tabitem}{\textbullet~~}
\usepackage{caption}

\usepackage{flowchart}
\usetikzlibrary{shapes,arrows,matrix,decorations.pathreplacing,shapes.geometric,positioning,calc}  

\usepackage{dsfont}
\usepackage{amssymb}
\usepackage{array}
\usepackage{xcolor}
\newcommand{\yasir}[1]{\textcolor{black}{#1}}
\newcommand{\rev}[1]{\textcolor{black}{#1}}
\newcommand{\revi}[1]{\textcolor{black}{#1}}
\newcommand{\addp}[1]{\textcolor{black}{#1}}

\newcommand{\comst}[1]{\textcolor{black}{#1}} % For COMST Revision 1
\newcommand{\revtwo}[1]{\textcolor{black}{#1}} % For COMST Revision 2
\newcommand{\mubcom}[1]{\textcolor{black}{#1}} 
\begin{document}

\title{Differential Privacy Techniques for Cyber Physical Systems: A Survey}

\author{Muneeb Ul Hassan, Mubashir Husain Rehmani, and Jinjun Chen
\thanks{M. Ul Hassan and J. Chen are with the Swinburne University of Technology, Hawthorn VIC 3122, Australia  (e-mail:  muneebmh1@gmail.com; jinjun.chen@gmail.com).}
\thanks{M.H. Rehmani is with the Department of Computer Science, Cork Institute of Technology, Rossa Avenue, Bishopstown, Cork, Ireland (e-mail: mshrehmani@gmail.com).}
\thanks{Please direct correspondence to M.H. Rehmani.}
\thanks{This paper is partly supported by Australian Research Council (ARC) projects DP190101893, DP170100136, LP140100816 and LP180100758.}
}

%\author{\IEEEauthorblockN{Muneeb Ul Hassan\IEEEauthorrefmark{1}, Mubashir Husain Rehmani\IEEEauthorrefmark{4}, Jinjun Chen\IEEEauthorrefmark{1}\\
%\IEEEauthorblockA{\IEEEauthorrefmark{1}Swinburne University of Technology, Hawthorn VIC 3122, Australia\\ \IEEEauthorrefmark{4} Waterford Institute of Technology (WIT), Waterford, Ireland}}
%Wherever\\ There will be one Bracket :-) For My info
%Email: \IEEEauthorrefmark{1}author.one@add.on.net,
%\IEEEauthorrefmark{2}author.two@add.on.net,}}

\maketitle

\begin{abstract}

Modern cyber physical systems (CPSs) \rev{has widely being used in our daily lives} because of development of information and communication technologies (ICT). With the provision of CPSs, the security and privacy threats associated to these systems are also increasing. Passive attacks are being used by intruders to get access to private information of CPSs. In order to make CPSs data more secure, certain privacy preservation strategies such as encryption, and k-anonymity have been presented in the past. However, with the advances in CPSs architecture, \revtwo{these techniques also need certain modifications.} Meanwhile, differential privacy emerged as an efficient technique to protect CPSs data privacy. In this paper, we present a comprehensive survey of differential privacy techniques for CPSs. In particular, we survey the application and implementation of differential privacy in four major applications of CPSs named as energy systems, transportation systems, healthcare and medical systems, and industrial Internet of things (IIoT). Furthermore, we present open issues, challenges, and future research direction for differential privacy techniques for CPSs. This survey can serve as basis for the development of modern differential privacy techniques to address various problems and  data privacy scenarios of CPSs. 

\end{abstract}


%% ###################### Tikz Package Shapes  ########################

\tikzstyle{decision} = [diamond, draw, fill=blue!50]
\tikzstyle{line} = [draw, -stealth, line width= 0.4mm]
\tikzstyle{elli}=[draw, ellipse, fill=red!50,minimum height=5mm, text width=5em, text centered]
\tikzstyle{block} = [draw, rectangle, fill=blue!20, rounded corners, minimum height= 10mm, minimum width = 20mm, text width=11em, text centered]

\tikzstyle{smallblock} = [draw, rectangle, fill=orange!20, rounded corners, minimum height= 10mm, text width=5.5em, text centered]

%% ###################### Tikz Package Shapes  ########################

\section{Introduction}
\rev{Previously, embedded computers were used to control and monitor the physical processes via feedback loop control~\cite{newmbref06}. With the passage of time, integration of computation technologies with traditional embedded physical systems lead the foundation of new type of systems named as cyber physical systems (CPSs)~\cite{newmbref07}.} The advances in CPSs  have gathered considerable attention over the last ten years~\cite{iotref07}. The major reason behind this stupendous attention is the dual nature of CPSs, via which they integrate the dynamic properties of embedded computers with those of information and communication technologies (ICT)~\cite{intref01}. Similarly, the merger of ICT and embedded systems spread to a number of physical domains of dynamic nature, including energy, transportation, healthcare, medical, industrial, and manufacturing systems~\cite{intref02}.~\rev{Majority of CPSs are deployed in life support devices, critical infrastructures (CI), or are very vital to our everyday lives. Therefore, CPSs users \mubcom{expect them to be} emancipated from every type of vulnerabilities. One of the critical issue in deployment of CPSs in real world is \mubcom{their} privacy, as any type of information leakage can result in very serious consequences~\cite{intref09}. Particularly, \rev{the complex architecture of CPSs} make it difficult to assess the privacy threats, and new privacy issues arises. It is also strenuous to trace, identify, examine, and eliminate privacy attacks that may target multiple components of CPSs \rev{such as real-time sensors, wearable health devices, industrial control systems, etc.}~\cite{intref09}.} Similarly, CPSs basically rely on diverse number of sensors and data centres containing very huge amount of personal and private data. \rev{For example, wearable devices of patients are continuously reporting their real-time data to consulting doctors~\cite{newmbref08}. \revtwo{However, if one does not use a strong privacy preservation scheme} during this communication, then any adversary can try to hack this personal information and can use it for illegal benefits such as blackmailing, false information injection, etc~\cite{newmbref09}. Therefore, there is a strong possibility of compromising the personal privacy of CPS users in the absence of proper privacy protection strategy~\cite{intref03}.}

% ===================== Privacy Table  ===========================
\begin{table*}[ht]
\begin{center}
 \centering
 \footnotesize
 \captionsetup{labelsep=space}
 \captionsetup{justification=centering}
 \caption{\textsc{\\Comparison of Privacy Preservation Strategies on the basis of Method, Merits, Weaknesses, and Computational Overhead.}}
  \label{tab:privtable}
  \begin{tabular}{|P{2cm}|P{4.5cm}|P{3.5cm}|P{4cm}|P{1.7cm}|}
  	\hline
  	
\rule{0pt}{2ex}
\centering \bfseries Privacy \newline Name  & \centering \bfseries Method to Protect Privacy & \centering \bfseries Merits \newline (Advantages) & \centering \bfseries Weaknesses \newline and \newline Challenges &  \bfseries Computational Overhead \\
\hline

\rule{0pt}{2ex}
\centering \textbf{Encryption} & Public and private keys are assigned to transmitted data that are used to decrypt data at receiving end & \tabitem Original data is not lost \newline \tabitem Data becomes inaccessible to unauthorized users & \tabitem Computationally complex \newline \tabitem Reduces system speed \newline \tabitem Not Suitable for public databases & Very high~\cite{sgref09} \\
\hline

\rule{0pt}{2ex}
\centering \textbf{Anonymization} & Personal identifiable information (e.g. name, date of birth, passport number etc.) is removed before query evaluation & \tabitem Limits disclosure risks \newline \tabitem Works with high dimensional data & \tabitem Original data is lost  \newline \tabitem  100\% privacy is not guaranteed \newline \tabitem Chances of re-identification exists in large data & High~\cite{surref16} \\
\hline

\rule{0pt}{2ex}
\centering \textbf{Differential Privacy\\ (focus of this survey article)} & Random noise is added using various mathematical algorithms, e.g. Laplace, Gaussian, etc. & \tabitem Low complexity \newline \tabitem Original data is not lost \newline \tabitem Privacy can be controlled according to need by varying privacy parameter & \tabitem Dimensionality curse \newline \tabitem Reduction in data utility \newline \tabitem Selecting desirable trade-off between privacy and accuracy is tough  & Low~\cite{newmbref22}\\
\hline

 \end{tabular}
  \end{center}
\end{table*}

% ===================== Privacy Table  ===========================

\rev{Attacks on CPSs can be classified into passive (privacy oriented) or active (security oriented). The basic objective of passive attacks is to access a certain amount of private data being shared in the network, or to infer about any critical information from public dataset~\cite{intref02}.} Many researchers proposed cryptographic techniques to preserve data privacy~\cite{intref04, newmbref01, intref05}. However, these cryptographic techniques are computationally expensive, because users’ needs to maintain the set of encryption keys. Moreover, it becomes more difficult to ensure privacy in a situation when public sharing of data is required. \rev{Similarly, anonymization techniques such as k-anonymity~\cite{newmbref03} \mubcom{are also} proposed by researchers to address privacy issues. \comst{However, these anonymization \revtwo{strategies do not guarantee} complete level of protection from adversaries because the \revtwo{chances of re-identification increase if size} of attributes in dataset increases~\cite{surref16}. An adversary trying to infer the data can match the non-anonymized data with anonymized data, which in turn will lead to privacy breach.} Another important privacy scheme named as differential privacy was introduced in 2006 to overcome these privacy issues. Currently, the use of differential privacy is emerging as a future of privacy~\cite{intref06}. \revi{Differential privacy protects statistical or real-time data by adding desirable amount of noise along with maintaining a healthy trade-off between privacy and accuracy.} \comst{In differential privacy, user can control the level of privacy or indistinguishability, which in turn will lead to protection of maximum possible privacy for any particular individual in dataset. For example, the value of privacy parameter can be used to control trade-off between utility and privacy, it can either be 100\% utility or 100\% privacy depending upon the requirement of system.} A detailed overview of these privacy preserving strategies along with their merits and demerits is presented in Table.~\ref{tab:privtable}.}

% ===================== Privacy Table  ===========================
\begin{table*}[ht]
\begin{center}
 \centering
 \footnotesize
 \captionsetup{labelsep=space}
 \captionsetup{justification=centering}
 \caption{\textsc{\\\comst{\mubcom{Utility-Privacy Comparison between Differential Privacy, Anonymization, and Encryption .}}}}
  \label{tab:privtable1}
 {\color{black} \begin{tabular}{|P{2cm}|P{2cm}|P{12.5cm}|}
  	\hline
  	
\rule{0pt}{4ex}
\centering \bfseries Privacy Technique  & \bfseries Major Constraint&  \bfseries Description \\
\hline

\rule{0pt}{3ex}
\centering \textbf{Anonymization} & Limited privacy & Anonymization does not guarantee complete privacy as it is claimed that anonymized data is either protected or either usable, because chances of re-identification always exists in anonymization~\cite{surref16}. \\
\hline

\rule{0pt}{3ex}
\centering \textbf{Encryption} & Third-party privacy & Protects data from third-party intruders along with providing complete privacy, but privacy is not protected from observing analyst~\cite{intref11}.\\
\hline

\rule{0pt}{3ex}
\centering \textbf{Differential Privacy} & Utility-privacy trade-off & Deals with utility-privacy trade-off that can be controlled by data provider according to privacy requirement~\cite{surref02}.\\
\hline

 \end{tabular}}
  \end{center}
\end{table*}

% ===================== Privacy Table  ===========================

\rev{Differential privacy has the capability to preserve large proportion of data from both, databases and real-time data~\cite{intref07}. Data perturbation is carried out in majority of differential privacy techniques.~\revi{In data perturbation, amount of noise is calculated using differential privacy algorithms and this noise is further added to query data to make it secure and indistinguishable for observer.} This perturbation \mubcom{has} direct effect with the accuracy of data being reported. On the other hand, the more \mubcom{perturbed} data ensures that privacy is strongly protected. Therefore, while using differential privacy, one needs to maintain an advantageous trade-off among accuracy and privacy. Due to this} privacy and accuracy trade-off, utilizing differential privacy in CPSs is a challenging task, \rev{because various CPSs applications require accurate reporting of data, for example health care and medical systems.} To efficiently use differential privacy techniques in CPSs, various \mubcom{techniques} in energy systems, transportation systems, healthcare, machine learning, and industrial systems have been proposed in literature. 
The actual goal is to improve privacy level along with minimizing the trade-off with accuracy.

\subsection{Motivation: Differential Privacy for Cyber Physical Systems}

To date, various privacy preservation strategies have been proposed by researchers to overcome certain privacy threats. \textit{Encryption} is one of the traditional privacy preserving technique used by majority of systems to protect the data from adversaries and unauthorized users~\cite{intref11}, \rev{because it provides feature of data inaccessibility to unauthorized users.} However, in modern CPSs, encryption can barely be applied, because of the sensors' limitation of computing capacity~\cite{sgref09}. \comst{For example in public key cryptography also called as asymmetric cryptography, the generation, and distribution of public and private keys is a computationally complex task and cannot easily be carried out with small devices having limited resources~\cite{comsttextref01}.} Furthermore, various attacks, such as brute-force attack may be used by any vulnerability against the encrypted CPS data~\cite{intref12}.
Similarly, in a network of multiple sensors, encryption strategies require the interconnection of every node for generation and transmission of private keys in the network. Therefore, if one node gets failed in a network of~\textit{n} number of nodes, then the decryption and collection of data from CPSs nodes becomes nearly impossible, because of missing keys in the network~\cite{sgref09}.

Another privacy preserving strategy used by researchers is \textit{data anonymization}~\cite{intref15}. However, recent literature work indicates that privacy of anonymized data can easily be compromised. For example, Montjoye~\textit{et al.}~\cite{intref16} collected and simply anonymized mobility data for 15 minutes of 1.5 million people. Even from this anonymized data, they identified a person with around 95\% accuracy via four temporal points only. Furthermore, the weaknesses of simply anonymized data was confirmed further by a test carried out on credit card transactions~\cite{intref17}, and researchers re-identified the individuals with 90\% accuracy by using only four temporal points.\\
\rev{Existing privacy preservation schemes being used in CPSs pose serious challenges to users’ privacy. Therefore, a perturbation technique that protects the private data with appropriate mechanism was the need.} In 2006, C. Dwork proposed the concept of differential privacy as an efficient privacy preserving approach to obstruct adversaries from recovering data~\cite{intref13}. Similarly, a statistical differential privacy interpretation was developed by Wasserman in 2010~\cite{intref18}. Continuing this research line, researchers proposed \textit{membership privacy}~\cite{intref19} and \textit{differential identifiability}~\cite{intref20} to address certain problems in differential privacy framework. \rev{In context of CPSs, Zhu~\textit{et al.} in~\cite{iotref07} suggested the usage of differential privacy in advanced CPSs.  Contrary to encryption, differential privacy provides a less complex, privacy preservation mechanism. \comst{The \mubcom{actual} reason is that computational cost of differential privacy only comprises of noise computation using a pre-defined probability distribution. However, the nodes using encryption has to carry out certain tasks of generation, and distribution of keys along with encrypting and decrypting data. Therefore, the computational complexity of differential privacy is fairly low as compared to encryption.} Furthermore, in differential privacy, \revtwo{CPSs users can control the level of privacy} according to the need by varying noise addition parameter $``\varepsilon"$. Unlike anonymization, original data of CPSs is not lost during query evaluation using differential privacy because data is protected using perturbation methodology. \comst{Various noise addition mechanisms of differential privacy provide strong privacy protection against numeric and non-numeric queries~\cite{comstref43}.} \mubcom{By keeping in view its tremendous benefits, mathematical and theoretical basis}, and easy realization, differential privacy has been applied extensively in CPSs to preserve individual privacy~\cite{intref14}.}
%======== Table 1 =============

\begin{table}[t]
\begin{center}
\small
 \centering
  \captionsetup{labelsep=space}
 \captionsetup{justification=centering}
 \caption{\textsc{\\ \footnotesize{List of Acronyms and Corresponding Definitions.}}}
  \label{tab:acrtable}
  \begin{tabular}{|P{1.5cm}|P{6.3cm}|}
  	\hline
  	\rule{0pt}{2ex}
	\textbf{Acronyms} &\textbf{Definitions}\\
  	\hline
  	\rule{0pt}{2ex} 
  	 AMI & Advanced Metering Infrastructure\\
  	\hline
  	\rule{0pt}{2ex}  
  	BLH & Battery Load Hiding \\
  	\hline
  	\rule{0pt}{2ex} 
  	CI & Critical Infrastructure\\
  	\hline
  	\rule{0pt}{2ex} 
  	CIDS & Collaborative Intrusion Detection Systems\\
  	\hline
  	\rule{0pt}{2ex} 
  	CPS & Cyber Physical System \\
  	\hline
  	\rule{0pt}{2ex} 
  	D2D & Device-to-Device \\
  	\hline
  	\rule{0pt}{2ex} 
  	DSM & Demand Side Management \\
  	\hline
  	\rule{0pt}{2ex} 
  	DSRC & Dedicated Short-Range Communication \\
  	\hline
  	\rule{0pt}{2ex}
  	EM & Exponential Mechanism \\
  	\hline
  	\rule{0pt}{2ex} 
  	EV & Electric Vehicle \\
  	\hline
  	\rule{0pt}{2ex} 
  	FHMM & Factorial Hidden Markov Model  \\
  	\hline
  	\rule{0pt}{2ex} 
  	HA & Hybrid Automation\\
  	\hline
  	\rule{0pt}{2ex} 
  	HetVNET & Heterogeneous Vehicular Networks\\
  	\hline
  	\rule{0pt}{2ex} 
  	ICS & Industrial Control Systems \\
  	\hline
  	\rule{0pt}{2ex} 
  	ICT & Information and Communication Technologies\\
  	\hline
  	\rule{0pt}{2ex} 
  	IIoT & Industrial Internet of Things\\
  	\hline
  	\rule{0pt}{2ex} 
  	IoT & Internet of Things\\
  	\hline
  	\rule{0pt}{2ex} 
  	IoV & Internet of Vehicle \\
  	\hline
  	\rule{0pt}{2ex} 
  	ITS & Intelligent Transportation System \\
  	\hline
  	\rule{0pt}{2ex} 
  	LM & Laplace Mechanism\\
  	\hline
  	\rule{0pt}{2ex} 
  	LPWA & Low Power Wide Area\\
  	\hline
  	\rule{0pt}{2ex} 
  	LTE & Long-Term Evolution\\
  	\hline
  	\rule{0pt}{2ex}
  	MAB & Multi-Armed Bandit \\
  	\hline
  	\rule{0pt}{2ex} 
  	MANET & Mobile Ad-Hoc Networks \\
  	\hline
  	\rule{0pt}{2ex} 
  	NCS & Networked Control System\\
  	\hline
  	\rule{0pt}{2ex} 
  	NILM & Non-Intrusive Load Monitoring\\
  	\hline
  	\rule{0pt}{2ex} 
  	PII & Personally Identifiable Information\\
  	\hline
  	\rule{0pt}{2ex} 
  	PKC & Public Key Cryptography\\
  	\hline
  	\rule{0pt}{2ex} 
  	RER & Renewable Energy Resource\\
  	\hline
  	\rule{0pt}{2ex} 
  	SG & Smart Grid\\
  	\hline
  	\rule{0pt}{2ex} 
  	UNB & Ultra Narrow Band\\
  	\hline
  	\rule{0pt}{2ex} 
  	V2D & Vehicle-to-Device \\
  	\hline
  	\rule{0pt}{2ex} 
  	V2V & Vehicle-to-Vehicle\\
  	\hline
  	\end{tabular}
  \end{center}
\end{table}

%======== Table 1 =============

% ===================== Survey Table  ===========================
\begin{table*}[htbp]
\begin{center}
 \centering
 \footnotesize
 \captionsetup{labelsep=space}
 \captionsetup{justification=centering}
 \caption{\textsc{\\Summary Comparison of Previous Survey Articles of Differential Privacy (DP) With Their Application Scenario, Year, Major Contributions, and Considered Factors. \newline \tick ~Indicates that the topic is covered, \cross ~indicates that the topic is not covered, and \mult ~indicates that the topic is partially covered.}}
  \label{tab:surveytab01}
  \begin{tabular}{|P{2.5cm}|P{0.8cm}|P{0.8cm}|P{6.7cm}|P{3.8cm}|P{1.2cm}|}
  	\hline
  	\rule{0pt}{2ex}
\bfseries Application Scenario  & \bfseries Ref No. & \bfseries Year & \bfseries Major Contribution & \bfseries Considered Factors & \bfseries Discussed \newline CPSs \\
\hline

\rule{0pt}{2ex}
\centering \textbf{Mobile Sensing System} & ~\cite{surref02} & 2017 & A comprehensive survey on improvement of data utility & \tabitem Distribution \tabitem Optimization \newline \tabitem Calibration \tabitem Transformation \tabitem Decomposition & \mult \\
\hline

\multirow{8}{*}{\parbox{2cm}{\centering \textbf{Social Networks}}}

& ~\cite{surref09} & 2011  & A comprehensive survey over privacy techniques in social networks including differential privacy & \tabitem Privacy breaches \newline \tabitem Anonymization in social networks & \mult \\
\cline{2-6}

\rule{0pt}{2ex}
& ~\cite{surref11} & 2014  & A comprehensive survey of differential privacy in social networks is carried out and after that two differential privacy based \textit{outlink privacy} and \textit{partition privacy} standards are proposed & \tabitem Triangle count \newline and distribution \newline \tabitem Graph modelling  & \mult \\
\cline{2-6}
\hline

\rule{0pt}{2ex}
\centering \textbf{Finite Precision Semantics} & ~\cite{surref06} & 2016 & A discussion about method of quantifying of data privacy by finite precision & \tabitem Computational error  & \cross\\
\hline

\centering \textbf{\comst{Sensitive Data Mining}} & ~\cite{surref07} & 2014 & An investigation is done over learning base data release mechanism & \tabitem Loss function  & \cross \\
\hline

\rule{0pt}{2ex}
\centering \textbf{Mobile Recommender System} & ~\cite{surref12} & 2016 & An overview of privacy preserving algorithms including differential privacy is given by focusing on collection, generation, and storage in mobile recommender systems & \tabitem Privacy risks  & \mult\\
\hline

\multirow{8}{*}{\parbox{2cm}{\centering \textbf{Statistical Databases}}}

\rule{0pt}{2ex}
& ~\cite{surref08} & 2008  & A presentation of two basic techniques of differential privacy & \tabitem Learning theory \newline \tabitem Statistical data inference & \cross \\
\cline{2-6}

\rule{0pt}{2ex}
& ~\cite{surref10} & 2012  & An investigation and approach on individual experience cost as a function of privacy loss is proposed & \tabitem Privacy and accuracy trade-off & \cross \\
\cline{2-6}

\rule{0pt}{2ex}
& ~\cite{surref13} & 2017  & An overview about differential privacy and its relevance with other data science topics & \tabitem Computational complexity \newline \tabitem Cryptography \tabitem Theoretical computer science  &  \cross \\
\cline{2-6}

\hline
\rule{0pt}{2ex}
\centering \textbf{Communication in Big Data} & ~\cite{surref16}  & 2018 & A comprehensive comparison of differential privacy with other privacy preservation approaches is carried out in scenario of big data from the perspective of communication & \tabitem Privacy attacks \newline \tabitem Privacy risks in big data  & \mult \\
\hline

\multirow{8}{*}{\parbox{2cm}{}}%\centering \textbf{Big Data}}}
\rule{0pt}{2ex}
& ~\cite{surref01} & 2016  & Overview of differential privacy methods to chose accurate epsilon value for a better trade-off & \tabitem Epsilon value & \cross \\
\cline{2-6}

\rule{0pt}{2ex}
\centering \textbf{\comst{Miscellaneous }} & ~\cite{surref04} & 2016  & A detailed study on data clustering and privacy framework & \tabitem Cryptography \newline \tabitem Data mining\newline \tabitem Biometric privacy \newline \tabitem Game theory & \cross \\
\cline{2-6}
\rule{0pt}{2ex}
& ~\cite{surref03} & 2017  & Analysis over data publishing and data analysis using differential privacy & \tabitem Data release mechanism \newline \tabitem Efficiency \tabitem Accuracy & \cross \\
\cline{2-6}

\rule{0pt}{2ex}
& ~\cite{surref05} & 2018  & A comparison of differential privacy with other big data privacy schemes & \tabitem  Execution time \newline  \tabitem Complexity \newline \tabitem Data utility & \cross\\
\cline{2-6}

\rule{0pt}{2ex}
& ~\cite{surref14} & 2018  & A survey on fundamental ideas of privacy budget and sensitivity in differential privacy & \tabitem  Noise calculation mechanism \newline \tabitem Fundamental architectures for differential privacy & \cross\\
\cline{2-6}

\rule{0pt}{2ex}
& \revtwo{~\cite{rev2ref02}} & \revtwo{2019}  & \revtwo{A detailed survey over integration of modern differential privacy algorithms with deep learning models} & \tabitem  \revtwo{Privacy attacks} \newline \tabitem \revtwo{Private data extraction} \newline \tabitem \revtwo{Deep learning layers} & \revtwo{\cross}\\
\cline{2-6}

\rule{0pt}{2ex}
& \revtwo{~\cite{rev2ref01}} & \revtwo{2019}  & \revtwo{In-depth analysis of more than 50 variants of differential privacy from perspective of methodology} & \tabitem  \revtwo{Privacy loss} \newline \tabitem \revtwo{Differential privacy definitions} \newline \tabitem \revtwo{Computational effect of variants} & \revtwo{ \cross}\\
\cline{2-6}

\hline

\rule{0pt}{2ex}
\centering \textbf{Cyber Physical Systems} & This Work & 2018 & An in-depth survey of differential privacy techniques in applications of cyber physical systems (energy, transportation, healthcare, medical, and industrial IoT systems) & \tabitem Privacy preservation \newline \tabitem CPSs applications \newline \tabitem Privacy attacks \newline \tabitem Design mechanism \newline \tabitem Technical challenges & \tick \\
\hline
 \end{tabular}
  \end{center}
\end{table*}

% ===================== Survey Table  ===========================

\subsection{Contributions of This Survey Article}

While few previous survey articles have highlighted some specific aspects of differential privacy techniques in certain CPSs domains, to the best of our knowledge there is no comprehensive survey over the implementation and applications of differential privacy techniques in CPSs. In this paper, we survey state-of-the-art work on differential privacy techniques in CPSs scenarios. In summary, following contributions are made in the article:
\begin{itemize}

\item We review previous survey articles on differential privacy and highlight important features of them.
\item \rev{We focus more on the presenting practical aspects of differential privacy in CPSs.}
\item We provide a thorough survey of differential privacy and its implementation in CPSs.
\item We provide an extensive survey of applications of differential privacy in CPSs.
\item We survey the work done over implementation of differential privacy in energy systems, transportation system, healthcare, and industrial IoT systems.
\item We outline certain open issues, challenges, and possible future research direction for differential privacy based CPSs.

\end{itemize}
\subsection{Review of Related Survey Articles}

Our present survey article on differential privacy in CPSs is distinct from all previous studies, as we extensively cover the area of differential privacy implementation in CPSs. There is \rev{a} comprehensive literature of previous survey articles that has focus on differential privacy, and few of them focused over differential privacy in \mubcom{big data analysis}. However, to the best of our knowledge, there is no prior detailed survey article that thoroughly addresses differential privacy strategies in CPSs. We categorize the previous survey literature work over differential privacy into seven major categories named as statistical databases, social networks, mobile sensing system, finite precision semantics, \mubcom{sensitive data mining}, machine recommender systems, miscellaneous. The application scenario, major contribution and considered factors about these survey articles is presented in Table~\ref{tab:surveytab01}.

\mubcom{An extensive literature on differential privacy in context of salient privacy related features of differential privacy from context of data analytic, big data, and privacy budget is presented in~\cite{surref01,surref03,surref04,surref05,surref14}.} \revi{The term big data refers to transmission, collection, storage, or usage of large amount of data collected from any \comst{source~\cite{comstref01}.}} Zhou~\textit{et al.}~\cite{surref01} discussed about different methods of calculation of accurate~($\varepsilon$) value to minimize privacy and accuracy trade-off. A brief analysis over data release mechanism, efficiency, and accuracy of differential privacy is presented in~\cite{surref03}. In~\cite{surref04}, a detailed study on data clustering and privacy framework is presented by focusing mainly over cryptography, data mining, biometric privacy, and game theory. A survey over comparison of execution time, complexity, and data utility of differential privacy with other privacy schemes of big data is given in~\cite{surref05}. While, Jain~\textit{et al.}~\cite{surref14} surveyed the fundamental ideas of sensitivity and privacy budget by focusing on fundamental architecture and noise calculation mechanism of differential privacy. 

A comprehensive survey on improvement of data utility of differential privacy in mobile sensing systems is presented in~\cite{surref02}. The discussion in~\cite{surref09} covers the aspect of privacy breaches for differential privacy techniques including differential privacy in social networks. Similarly, in~\cite{surref11} authors carried out a comprehensive survey of differential privacy in social networks and then presented two privacy techniques \textit{(outlink privacy and partition privacy)} based on the concept of differential privacy. A method of quantifying data privacy and reducing computational error by finite precision based differential privacy is discussed in~\cite{surref06}.\\
The detailed investigation about sensitive data mining, loss function, and learning from data bases by focusing on differential privacy is carried out in~\cite{surref07}. Furthermore, the implementation of various privacy preserving algorithms of mobile recommender system is compared with differential privacy by Xu~\textit{et al.}~\cite{surref12} in context of privacy risks. 
Studies on privacy preservation of statistical databases using differential privacy has been presented in~\cite{surref08,surref10,surref13}. C. Dwork in~\cite{surref08}, laid the foundation of differential privacy for statistical data inference and surveyed two basic techniques of differential privacy. Similarly, an investigation and approach on individual experience cost as a function of their privacy and accuracy trade-off is proposed in~\cite{surref10}. \rev{Furthermore}, the comparison of differential privacy with other data science privacy strategies by focusing on computational complexity, and theoretical basis is carried out in~\cite{surref13}. \rev{The field of privacy preservation in big data from the perspective of communication is analysed by Wang~\textit{et al.} in~\cite{surref16}. \revi{The paper} presents the comparison of differential privacy with other privacy preservation approaches in context of framework and preserving technique. Furthermore, it also highlights certain privacy attacks that needs to be analysed in the mentioned privacy preserving approaches.} \revtwo{A detailed discussion about the integration of differential privacy with various deep learning models is presented in~\cite{rev2ref02}. The authors first discusses three aspects of deep learning models along with the possible privacy attacks. Afterwards, authors demonstrated the integration of differential privacy in these models from perspective of layer-wise implementation. Similarly, an in-depth technical analysis over extensions and variants of differential privacy is covered by authors in~\cite{rev2ref01}. The paper presents dimension, axioms, and relation based analysis of 54 differential privacy variants and analysed them from the point-of-view of computational complexity, and privacy loss.}\\ However, the privacy topics of all differential privacy surveys do not address the applications and implementation of differential privacy in CPSs from any perspective.

\subsection{Article Structure}

A list of acronyms that are used throughout our survey paper is presented in Table.~\ref{tab:acrtable}. The remainder of this paper is organized as follows: Section 2 provides an overview of differential privacy and CPSs, while Section 3 surveys differential privacy techniques in energy systems. Section 4 provides a detailed survey of privacy preservation of transportation systems using differential privacy. Section 5 surveys implementation of differential privacy in healthcare and medical systems. Similarly, Section 6 surveys differential privacy approaches in industrial Internet of things systems. In Section 7, we outline certain open issues, challenges, and future research directions. Finally, Section 8 concludes the survey article.

\section{Differential Privacy and Cyber Physical Systems: An Overview}

\revi{Privacy can be defined as a method of protecting information that can be sensitive to any individual. The basic reason of privacy preservation is to prevent an intruder from learning more than minimum required information regarding any specific individual either in case of real-time or statistical data.}

\subsection{Privacy Attacks}

\revi{Adversaries always try to attack crucial systems in order to get complete or partial access to information. In this section, certain privacy attacks closely related to CPSs and differential privacy are discussed.}

\subsubsection{Disclosure Attack}
\revi{ Disclosure attack is a traffic pattern analysis attack, and in this type of attack, adversary is able to recognize the defined set of receivers on the basis of observed traffic~\cite{addpaper53, addpaper54}. Adversaries use this attack method to identify the specific receiver and compromise its communication. However, disclosure attacks are not easy to implement because they require a certain level of computational efficiency. This is because of fact that adversary has to scan the whole network several times in order to get accurate receivers’ identity that makes it difficult to launch~\cite{addpaper55}. Still, in order to overcome this attack, the real-time information being communicated between sender and receiver needs to be protected; that even if adversary is able to compromise the receiver, he will not be able to judge the accurate information. }

\subsubsection{Linking Attack}
\revi{ The type of attacks in which external data is combined with anonymized or protected data in order to infer critical information is known as linking attack~\cite{addpaper56}. For example, two anonymized datasets are linked together having different types of data about same individuals; re-identification can easily be carried out using linking attack. In the age of big data, launching an effective linking attack can be quite easy for any adversary. Thus, even anonymized \revtwo{data is not safe, and it can be used} to breach any individual privacy. Therefore, a privacy protection strategy to efficiently protect query evaluation over statistical data is required. }

\subsubsection{Differencing Attack}

\revi{Direct queries about any individual are usually blocked during a query evaluation to avoid any privacy breach. For example, in a hospital database, queries like “does Mark have diabetes” are restricted because such queries directly violate privacy of individuals. Instead of these, queries for aggregated results are usually allowed, for example “how many men in a specific region have diabetes”. However, an adversary can submit multiple queries to get personal information about certain individual. For example, an intruder can first ask “How many individuals in the dataset have diabetes”, and then it can submit query as “How many people in the diabetic dataset, not named Mark”. Thus, by repeating such queries, adversary will be able to determine the diabetes status of Mark. This type of repetitive query evaluation attack is known as differencing attack~\cite{surref16}. A privacy mechanism that does not give 100\% accurate output to adversary for such statistical queries is required to protect secret data of individuals. }

\subsubsection{Correlation Attack}
\revi{In real-world data, strong \revtwo{correlation may exist} such as shared relationships and family members share attributes in various social networking datasets. If an adversary tries correlation attack using similar datasets, then this existing correlation may lead to disclosure of more than expected information~\cite{addpaper57, addpaper58}. An intruder having different anonymized datasets for comparison can obtain private information of individuals in datasets by performing correlation attack, thereby it directly violates the principles of privacy. For example, anonymized datasets of a hospital can be merged and correlated to find the presence of any specific disease in members of family. This certain type of attack is called as correlation attack, and adversaries can easily launch correlation attack to identify certain details if they have rich data about their targets. In order to prevent correlation attack, a privacy preserving mechanism with efficient data handling is required that reduces the risk of information leakage even in case of public query evaluation.
}

%--------------------- FIGURE ------------------------

\begin{figure*}[t]        
\centering
\includegraphics[scale = 0.8]{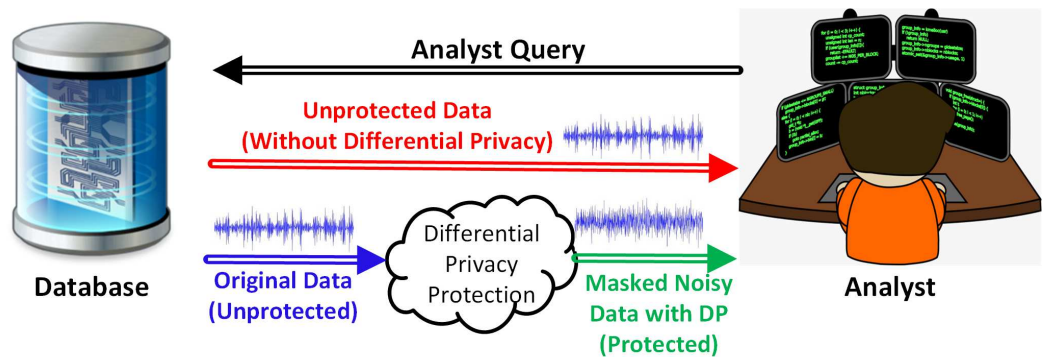}
 \caption{Analyst query evaluation scenario explaining data output with Differential Privacy (DP) preservation (protected data) and without DP preservation (unprotected data).}
  \label{fig:fig01}   
\end{figure*}

%--------------------- FIGURE ------------------------

\subsection{Differential Privacy}

To date, most of work on \mubcom{privacy preservation} is done in perspective of databases. \comst{This work can be categorized into two major domains; first domain involves the protection of complete data from database in which anonymization techniques play a major roles, and the second one involves development of a good theoretical framework on the basis of privacy requirement in which differential privacy came up as a viable solution. Anonymization techniques initiated from \textit{k-anonymity}, to \textit{l-diversity}, and then to \textit{t-closeness} method~\cite{dpcps01}. Similarly, theoretical framework based differential privacy strategies are further divided into \textit{differential identifiability} and \textit{membership privacy} ~\cite{surref04}. The detailed discussion about these derivatives is out of the scope of this article. \revtwo{However, detailed discussion about differential privacy and its integration in different CPSs domain is presented.}} 
% This work can further be dissected into two major categories: data clustering and theoretical framework. The data clustering techniques initiated from \textit{k-anonymity}~\cite{newmbref03}, to \textit{l-diversity}~\cite{newmbref05}, and then to \textit{t-closeness} methods~\cite{dpcps01, newmbref04}. Similarly, differential privacy also emerged as an optimal method to preserve statistical data privacy. 
%These theoretical models mainly include the modification of framework step by step. Further development in this framework mechanism lead the foundation of two algorithms from differential privacy named as \textit{differential identifiability} and \textit{membership privacy}  \cite{surref04}. However,
In this section,~\revi{classification, comparison, and application scenarios of differential privacy is discussed}. The concept of differential privacy technique on the basis of probability model was first introduced by C. Dwork~\cite{intref13}. \revi{The model} was totally independent of the prior knowledge of adversary~\cite{intref13},~\cite{dpcps02}. \comst{The aim of differential privacy is to make sure that the output result of any query should not reveal enough information about any individual that leads to its identification.} The randomized algorithm $\Re$ of differential privacy ensures that the output values of query cannot be distinguished irrespective of absence or presence of a specific member in the database ${\ss}$. This means, that query results of neighboring datasets are indiscernible by introducing some randomized value of noise~\cite{surref02}. This sums up the conclusion, that adversary will not be able to presume sensitive information of any dataset with confidence. Differential privacy can formally be defined on the basis of two neighboring databases ${\ss}$ and ${\ss}'$ that differs from each other in only one single member.\\

\noindent \underline{\emph{\textbf{Definition 1 (\mubcom{Neighboring} Datasets):}}}
A randomized algorithmic function $\Re$ satisfies $\varepsilon$-differential privacy condition $\mathbb{P_R}$ if for any two adjacent datasets ${\ss}$ and ${\ss}'$, and for any sort of possible outcome $\xi \in Range(\Re)$, we get:

\begin{equation}
\mathbb{P_R}[\Re(\ss) \in \xi] \leq exp(\varepsilon) \times \mathbb{P_R}[\Re(\ss') \in \xi]
\label{eqn:eqn1}
\end{equation}

In above equation, $ Range(\Re)$ is the range of resultant output function $\Re$. Similarly, $``\varepsilon"$ is the epsilon privacy parameter, that determines the actual level of privacy for proposed mechanism. The lower value of $\varepsilon$ is desired in order to have stronger privacy and vice versa~\cite{newmbref10}.\\

\noindent \underline{\emph{\textbf{Definition 2 (Global Sensitivity):}}}
\comst{The value of sensitivity actually determines the required amount of perturbation in differentially private mechanism. Similarly, the term global sensitivity works over the phenomenon of maximum possible difference between query outputs from two datasets differing with each other by only one element (also known as neighboring datasets). For a randomized query $f: \ss \rightarrow \Re$, the value of global sensitivity $\Delta f_{gs}$ can be found using the following formula~\cite{comstref43}:}
\begin{equation}\color{black}
\Delta f_{gs} = \max_{{\ss}, {\ss}'} ||f ({\ss}) - f ({\ss}')|| 
\label{eqn:eqn2}
\end{equation}
The basic noise addition mechanism for differential privacy is shown in Fig.~\ref{fig:fig01}.
Discussion about differential privacy can be divided into two major branches; differential privacy existing methods, and noise addition mechanisms.
\vspace{3mm}
%\hspace*{\fill}

%--------------------- FIGURE ------------------------

\begin{figure*}[t]        
\centering
\includegraphics[scale = 0.65]{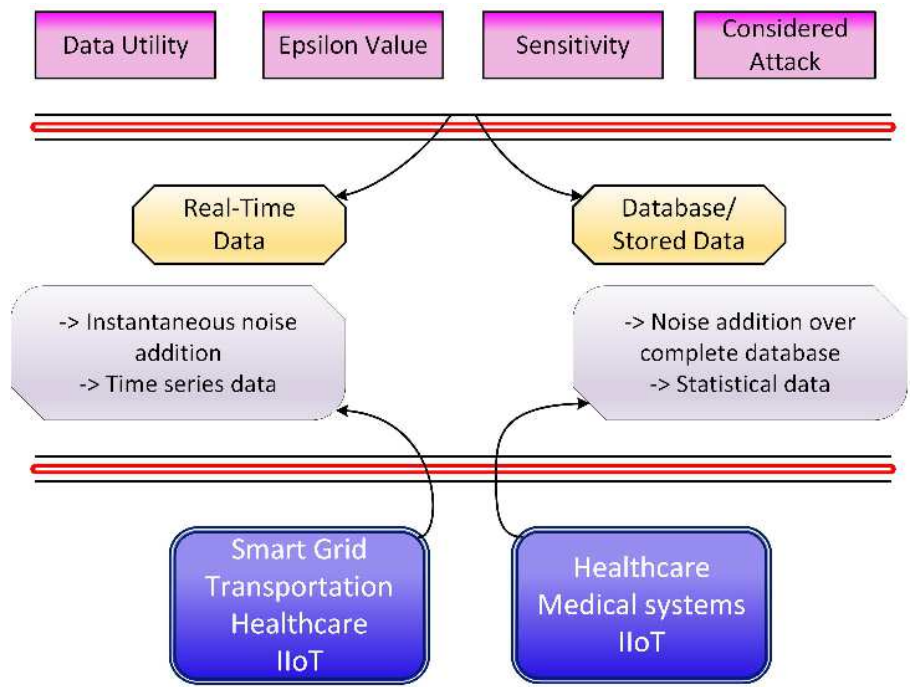}
 \caption{\comst {Differential privacy implementation in cyber physical systems can broadly be classified into two scenarios (real-time data and stored data), however both the categories face some similar technical challenges such as selection of value of epsilon, sensitivity, data utility, and attack protection. }}      
  \label{fig:refdptech}   
\end{figure*}

%--------------------- FIGURE ------------------------

% ===================== Privacy Table  ===========================
\begin{table*}[ht]
\begin{center}
 \centering
 \footnotesize
 \captionsetup{labelsep=space}
 \captionsetup{justification=centering}
 \caption{\textsc{\\\comst{Medical Record based Demonstration of Exponential Mechanism of Differential Privacy.} \newline \newline \mubcom{The table shows the trade-off between privacy and utility such that 0.2 at $\varepsilon=$0 demonstrates that there are 20\% chances of that specific answer being picked which means 100\% privacy, and 0.72 in $\varepsilon=$1 demonstrates the chances for headache  are 72\%, which means minimum privacy.}}}
  \label{tab:expotable}
  {\color{black}\begin{tabular}{|P{3cm}|P{3.5cm}|P{2cm}|P{2cm}|P{2cm}|}
  	\hline
  	
\rule{0pt}{3ex}
\bfseries Analyst Query  & \bfseries No. of Applicant & \bfseries $\varepsilon$ = 0 & \bfseries $\varepsilon$ = 0.1 &  \bfseries ~~~~~$\varepsilon$ = 1 \\
\hline

\rule{0pt}{3ex}
\textbf{Cough} & 32 & 0.2 & 0.275 & 0.2675 \\
\hline

\rule{0pt}{3ex}
\textbf{Asthma} & 18 & 0.2 & 0.136 & 2.4 x $10^{-4}$ \\
\hline

\rule{0pt}{3ex}
\textbf{Cholesterol} & 24 & 0.2 & 0.185  & 4.9 x $10^{-3}$\\
\hline

\rule{0pt}{3ex}
\textbf{Dehydration} & 12 & 0.2 & 0.101  & 1.2 x $10^{-5}$\\
\hline

\rule{0pt}{3ex}
\textbf{Headache} & 34 & 0.2 & 0.303 & 0.727\\
\hline

 \end{tabular}}
  \end{center}
\end{table*}

% ===================== Privacy Table  ===========================

\subsubsection{Differential Privacy Existing Methods} \label{existmethod}
Differential privacy existing protocols can be divided into two major categories, one according to the perspective of differential privacy optimization, and other according to the perspective of datasets. \comst{The integration of differential privacy in any data can further be categorized on the basis of two categories:} $(i)$ \textit{Distribution Optimization}~\cite{dpcps03, dpcps04,dpcps05,dpcps06,dpcps07,dpcps08}, in this branch, the probability density function of differential privacy is optimized without taking in account the dataset. In these techniques, differential privacy is generally achieved by adding randomized noise calculated via Laplacian or Exponential mechanism. The probability distribution of these schemes is further partitioned into centralized and distributed schemes. \addp{$(ii)$ \textit{Sensitivity Calibration}~\cite{dpcps09,dpcps10,dpcps11,dpcps12,dpcps13,dpcps14,addpaper13}, in these techniques, data utility is improved by calibrating the sensitivity value to an optimal state. \revtwo{The sensitivity is further smoothed and lowered} in order to enhance the data utility.} \comst{These two mentioned categories perturbs data on the basis of mentioned criteria, \yasir{for example if one needs to preserve} data according to a certain probability distribution, then it will use the first “distribution optimization”, or if one is interested to protect data according to its sensitivity value then “sensitivity calibration” is the most suitable option for that. However, these both types of integrations are also interlinked with each other in a manner that one can adjust both values at the same time to achieve desirable results. \revtwo{For example, a person wants to protect its data} can use any probability distribution (such as Exponential, Laplacian, etc.) along with its required sensitivity value. }

Similarly, the branch of differential privacy in perspective of dataset is also divided into two major subcategories: $(i)$ \textit{Synopsis of Original Databases and Datasets}~\cite{dpcps15,dpcps16,addpaper48,dpcps18,dpsg20,dpcps20,dpcps21,dpcps22,dpcps23,dpcps24,dpcps25,dpcps26}. In this method, the synopsis of datasets is built by applying various techniques of decomposition, transformation, or compression. The noise in these techniques is added in such a way that it minimizes rate of error and improves utility of data, along with satisfying $\varepsilon$-differential privacy value; $(ii)$ \textit{Correlation Exploitation}~\cite{dpcps09},~\cite{dpcps15}~\cite{dpcps27,dpcps28,dpcps29}, in these proposed strategies of differential privacy, the correlation among attributes and data records is exploited to reduce the effect of noise and redundancy, that in return preserves data privacy more effectively.
\vspace{3mm}
%\hspace*{\fill}

\subsubsection{Data Perturbation Mechanisms} \label{noisesec}
\comst{In differential privacy, noise addition mechanism is referred as a way to protect data by perturbing it via pre-defined mechanisms.} Three noise adding mechanisms are generally used in differential privacy approaches. They are named as Laplace mechanism (LM)~\cite{dpcps31}, Gaussian mechanism~\cite{addpaper06}, and Exponential mechanism (EM)~\cite{surref14}. The actual magnitude of added noise directly depends upon the global sensitivity and privacy budget~\cite{dpcps30}. \mubcom{Another term named as privacy bound do also contribute while noise addition in some circumstances. \mubcom{Generally,} privacy bound is referred as privacy budget unless a specific bound is required~\cite{comstref62}.}

\paragraph{\textbf{Laplace Mechanism}}

In Laplace mechanism, the noise is computed using the Laplacian function, and each coordinate of data is perturbed using the calculated Laplacian noise from LM distribution. The sensitivity of the differential privacy function determines the scale of noise being added.\\
In a given dataset $\ss$, function $\Re$, and the global sensitivity $\delta f_s$, the randomized algorithm $\AA$ in Eq. \ref{eqn:eqn3} satisfies $\varepsilon$-differential privacy parameter, if the calculated noise value complies with the actual value of Laplace distribution; which is, $noise\thicksim Lap(\delta f_s / \varepsilon$). LM is generally used in case of numerical output results~\cite{dpcps31}.
\begin{equation}
\AA = \Re(\ss) + Lap(\delta f_s / \varepsilon)
\label{eqn:eqn3}
\end{equation}
\paragraph{\textbf{Exponential Mechanism}}
A method to implement differential privacy in case of non-numerical outputs is Exponential mechanism. Exponential mechanism was specifically developed for certain conditions in which the best response was required to be picked. 
In a given dataset $\ss$, $\l \in \L$ denotes the possible answer, in a score function $u: \ss \times \L \to \L$; and a randomized algorithm $\AA$ selects a probability based answer, then the given randomized algorithm $\AA$ will satisfy the $\varepsilon$-differential privacy according to Eq. \ref{eqn:eqn4} ~\cite{surref14}.
\begin{equation}
\AA (\ss , u) = {\l : |\mathbb{P_R} [\l \in \L] \infty exp (\varepsilon u(\ss , \l)/2 \Delta u)}
\label{eqn:eqn4}
\end{equation}
In the above presented equation, $\Delta u$ denotes the sensitivity of exponential score function. The value of $\Delta u$ varies according to the requirement of user.\\
\comst{For example, Table~\ref{tab:expotable} discusses an example of exponential mechanism, in which we take the value of sensitivity $\Delta u$ = 1, and the dataset of medical records is evaluated on the basis of different epsilon values. The third column of table shows that when $\varepsilon$ is taken as 0, then the mechanism can uniformly pick any option from all five options because the probability of selection of all possible outcomes becomes the same. Hence, it guarantees 100\% privacy, but the utility is minimum. \yasir{Similarly, in case of $\varepsilon$ = 0.1, }Headache has the maximum probability of being picked from the samples and Dehydration has minimum probability. Although, the gap between values is not very large, thus the \yasir{value of $\varepsilon$ = 0.1 can} provide a considerable level of privacy. Finally, when $\varepsilon$ = 1, the gap of probability between Headache and Dehydration is very significant which indicates a very high utility value, but the level of privacy reduces via considerable amount.}
%that is defined as follows in Eq.~\ref{eqn:eqn5}:
%\begin{equation}
%\Delta u = max(\l \in \L)  max((||\ss \Delta \ss’||) = 1)|u(\ss,\l) – u(\ss’, \l)|
%\label{eqn:eqn5}
%\end{equation}

\paragraph{\textbf{Gaussian Mechanism}}
\addp{
Gaussian mechanism is another essential block that is currently being used in implementation of differential privacy algorithms. Similar to Laplace mechanism, noise in Gaussian mechanism is calculated using normal (Gaussian) distribution~\cite{addpaper06, newmbref10}. In a query function $f$, let value of $\varepsilon$ be between 0 and 1, then the output for Gaussian perturbation $\sigma$ will be as follows:
\begin{equation}
\sigma = \frac{\Delta_2 f}{\varepsilon}
\sqrt{2log(1.25/ \varepsilon)}
\label{eqn:eqn6}
\end{equation}
}

\noindent
\textbf{Composition Theorem}~\cite{dpcpsnew01}: In addition to noise addition mechanisms, differential privacy also has two composition theorems, which can be defined formally as follows:\\
\indent \textbf{Theorem \# 01} (Sequential Composition): 
\comst{The basic concept of sequential composition theorem states that if we have n-algorithms that are differentially private individually, and we want to feed results of first algorithm to second one, and so on, without sacrificing the complete privacy of output results. Then, sequential composition theorem allows \mubcom{such operations.} Sequential composition theorem is usually beneficial for algorithms involving multiple iterations over same dataset.} \\

\noindent \underline{\emph{\textbf{Proof:}}}
\comst{A mechanism M($\mathcal{B}$) follows $n\varepsilon$-sequential composition differential privacy theorem if it obeys following property.\\
Let $\mathcal{B}_1 \& \mathcal{B}_2$ be two neighboring datasets then:}

\begin{equation*}
\begin{split}
\color{black} P_r[M(\mathcal{B}_1) = x_n] & \color{black} = P_r[M_1(\mathcal{B}_1) = x_1] P_r[M_2(\mathcal{B}_1;x_1) = x_2] ...\\
& \color{black} ~~~~P_r[M_n(\mathcal{B}_1; x_1,...x_{n-1}) = x_n] \\
& \color{black} \leq exp(n\varepsilon) \prod_{k=1}^{n} P_r[M_k(\mathcal{B}_2; x_1,...x_{k-1}) = x_k]\\
& \color{black} = exp(n\varepsilon) P_r[M(\mathcal{B}_2) = x_n]
\end{split}
\label{eqn:eqnc1}
\end{equation*}

\indent \textbf{Theorem \# 02} (Parallel Composition): 
\comst{Parallel composition applied in a condition when there is a single dataset which is further partitioned into n disjoint subsets.} Privacy bound can be improved when queries are applied to disjoint subset of data. Primarily, if we partition input records into disjoint sets that are independent of actual data, then ultimate privacy guarantee of differential privacy depends only over worst guarantees of each differential privacy analysis subjected to data, not the sum.\\
\noindent \underline{\emph{\textbf{Proof:}}}
\comst{A mechanism M($\mathcal{B}$) follows parallel composition differential privacy theorem if it obeys following property.\\
Let $\mathcal{B}_1 \& \mathcal{B}_2$ be two neighboring datasets then:}
\begin{equation*}
\begin{split}
\color{black} P_r[M(\mathcal{B}_1) = x_n] & \color{black} = \prod_{k=1}^{n} P_r[M_k(\mathcal{B}_2; x_1,...x_{k-1}) = x_k]\\
& \color{black} \leq exp(\varepsilon) P_r[M_L(\mathcal{B}_2; x_1,...x_L) = x_L]\\
&~~~ \color{black} \prod_{k \neq L}^{n} P_r[M_L(\mathcal{B}_1; x_1,...x_{k-1}) = x_L]\\
& \color{black} = exp(\varepsilon) P_r[M(\mathcal{B}_2) = x_n]
\end{split}
\label{eqn:eqnc2}
\end{equation*}

%--------------------- FIGURE ------------------------

\begin{figure*}[t]        
\centering
\includegraphics[scale = 0.7]{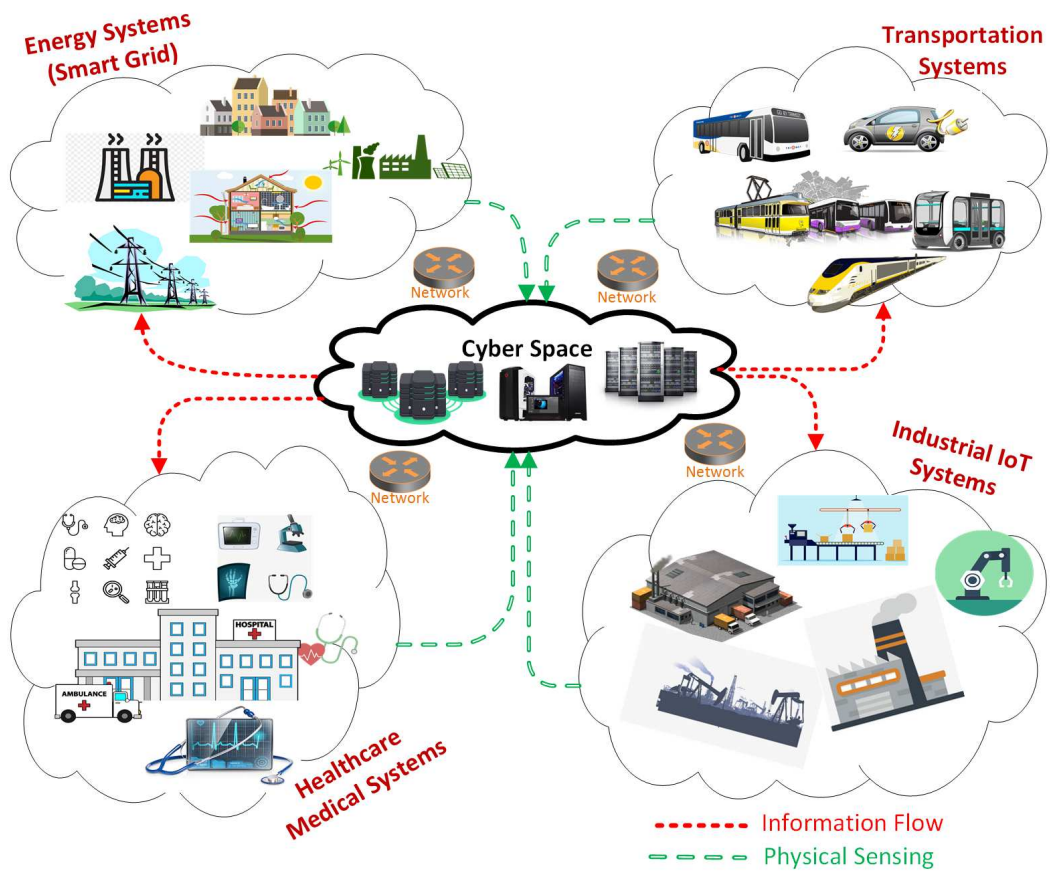}
 \caption{Application Scenarios of cyber physical systems describing the information flow and physical sensing between cyber space and connected physical devices.}      
  \label{fig:fig02}   
\end{figure*}

%--------------------- FIGURE ------------------------

\subsection{Technical Challenges in Application of Differential Privacy}

\revi{Although the basic privacy preservation concept of differential privacy is not much complex and most of the techniques require only data perturbation, but its implementation in certain applications faces numerous technical challenges. In this section, we highlight few challenges that researchers and industries do face while implementing differential privacy in emerging applications. \comst{ A graphical illustration of some technical challenges in application of differential \yasir{privacy in different} scenario is presented in Fig. \ref{fig:refdptech}}}

\subsubsection{Sensitivity}

\revi{
Differential privacy was actually introduced to maintain the level of indistinguishability between certain cases, e.g., the presence or absence of record of some specific individual in dataset. In practical datasets, statistical queries are evaluated with low sensitivity using differential privacy~\cite{addpaper34}. For example, any random value $x$ may be 1, or may even be 10,000,000 in a statistical record, thus, its domain will be $x \in$ [1, $10^7$]. \revtwo{A malicious administrator submits a query} to aggregate all values of $x$ such as $SUM(x)$ which also involve the sensitive value of a participant. In this case, differential privacy algorithm calculates and adds Laplace noise in accordance to its standard deviation that will be directly proportional to range $10^7$. In such cases, noise potentially hides the useful information and does not let intruders to know the exact sensitive protected value of individual participant. However, high value of sensitivity in a differential privacy algorithm will lead to undesirable loss of utility in various queries related to aggregation. Therefore, in order to maintain a certain utility level, privacy is relaxed by service providers or individuals and a degree of information is allowed to get leaked~\cite{surref16}. For example, a case in which an adversary is able to estimate salary of a certain participant in a certain range is better than the case in which adversary will not be able to infer any participant salary while performing statistical analysis. In this way, relaxation in differential privacy algorithms allow highly private output but along with privacy loss. That is why trade-off is always there that needs to be maintained between utility and privacy. Several emerging services and applications are using diverse sensitivities~\cite{addpaper35, addpaper36, addpaper37}. Therefore, it is still a challenge for researchers to propose optimal algorithms which overcome the privacy-utility trade-off by using optimal sensitivity value in the most efficient way.
}

\subsubsection{Choosing Epsilon Value}
\revi{Despite of being mathematically sound, still there is no rigorous method that explains choosing and generation of optimal value of $\varepsilon$. Epsilon parameter serves to be one of the most important factor in controlling the trade-off between utility and privacy~\cite{addpaper38}. A smaller value of $\varepsilon$ indicates quite high level of noise in the mechanism; thus, adversary will not be able to attack individual privacy. However, this high value of $\varepsilon$ results in loss of utility or accuracy in output data. Therefore, choosing the optimal value of $\varepsilon$ in various practical scenarios is a challenging task and is not completely addressed by researchers yet. In literature, certain researches have been carried out to find most optimal value of $\varepsilon$. In~\cite{addpaper38, addpaper39, surref08, addpaper41, addpaper42, addpaper43}, various approaches have been proposed to calculate optimal $\varepsilon$ value on the basis of certain criteria such as success probability, largest affordable value, etc. However, choosing an accurate $\varepsilon$ is still a considerable technical challenge for researchers.
}

\subsubsection{Data Correlation}

\revi{
Strong coupling correlation often exhibits in real-world datasets, and certain records can be correlated with each other; that results in disclosure of information~\cite{addpaper44}. For instance, differential privacy assures that modifying, adding, or removing any individuals’ record from a set of data will have no effect over aggregated data. However, presence of data correlation can benefit adversary by providing information that may help to infer private data of any specific individual. For example, in case of any social networks data, the presence of any specific disease in a specific family, and adjacent spatio-temporal location continuity can help the adversary to infer private individual information; this directly violates differential privacy definition. In~\cite{ dpcps09, addpaper46, addpaper47}, researchers developed model-based approaches, based over sensitivity weights, correlation degree, and correlated sensitivity to overcome this issue. Similarly, transformation-based approaches working over principle of \yasir{transformation of} actual correlated data to independent data while maintaining key information have also been proposed by researchers in ~\cite{addpaper48, addpaper49, dpsg20, addpaper51, addpaper52}. However, experimental results evaluated that model-based approaches do not cover all aspects of correlation, and transformation-based approaches actually damages the correlated data up to some extent~\cite{surref16}. Therefore, overcoming correlation in differential privacy algorithm is currently among few biggest challenges for differential privacy researchers.
}

% ===================== Privacy Table  ===========================
\begin{table*}[ht]
\begin{center}
 \centering
 \footnotesize
 \captionsetup{labelsep=space}
 \captionsetup{justification=centering}
 \caption{\textsc{\\\comst{\mubcom{Comparison of Differential Privacy and Information Theoretic Privacy from Technical Perspective.}}}}
  \label{tab:infocomp}
 {\color{black} \begin{tabular}{|P{3cm}|P{6.5cm}|P{6.5cm}|}
  	\hline
  	
\rule{0pt}{4ex}
\centering \bfseries Privacy Factor  & \bfseries Differential Privacy &  \bfseries Information-Theoretic Privacy \\
\hline

\rule{0pt}{3ex}
\centering \textbf{Privacy Bound} & Privacy bound controlled by $e^\varepsilon$ sensitivity and $\Delta$ factor~\cite{intref07} & Privacy bound controlled by entropy mechanism (such as Shannon entropy)\mubcom{~\cite{comstref55}} \\
\hline

\rule{0pt}{3ex}
\centering \textbf{Privacy Guarantee} & Provides guarantee for users' privacy~\cite{surref02} & Characterises confidentiality property of program~\cite{comstref57} \\
\hline

\rule{0pt}{3ex}
\centering \textbf{Privacy Definition} & Can be added by adding appropriate random noise~\cite{dpcps03} & It limits the probability of inferring secret information via entropy~\cite{comstref53}\\
\hline

\rule{0pt}{3ex}
\centering \textbf{Privacy Strength} & Differential privacy is termed as a stronger privacy as it implies bound on mutual information~\cite{comstref54} & Restricts the probability of guessing but does not have bound over differential privacy mechanism~\cite{comstref54} \\
\hline

 \end{tabular}}
  \end{center}
\end{table*}

% ===================== Privacy Table  ===========================

\subsection{\comst{Effect of Differential Perturbation over Analyst and Adversary}}

\comst{Differential privacy mechanism ensures that the privacy of user gets protected from adversary, and in order to do so, it perturbs data using various differential mechanisms. However, choice of sensitivity and epsilon value play an important role in determining the trade-off between accuracy and privacy. One important factor that differential privacy ensures is that the adversary will not be able to judge with confidence about presence or absence of any individual in a dataset. In order to analyse it, let’s imagine two datasets D and D$^\prime$ differing with each other by just one record. If an adversary makes a query “F” on both presented datasets, then there is a very high probability that adversary will get same result “R” in both cases. On the basis of this result, the adversary will not be able to differentiate that a specific person “X” is present or absent in the dataset. However, on the other hand, a genuine analyst who wants to analyse the data in a legitimate way and do not have any intention to intrude into details of any specific individual will not feel much difference in the results. In this manner, differential privacy ensures that the output result should not disclose too much statistical information about any specific individual of that dataset.}

\subsection{\comst{Comparison Between Differential Privacy and Information-Theoretic Privacy}}

\comst{In recent years, another modern privacy preservation strategy named as “information-theoretic privacy” is being used by researchers to protect users’ data which is emerged from information-theoretic flow security. Similar to differential privacy, information-theoretic privacy do also work over preserving the private information by using statistical and analytical tools by leveraging the concept of data disclosure~\cite{comstref51, comstref52}. Information-theoretic privacy quantifies and characterises confidentiality of query evaluation and limits the probability of disclosure of secret information using certain entropy based techniques such as Shannon entropy~\cite{comstref53, comstref54}. \mubcom{Where Shannon entropy can be termed as average rate via which the information is generated by an available data source. Formally, Shannon entropy is defined as a negative log function of the product of probabilities of bunch of independent events~\cite{shannref01}.} Another important concept in information-theoretic privacy mechanism is mutual information which is referred as a parameter that signifies the relation between two random variables within dataset~\cite{comstref57}. For example, within a dataset, the amount of information that a variable “A” can provide about variable “B” is termed to be its mutual information parameter. Similarly, this parameter will be zero if these variables are totally independent and irrelevant of each other. This parameter also serves as a foundation of many information theoretic approaches and researches are being carried out actively to enhance this parameter to an optimal level in order to provide maximum privacy guarantee.\\
In order to compare information theoretic privacy with differential privacy, first we need to consider their theoretical bounds and strength. Firstly, differential privacy works over the phenomenon of noise addition which depends upon $\varepsilon$ and $\Delta$ factor that are referred as privacy parameter and sensitivity respectively. \yasir{However, contrary to differential privacy, information} theoretic privacy mechanisms are controlled by entropy based mechanisms that controls the level of indistinguishability for quantitative analysis~\cite{comstref55}. Furthermore, differential privacy guarantees privacy of user and by only providing analytical proof for private answer of query. However, information theoretic privacy protects data by varying the difficulty of guessing the correct answer which can also be termed as leakage~\cite{comstref56}. Finally, in order to compare privacy strength, certain researches have showed that differential privacy is a stronger privacy guarantee as it implies bound on Shannon mutual information \\revtwo{and this bound approach value of “0”} as the privacy controlling parameter $\varepsilon$ approaches “0”. (A detailed discussion about this is elaborated by Alvim~\textit{et al.} in~\cite{comstref54}.) Contrary to this, information theoretic privacy \revtwo{does not impose any bound} over any parameter of differential privacy. Keeping in view all the discussion it can be summarized that both information theoretic privacy and differential privacy provides strong privacy guarantee to their respective usage application and both have their advantages and limitations that can be adjusted according to requirement. A detailed technical comparison between information theoretic privacy and differential privacy is given in Table~\ref{tab:infocomp}. \mubcom{Moreover, since information theoretic privacy is in itself a huge topic, therefore, in this paper, we only focused on differential privacy. However, interested readers can further understand information-theoretic privacy from the following references ~\cite{itpref01, comstref52, itpref02, itpref03, comstref53}}. }

\subsection{Cyber Physical Systems}

In 1946, first computer was invented to perform ballistic calculations~\cite{dpcps32}. With the passage of time, the position of computers got strengthened in various close loop controls around different physical systems. From this motivation, in 1973, the first computer which was capable of real-time computations was developed~\cite{dpcps33}, which addressed the problem of scheduling multiple tasks so that every job gets completed before the deadline. Since then, the interest initiated in CPSs, though the actual name invented quite late. In the late 90’s, the interest between interaction of physical and computational systems nourished when industries started modelling physical plants using differential equations which was named as hybrid systems~\cite{dpcps34}.

In the meantime, another path named networks was also leading towards this area from the origin of Internet and cellular telephony~\cite{dpcps35}. This field of network control paved its way with the development of Smart Dust Project in 1998~\cite{dpcps36}, in which the nodes connected with the network can bring information regarding the physical environment around the devices. This evolutionary path initiated from communication, computation, and network control merged into a broader domain called as $networked$ CPS~\cite{dpcps34}. Another path that have traces in the present CPSs is control system, which paved its path towards modern CPSs by invention of digital control approximately 50 years ago~\cite{dpcps34}. Around 2006, the researchers working in hybrid systems, control systems, and real-time systems devised the name \textit{Cyber Physical Systems} to describe the systems incorporating cyber and physical worlds. Currently, the modern CPSs that we see all around us are basically the merger of communication, computing, and control technologies.

In the modern world, CPSs are closely linked with certainly every field of life including healthcare, energy, automotive, civil infrastructure, transportation, and industry. The information is being generated, sensed, and transmitted from all these technologies, and is being stored in certain databases and servers~\cite{jinref01, jinref07}. Securing this transmission and data sensing gives birth to another critical domain of privacy and security in CPSs. In this section, we will be discussing system architectures, and applications of CPSs. While the privacy of CPSs and implementation of differential privacy will be discussed in Section~\ref{newlabel01}.

%\hfill
%\hspace*{\fill}
\subsubsection{System Architectures}

Generally, CPSs architectures consists of two major functional components~\cite{dpcps45}: $(i)$ \textit{Advanced Connectivity}, in which the real-time data transmission and reception is taking \mubcom{part between physical} world and cyber space; and $(ii)$ \textit{Intelligent Data Management}, in which computational, and analytical capabilities are developed that is the core part of cyber space. Researchers have divided CPSs into certain models on the basis of their architecture. In this section, we review three major architectures of CPSs. 

\paragraph{\textbf{Networked Control Systems (NCS)}}
\yasir{The mediation of} communication network between computing and physical devices is one of the major attribute of modern CPSs, and these types of control systems are called as networked control systems (NCSs)~\cite{dpcps34}. In NCSs, the control signals from devices and feedback signals from output can be exchanged in between system components. Owning this communication network based control, NCSs have many benefits as compared to traditional control systems. NCSs eliminate the use of unnecessary wires, they are less complex, cost friendly in designing and implementation, and can be upgraded or modified without major reformations in the basic structure~\cite{dpcps46}. On the other hand, NCSs do also have certain issues as well. The major issues NCSs face are: reduction of network traffic load due to sampling~\cite{dpcps37, dpcps38, dpcps39}, network induced delays~\cite{dpcps40}, packet dropout phenomenon~\cite{dpcps41}, and quantization errors~\cite{dpcps42}. 

\paragraph{\textbf{Hybrid Systems}}
Systems which work on the switching phenomenon between multiple operation modes are known as switching systems. Similarly, the framework of CPSs that is able to capture both, transitions between various continuous and discrete states over time is known as hybrid automation (HA)~\cite{dpcps34}. HA is generally used to model the complex dynamic nature of CPSs via several mathematical and computational formalisms~\cite{dpcps43, dpcps44}. 

\paragraph{\textbf{Distributed Hybrid Systems}}
These types of systems are quite complex, because they involve the differential equations based dynamic, discrete models, and real-time communication and computation technologies~\cite{dpcps34}. \revtwo{These types of systems are} basically the composition and combination of several cyber and physical systems. Researchers are working in proving correctness of these systems by developing certain systematic methods and compositional frameworks.

%\hfill
%\hspace*{\fill}
\subsubsection{Applications of CPSs}
The research spectrum of CPSs is quite broad and it is hard to limit CPSs to few selected domains. CPSs are present in approximately every phase of life, from our homes to offices, and from our public life to personal life. The applications of CPSs with respect to information transmission and physical sensing is shown in Fig.~\ref{fig:fig02}. In this section, we discuss few benefits that CPSs research is providing and can provide in some selective application areas.

\paragraph{\textbf{Energy Systems (Smart Grid)}}
Traditional energy systems can be summed up into generation, transmission, and distribution of electricity. But the modern energy systems named as \textit{smart grids} are a combination of energy and ICT. Smart grid is said to be next generation infrastructure that will be capable of managing all our energy, and environmental needs, by providing us \yasir{un-interrupted,} cost-effective, and environment friendly electricity~\cite{dpcps47}. The production, transmission, and distribution efficiency of electric system can be optimized by using real-time measurements, analysis, and sensing techniques. Moreover, cyber and physical interactions are playing a vital role in advancement of efficient smart grid, and various new technologies, and certain methods are being developed to facilitate smart grid users. Few of them are real-time pricing, demand response, load forecasting, real-time load monitoring, etc~\cite{dpcps48}.
\paragraph{\textbf{Transportation Systems}}
Intelligent transportation systems (ITSs) is one of the emerging field of CPSs. In ITSs, development of traffic systems, vehicles, mass transit, and other similar factors are being addressed in order to enhance efficiency, congestion, sustainability, and safety~\cite{dpcps34}. A new terminology named as Internet of Vehicles (IoV) is introduced by researchers, in which every vehicle travelling in a certain radius will be connected with each other via device-to-device (D2D) and vehicle-to-vehicle (V2V) communication~\cite{dpcps49}. By using these capabilities, intelligent vehicles can aid the drivers or can even drive intelligently by sensing, estimating, and monitoring their surrounding circumstances and conditions. Moreover, the electric vehicles (EV) that are considered to be future of cars are being made fully intelligent, autonomous, and \yasir{environmental friendly.} ITSs are considered to be future of transportation systems and CPSs technologies are playing a vital role in practical implementation of this system in our daily life.
\paragraph{\textbf{Healthcare and Medical Systems}}
Healthcare and medical systems are considered to be one of the most sensitive systems because they are directly linked with the wellbeing of people connected to them. Similarly, the evolutionary technologies of CPSs being implemented in healthcare systems are making them efficient, reliable, and intelligent. Healthcare systems generally contain wearable body sensors, OT (operation theatre) equipment, and physiological hardware devices. In modern era, these devices are made smart by connecting them with Internet and naming these things as \yasir{e-Health.} Wireless sensors are also being implanted in majority of these devices for real-time monitoring, but wireless devices have their own limitations and constraints~\cite{dpcps34}. Similarly, real-time response and user personal data protection are few major issues of these devices that are currently being observed and improved by medical researchers and biomedical engineers. 
\paragraph{\textbf{Industrial Internet of Things (IIoT)}}
With the evolution of fourth generation of industry (Industry 4.0), the industrial systems are becoming smarter day by day. This term of Industrial Internet of Things (IIoT) flourished when researchers started integrating concepts of Internet of Things (IoT) with the environment of industrial control systems (ICS)~\cite{dpcps50}. The rise of IIoT is expected to enhance the process optimization, workers safety, factory management, etc~\cite{jinref04}. At the same time, the practical implementation of IIoT in industries is facing certain difficulties, e.g., developing efficient communication protocols, security and privacy issues of massive datasets, and formulation of efficient design patterns~\cite{dpcps51}. Researchers on the other hand are also focusing over the merger of various technologies such as edge computing, fog computing, and cloud computing with IIoT to make these systems more advance and autonomous.

% ===================== Privacy Table  ===========================
\begin{table*}[ht]
\begin{center}
 \centering
 \footnotesize
 \captionsetup{labelsep=space}
 \captionsetup{justification=centering}
 \caption{\textsc{\\Design Requirements of Differential Privacy in Applications of Cyber Physical Systems}}
  \label{tab:designtable}
  \begin{tabular}{|P{2cm}|P{1.2cm}|P{1.5cm}|P{3cm}|P{3.2cm}|P{2.3cm}|P{1.5cm}|}
  	\hline
  	
\rule{0pt}{2ex}
\centering \bfseries Application \newline Name  & \centering \bfseries Level of Privacy Required & \centering \bfseries Data \newline Types & \centering \bfseries Sensitivity \newline Calibration & \centering \bfseries Common Attacks & \centering \bfseries Considered Factors & \bfseries 3$^{rd}$ Party Aggregation \\
\hline

\rule{0pt}{2ex}
\centering \textbf{Smart Grid} & Medium & Time-Series & Instance / Point-wise sensitivity & \tabitem Non-intrusive load monitoring (NILM) attacks & Cost effectiveness & Yes \\
\hline

\rule{0pt}{2ex}
\centering \textbf{Transportation Systems} & Medium & Time-Series  & Instance / Point-wise sensitivity & \tabitem Correlation attacks & Timeliness & Yes\\
\hline

\rule{0pt}{2ex}
\centering \textbf{Healthcare \newline \& \newline Medical Systems} & Extremely High & \centering Time-Series \newline and \newline Statistical & Instantaneous sensitivity \newline and global sensitivity & \tabitem Inference attacks & Data accuracy & No \\
\hline

\rule{0pt}{2ex}
\centering \textbf{Industrial Internet of Things} & High &  \centering Time-Series \newline and \newline Statistical & Instantaneous sensitivity \newline and global sensitivity  & \tabitem Stealthy attacks & Secure communication & No\\
\hline

 \end{tabular}
  \end{center}
\end{table*}

% ===================== Privacy Table  ===========================

\subsection{Motivation of Using Differential Privacy in CPSs}
\label{newlabel01}
Differential privacy has the potential to provide substantial amount of privacy to majority of application of CPSs and it may be a good choice where critical or public information needs to be preserved~\cite{dpcps52}.~\comst{Similarly, integration of differential privacy with modern CPSs has emerged as a hot-topic \yasir{not only in academic} field but it is also paving its paths in industry~\cite{comstref02}.} Most of CPSs applications are real-time, that generates large amount of data named as big data. In order to handle and protect~\revi{transmission, communication, and storage of} this big data of CPSs, a strong privacy preservation approach such as differential privacy is required. 
In this subsection, we provide few of the major stimulating causes for applying differential privacy in CPSs:

\begin{itemize}

\item Various privacy techniques, such as encryption, k-anonymity, \yasir{l-diversity, and t-closeness} have been proposed for big data. At the same time, certain applications of CPSs are also being evaluated over these privacy techniques. \comst{However, differential privacy is one of the most suitable option to preserve privacy because it does not degrade systems' speed as compared to other techniques because of the light-weight nature of algorithms of differential privacy ~\cite{surref05}. For example, in encryption, generation and distribution of public and private cryptographic keys in the network becomes a hectic task and if one node in the network of “n” nodes gets failed then the aggregation of data becomes impossible due to missing distributed keys in that network ~\cite{dpsg14}. Similarly, anonymizing complete dataset that contains millions of records is not feasible option for certain service provides because of limited computational capacity. Furthermore, in order to keep original records, the database \mubcom{companies have} to store both \mubcom{datasets along with; one} anonymized dataset for query evaluation, and one original dataset for their internal use. \revtwo{However, differential privacy eradicates both} of the mentioned issues as it only protects data at run time by using basic low-complex algorithms which require calculations using Laplacian, exponential, and Gaussian distributions.}
\item Differential privacy provides enough quantitative \mubcom{theoretical basis} \yasir{which provides researchers} an exact information that how much statistical CPSs data is safe to release along with what amount of accuracy~\cite{surref13}.
\item The original data of CPSs applications is very critical and the owner of data cannot take risk of losing that data. Therefore, in differential privacy preservation, the original dataset remains the same and is not modified at all. Irrespective of k-anonymity, \yasir{l-diversity, or t-closeness} schemes, where original values of data are manipulated to preserve identity~\cite{dpcps54, dpcps55, dpcps56}.
\item \addp{Differential privacy perturbs data by adding noise in such a way that the preserved CPSs statistical or real-time data can still be used by analysts according to their required needs~\cite{surref05, comstref02}.}
\item Differential privacy provides substantial protection even in distributed CPSs environment, irrespective of other privacy preservation schemes that cannot provide efficient results because of correlation issues among attributes~\cite{dpcps57}.
\item Most of the encryption strategies used by real-time devices are computationally complex and need the generation of private encrypted keys at \mubcom{every node.} But differential privacy provides a light-weight solution to preserve privacy for CPSs devices as compared to computationally complex encryption schemes \mubcom{as it only perturbs} the data with certain calculated amount of noise~\cite{dpcps58, newmbref20}.
\item \comst{If the list of queries is large in any CPSs database, traditional differential privacy suffers from utility loss, however integration of modern machine learning and deep learning algorithms with differential privacy is turning out to be a feasible solution. Researches have proved that differential privacy integrated with state-of-the-art machine learning and deep learning algorithms can effectively meet the demands of listing, perturbing, and query evaluation in large databases~\cite{iotref07}.}
\item In social CPSs, differential privacy provides both \textit{node privacy} and \textit{edge privacy}, by protecting individual information and any specific relationship information respectively~\cite{surref11}.

\end{itemize}

\subsection{Design Requirements of Differential Privacy in CPSs Applications}

%--------------------- FIGURE ------------------------

\begin{figure*}[t!]        
\centering
\includegraphics[scale = 0.7]{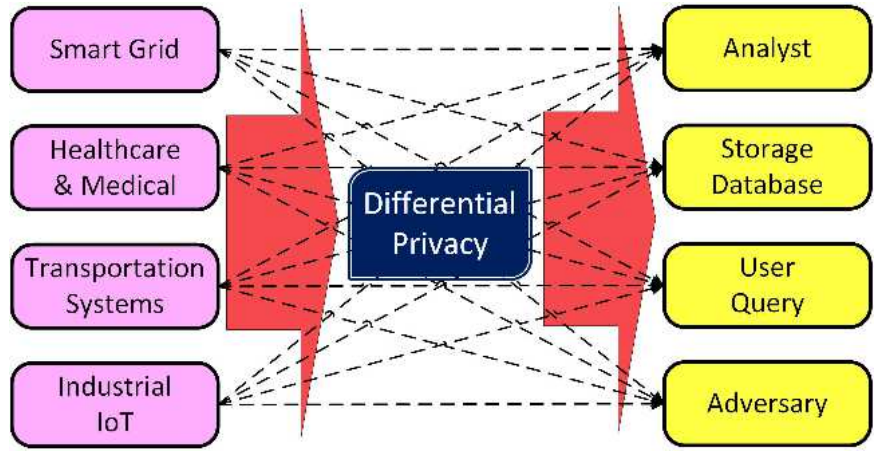}
 \caption{\small{ \comst {Use case scenarios of differential privacy in applications of cyber physical systems such data analysing by analyst, storage databases, user query evaluation, and adversary query request.}}}      
  \label{fig:refdpcps}   
\end{figure*}

%--------------------- FIGURE ------------------------

%=============================== TAXONOMY TABLE ====================

\begin{table*}[t!]

 \centering
 \captionsetup{labelsep=space}
 \small
 \captionsetup{justification=centering}
 \caption{\textsc{\\\comst{\mubcom{Milestones achieved in differential privacy from perspective of cyber physical systems}}}}%
\label{taxonomycps}
{\color{black}\begin{tabular}{|r |@{\foo} l|}

\hline
\midrule
\rule{0pt}{2.5ex}
\tikzmark{a} 2006  & \cite{intref13}~C. Dwork proposed the concept of differential privacy to obstruct adversaries from recovering private data.\\
%\hline
\rule{0pt}{2.5ex}
2011 &  \cite{sgref05}~G. Acs for first time integrated differential privacy with private smart metering data.\\
%\hline

\rule{0pt}{2.5ex}
2013 & \cite{tsref06}~F. Kargl implemented differential privacy with policy enforcement framework to protect transportation data.\\
%\hline
\rule{0pt}{2.5ex}
2014 & \cite{sgref03}~J. Zhao introduced the concept of integration of battery load balancing with differential privacy protection. \\
%\hline
\rule{0pt}{2.5ex}
2015 & \cite{hmref02}~H. Li developed private partition algorithm for electronic health record protection using differential privacy.\\
2015 & \cite{hmref04}~N. Mohamed used differential privacy to protect cancer database from SQL queries. \\
2015 & \cite{tsref02}~S. Han used concept of join differential privacy to protect real-time data reported by EVs.\\
2015 & \cite{sgref07}~M. Savi conducted experiment to protect smart metering data by using \yasir{Gaussian white} \& colored noise.\\
%\hline
\rule{0pt}{2.5ex}
2016 & \cite{iotref01}~G. Rodriguez worked over integration of differential privacy \& k-anonymity for industrial systems.\\
%\hline
\rule{0pt}{2.5ex}
2017 & \cite{sgref08}~G. Eibl proposed the notion of pointwise differential privacy for smart metering real-time data.\\
2017 & \cite{tsref01}~H. Zhai carried out differentially private auction to protect EV identities from swap stations during bidding. \\
2017 & \cite{tsref04}~Y. Shi protected railway freight data using apriori and differential privacy algorithm.\\
2017 & \cite{hmref01}~J. Zhang examined use of differential privacy in real-time health data via adaptive sampling.\\
2017 & \cite{iotref03}~Y. Wang presented the use of differential privacy to protect privacy of linear distributed control systems.\\
%\hline
\rule{0pt}{2.5ex}
2018 & \cite{sgref10}~H. Cao carried out private differentially private aggregation of smart grid over fog nodes.\\
2018 &\cite{tsref03}~T. Zhang proposed differentially private machine learning approach for vehicular networks\\
2018 & \cite{addpaper14}~L. Raisaro presented the use of encryption in combination with differential privacy for genomic \& clinical data. \\
\tikzmark{b} 2018  & \cite{iotref04}~L. Ni designed a differentially private algorithm for scanning in multi-core data clustering in industrial database systems.\\%
\bottomrule
\hline
\end{tabular}}
\tikz[remember picture,overlay] \draw[->] (a.center -| b.center) -- (b.center);
\end{table*}

%=============================== TAXONOMY TABLE ====================

\revi{
Differential privacy is a light-weight privacy preservation strategy that does not require complex hardware to run upon~\cite{newmbref22}. However, design and efficiency requirements vary in different applications of CPSs. \comst{Moreover, the optimality of \revtwo{differentially private mechanism also varies according to the application requirement.} For example, in some cases the optimal solution will be complete preservation of privacy, however in some \mubcom{cases providing} a certain \mubcom{level of utility can} be termed as an optimal solution. Similarly, one cannot determine any definition of optimal solution of differential privacy, therefore researchers working in differential privacy of use the term “approximate optimal solution ” while referring towards the most suitable solution according to requirement.}~Moving towards applications of CPSs, in context of smart grid, devices enabling differential privacy usually deal with communication of time-series data from one destination to another (mainly from smart meter to grid utility and vice versa). Therefore, point-wise privacy is generally considered in most of smart grid applications. Similarly, the most common attack in energy systems is non-intrusive load monitoring (NILM) attack, whose purpose is to identify the routine and appliance usage of smart home users~\cite{sgref10}. Therefore, implementation of differential privacy techniques in smart grid usually consider overcoming \mubcom{these attacks.} The aim of implementation of differential privacy strategies in energy systems is to provide a cost-effective medium level privacy protection by considering a healthy trade-off between accuracy and privacy~\cite{sgref02}. Moving further to transportation systems, the V2V and D2D communication also requires a privacy strategy to ensure time-series data protection at a specific instant of time without delay. The major task in ITSs is to protect the real-time location data co-ordinates that are being transmitted between different vehicles and devices in the network. This type of communication is vulnerable to correlation attacks, in which the correlation between real-world data may reveal more than expected information~\cite{dpcps57, newmbref24}. Therefore, most of differential privacy techniques being implemented over ITSs consider overcoming this attack in order to provide secure real-time location transmission. Another important aspect to consider in ITSs implementation is timeliness, the privacy schemes should not be complex enough to cause delays during protection. Therefore, this aspect cannot be neglected while designing ITSs privacy algorithms.\\
In healthcare and medical systems, the most important design requirement is to provide extremely high level privacy along with maximum accuracy, because these systems are directly linked with lives of patients and concerned people. That is why there is no chance of considering any less that optimal trade-off between accuracy and privacy~\cite{newmbref25}. Similarly, because of presence of both type of (statistical and time-series) data in these systems, \revtwo{differential privacy mechanisms need to consider} instantaneous and global privacy both for real-time reporting, and query evaluation mechanisms respectively. Finally, in case of industrial IoT systems, high level privacy is required to secure statistical and time-series industrial data. Various IIoT systems are vulnerable to stealthy attacks that may cause certain privacy harms such as false data injection~\cite{newmbref26}. Differential privacy strategies in IIoT systems needs to overcome such stealthy attacks in order to make these systems resilient from adversary interference. \comst{Use case scenarios after integration of differential privacy in CPSs applications is presented in Fig.~\ref{fig:refdpcps}. Similarly, taxonomy diagram of differential privacy from CPS perspective is given in Table~\ref{taxonomycps}.} The detailed design requirements of differential privacy implementation in different CPSs applications is presented in Table~\ref{tab:designtable}.
}

\section{Differential Privacy in Energy Systems (Smart Grid)} \label{dprefes}

%--------------------- FIGURE ------------------------

\begin{figure*}[t]        
\centering
\includegraphics[scale = 0.65]{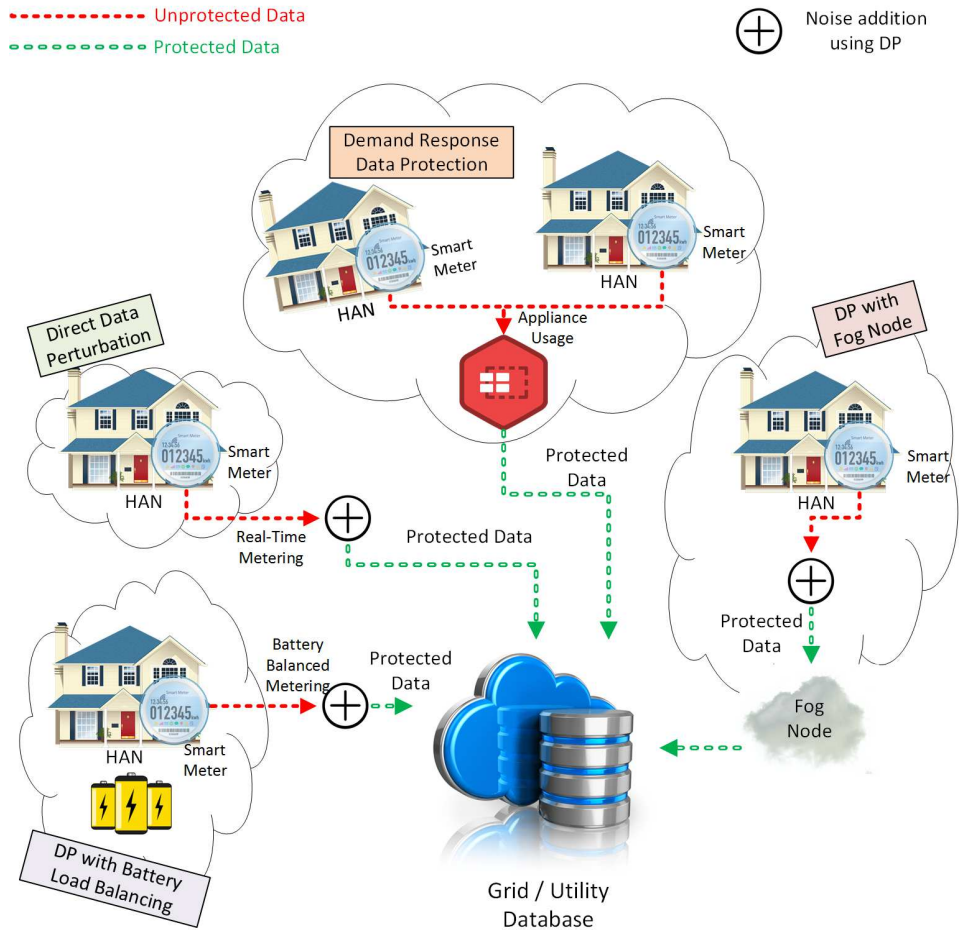}
 \caption{\small {Illustration of organization of differential privacy (DP) implementation in energy systems into four scenarios: direct data perturbation, battery load balancing, demand response protection, and DP with fog node. When DP is incorporated into these scenarios, real-time smart metering data of smart meter users is protected and the data stored by database can be used for query evaluation.}}  
  \label{fig:newgrid}   
\end{figure*}

%--------------------- FIGURE ------------------------

\begin{figure*}[]
     \centering

\begin{tikzpicture}

\node [block,  text centered, fill=orange!50, minimum width = 25em,  text width=25em] (a1) {Differential Privacy Approaches in Energy Systems \\ (Smart Grid)\\ Sec.~\ref{dprefes}};
\node[block, below of=a1,xshift = 4em , yshift=-4em, fill=red!30, minimum width = 15em,  text width=15em](b1){Smart Meter Load Monitoring \\ Sec.~\ref{dprefes02}};

\node [block, below of=b1, xshift = -5em, yshift=-2.5em, minimum width = 20mm,  text width=8em] (b1c1) {DP with \\ Battery Load\\ Balancing~\cite{sgref01},\\~\cite{sgref02},~\cite{sgref03},~\cite{sgref04}};

\node [smallblock, below of=b1c1, xshift = 4em, yshift=-2.5em,] (c1d1) {MAB\\ Algorithm ~\cite{sgref01},~\cite{sgref02}};
\node [smallblock, below of=b1c1, xshift = 4em, yshift=-7em] (c1d2) {State Aware\\ Scheme\\~\cite{sgref03},~\cite{sgref04} };

\node [block, below of=b1, xshift = 5em, yshift=-2.5em, minimum width = 20mm,  text width=08em] (b1c2) {Direct Data\\ Perturbation~\cite{sgref05},\\~\cite{sgref06},~\cite{sgref07},~\cite{sgref08}};

\node [smallblock, below of=b1c2, xshift = 11em, yshift=-2.5em, ] (c2d1) {Gamma\\ Distribution ~\cite{sgref05}};
\node [smallblock, below of=b1c2, xshift = 4em, yshift=-5em,] (c2d2) {\mubcom{Empirical\\ Model}\\~\cite{sgref06}};
\node [smallblock, below of=b1c2, xshift = 11em, yshift=-8em] (c2d3) {\comst{Gaussian\\ Noise ~\cite{sgref07},~\cite{comstref61} }};
\node [smallblock, below of=b1c2, xshift = 4em, yshift=-11em] (c2d4) {Laplacian\\ Noise \\~\cite{sgref08},~\cite{addpaper03}~\newline~\cite{comstref19} };
\node [smallblock, below of=b1c2, xshift = 11em, yshift=-14em] (c2d5) {$K$-means\\ Algorithm\\~\cite{addpaper05} };

\node[block, below of=a1, yshift=-4em, xshift = -20em,  text width=8em, fill=red!30 ](b2){Grid Demand Response\\ Sec.~\ref{dprefes01}};
\node[block, below of=b2, yshift=-3em, text width=8em ](b2c1){Used Appliances Data and Worked over Demand Response \cite{sgref09}};

\node[block, below of=a1, yshift=-4em, xshift = -10.5em, fill=red!30, ,  text width=08em ](b4){\comst{Smart Buildings\\ Sec.~\ref{dprefes04}}};
\node[block, below of=b4, yshift=-3em, text width=08em ](b4c1){\comst{Real-time Data\\ Analysis ~\cite{comstref47}, \cite{comstref21}}};

\node [smallblock, below of=b4c1, xshift = 4em, yshift=-3em,] (c1d4) {\comst{Sensors Data Analysis \\~\cite{comstref47}}};
\node [smallblock, below of=b4c1, xshift = 4em, yshift=-8em] (c1d5) {\comst{Internet Traffic Analysis~\cite{comstref21}}};

\node[block, below of=a1, yshift=-4em, xshift = 20em,  text width=10em,  fill=red!30 ](b3){Grid Data Collection using Fog Computing \\ Sec.~\ref{dprefes03}};
\node[block, below of=b3, yshift=-2em,  text width=10em, ](b3c1){Markov Model use for Noise Addition~\cite{sgref10}};

\path [line] (a1)-- ($(a1.south)+(0,-0.25)$) -| (b1);
\path [line] (a1)-- ($(a1.south)+(0,-0.25)$) -|(b2);
\path [line] (a1)-- ($(a1.south)+(0,-0.25)$) -|(b3);
\path [line] (a1)-- ($(a1.south)+(0,-0.25)$) -|(b4);

\path [line] (b1)-- ($(b1.south)+(0,-0.25)$) -|(b1c1);
\path [line] (b1)-- ($(b1.south)+(0,-0.25)$) -|(b1c2);

\path [line] (b1c2)-- ($(b1c2.south)+(0,-0.25)$) |-(c2d1.west);
\path [line] (b1c2)-- ($(b1c2.south)+(0,-0.25)$) |-(c2d4.west);
\path [line] (b1c2) -- ($(b1c2.south)+(0,-0.25)$) |-(c2d2.west);
\path [line] (b1c2) -- ($(b1c2.south)+(0,-0.25)$) |-(c2d3.west);
\path [line] (b1c2) -- ($(b1c2.south)+(0,-0.25)$) |-(c2d5.west);

\path [line] (b1c1)-- ($(b1c1.south)+(0,-0.25)$) |-(c1d1.west);
\path [line] (b1c1) -- ($(b1c1.south)+(0,-0.25)$) |-(c1d2.west);

\path [line] (b4c1)-- ($(b4c1.south)+(0,-0.25)$) |-(c1d4.west);
\path [line] (b4c1) -- ($(b4c1.south)+(0,-0.25)$) |-(c1d5.west);

\path [line] (b2)--(b2c1);
\path [line] (b3)--(b3c1);

\path [line] (b4)--(b4c1);

\end{tikzpicture}

	\small \caption{The differential privacy approaches implemented in energy systems (smart grid) can be mainly classified into real-time data monitoring, demand response, and combination with fog computing. }
     \label{fig:tnfig01}
\end{figure*}
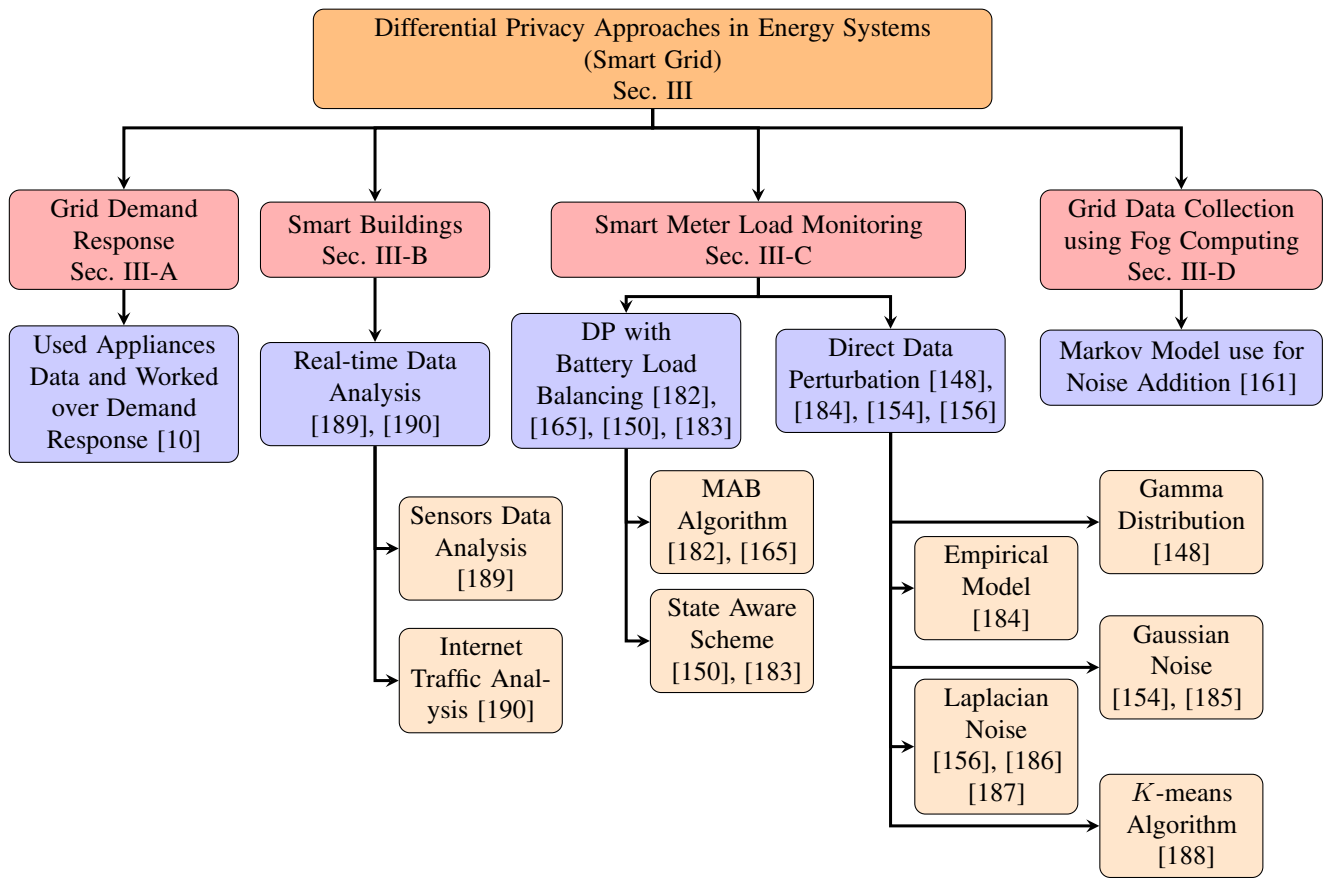

The term smart grid (SG) is traditionally used for electrical grid which is the enhanced version of power grid of 20th century~\cite{dpsg01}. Traditional forms of power grids are generally used to perform basic tasks of transmitting energy from generating station to customers. On the other hand, modern power grid or SG uses bidirectional flow of information and electricity~\cite{dpsg02}. \revi{This bidirectional communication in SG is usually carried out using modern communication technologies, such as ZigBee, Bluetooth, wireless LAN, powerline communication, optical networks, and cognitive radio communication~\cite{addpaper27}.} Because of this two-way flow of information and electricity, smart grid outperforms traditional energy grid by delivering energy in more efficient ways and by tackling large number of drawbacks of traditional power grid~\cite{dpsg01}. \\
Along with the benefits of SG, the efficient way of communication and SG data storage also paved the path towards certain security and privacy issues~\cite{newmbref12}. For instance, leakage of real-time energy reading of SG user can become a serious threat towards that individual’s personal life~\cite{dpsg03, comstref05}. Moreover, certain non-intrusive load monitoring (NILM) techniques have been designed to know exact appliance usage of any specific home during specific interval of time~\cite{dpsg04}. \comst{George W. Hart introduced the concept of NILM for the first time which further lead to the development of modern NILM techniques we have nowadays~\cite{comstref38}. NILM is referred as a technique to determine every minute detail about energy consumed inside the targeted area, e.g. usage of any specific appliance in a certain time slot can be extracted using such NILM techniques~\cite{comstref39}. Furthermore, the key steps included in NILM techniques can be termed as event detection, feature extraction, and load identification. Since 2010, researchers are actively working to develop more advanced NILM techniques to extract every minute detail from load data. However,} \revi{getting exact usage pattern of smart home appliances can cause certain privacy concerns for people living inside the house. For instance, any intruder can detect the routine of residents and can plan a theft, or advertising companies can do targeted advertisements by detecting the missing appliance in home.} Therefore, the privacy protection of SG users has always been the most crucial point among researchers working over SG. \mubcom{Majority of smart grid scenarios come under real-time data monitoring because smart grid devices are transmitting real-time data after a specific interval and therefore point-wise differential privacy protection is \revtwo{normally used to protect user's privacy} in these scenarios.} In order to protect this critical data, researchers have proposed numerous \mubcom{techniques to overcome} privacy issues in SG. \revi{Here, benefits and trends of differential privacy in SG have been discussed.} We divide differential privacy implementation in energy systems into three subclasses named as grid demand response, grid load monitoring, and grid data collection using fog computing, as illustrated in Fig.~\ref{fig:newgrid}. %Implementation of differential privacy in energy systems can be divided on the basis of various protection parameters, 
Furthermore, the detailed taxonomy of energy systems is given in Table~\ref{tab:sgtab01} and Fig.~\ref{fig:tnfig01}. In this section, we discuss four major scenarios of SG over which researchers have applied differential privacy.

\subsection{Grid Demand Response} \label{dprefes01}
One of the important goal of SG is to make energy use more efficient~\cite{dpsg21}. While, in order to obtain energy efficiency, management of volatile energy demands using scalable information is very important. The term demand side management (DSM) covers all aspects of demand response according to customer needs. DSM is quite important in operational cost reduction, elimination of blackouts, and reduction in emission of CO$_2$~\cite{dpsg22}. Generally, smart meter data is used for calculation of demand response. Contrary to this, if any intruder gets the high resolution demand response data, then this data can be used for various monitoring and unethical purposes~\cite{dpsg23}. Therefore, this data requires protection in order to secure individual private information~\cite{dpsg24}. As discussed in the prior section, the real-time data is protected using data perturbation strategy of differential privacy. But the calculation of demand response from this perturbed data is actual problem that arises. To resolve this problem, P. Barbosa~\textit{et al.}~\cite{sgref09} masked the data using Laplacian noise and after that worked over demand response calculation by dealing with individual appliance data. Similarly, they also evaluated two types of privacy attacks and showed that differential privacy is an optimal solution to overcome privacy risks. Therefore, the proposed differential privacy scheme efficiently protects demand response data by perturbing required features.

% ################ Flow Chart 1 #################

% ################ Flow Chart 1 #################

\subsection{\comst{Smart Buildings}}\label{dprefes04}

\comst{Recent facts and figures showed that more than 54\% of population of world is living in modern cities and urban areas, and it is predicted that by 2050 this ratio will reach  up to 66\%~\cite{buildref01}. This rapid increase in urban population has raised certain social, economic, organizational, and technical issues, which can be a harmful threat to economical and environmental situation of these areas. In order to overcome such situation, majority of governments are taking steps towards development and integration of “smart” concepts in all possible domains. Similarly, the concept of “smart city” refers to \mubcom{implification of all possible available resources and technologies} in a coordinated and intelligent manner with an aim to develop more sustainable and habitable urban centers~\cite{buildref02, comstref32}. One important sub block \yasir{that requires} considerable attention in such advancements are smart buildings. The concept of smart building refers to a modern building/home that is capable of performing measuring, controlling, monitoring, and optimizing operations without any external support. Smart buildings can further be classified into commercial and residential homes in which latest communication technologies are integrated to make them more suitable to live. Modern ICT technologies play an important role in such automation by providing a platform to carry out such real-time operations. The data stream from such smart buildings can be analysed to carry out certain automation tasks in order to make the city smart~\cite{buildref03}. For example, smart buildings can regulate their own heating and lighting based on the presence or absence of habitants~\cite{buildref04}, can monitor the quality and structural health of their own~\cite{buildref05}, and can make use of intelligent and smart appliance in order to automate daily tasks~\cite{buildref06}. In order to \mubcom{summarize, it can} be said that multiple number of sensors, actuators, and controllers installed in a smart building work together to provide a comfortable, and energy efficient \yasir{living to its habitants.} Despite of all these advantages, smart buildings are not 100 \% perfect and are prone to many security and privacy threats. Many service providers, involved parties, and third parties dealing with smart buildings at different stages have access to plenty of private data which can further be used to infer into personal information of its users~\cite{comstref48}. The data that can be used to attack privacy can be categorised into four sources: published data, observable data, leaked data, and repurposed data. The adversaries can use one of these four types of data to infer into personal life of habitants of smart buildings, which in turn can cause severe implications if not protected on time~\cite{buildref07}. One of the prospective method to preserve privacy of such data is to integrate differential privacy protection before storage or transmission of data. In this section, we discuss the integration of differential privacy in two major scenarios of smart buildings.}

% ===================== Smart Grid Table  ===========================

%This is related work Table
\begin{table*}[htbp]
\begin{center}
 \centering
 \scriptsize
 \captionsetup{labelsep=space}
 \captionsetup{justification=centering}
 \caption{\textsc{\\Comparative View of Differential Privacy Techniques in Energy Systems (Smart Grid) with their Specific Technique, Optimized Parameters, Privacy Criterion, Scenario, and Experimental Platform.}}
  \label{tab:sgtab01}
  \begin{tabular}{|P{1cm}|P{0.55cm}|P{0.5cm}|P{2.1cm}|P{2.7cm}|P{3cm}|P{1.1cm}|P{1.2cm}|P{0.85cm}|P{1cm}|}
  	\hline
\rule{0pt}{2ex}
\bfseries Main Category & \bfseries Ref No. & \bfseries Year & \bfseries Privacy Mechanism & \bfseries Technique of DP Used & \bfseries Enhancement due to Differential Privacy & \bfseries \centering Privacy \newline Criterion & \bfseries Platform Used &\bfseries \comst{Scenario} & \bfseries Time \newline Compl- \newline exity \\
\hline
\multirow{8}{*}{\parbox{2cm}{\centering \textbf{}}}%Real-Time Data Monitoring}}}

\rule{0pt}{2ex}
&~\cite{sgref05}& 2011 & DREAM (Differentially private Smart Metering) & Perturbed noise using Gamma distribution and encryption is used for aggregation & \tabitem Reduced cluster error \newline \tabitem Preserved appliances multiple slot privacy & \centering $\varepsilon$-differential privacy & Electricity Trace Simulator & \comst{Real-time} & $ O(n) $\\
\cline{2-10}

\rule{0pt}{2ex}
&~\cite{sgref06} & 2014 & Light weight privacy for smart metering data & Random masking value generated using Empirical model and error value & \tabitem Masked residential and industrial load profiles & \centering $ln$-mechanism privacy & N/A & \comst{Real-time} & $ O(n^2) $\\
\cline{2-10}

\rule{0pt}{2ex}
&~\cite{sgref03} & 2014 & Multitasking - BLH - Exp3 & Context aware battery load hiding strategy & \tabitem Enhanced mutual information sharing \newline \tabitem Optimized event detection accuracy & \centering $(\varepsilon, \delta)$-differential privacy & N/A & \comst{Real-time} & $O(log \; n)$ \\ \cline{2-10}

\rule{0pt}{2ex}
\centering \textbf{Smart Meter Load Monitoring} &~\cite{sgref04} & 2015 & Stateless and stateful privacy protection schemes & Proposed relaxed differential privacy strategy by adjusting noise distribution along with battery capacity & \tabitem Mutual information sharing optimized in differential capacities of battery & \centering $(\varepsilon, \delta)$-differential privacy & N/A & \comst{Real-time} & $- $\\
\cline{2-10}

\rule{0pt}{2ex}
&~\cite{sgref07} & 2015 & Precision-Privacy trade-off data perturbation technique for Smart Metering & White and colored Gaussian noise used for data perturbation & \tabitem Enhanced aggregated data privacy & \centering $\varepsilon$-differential privacy & N/A & \comst{Real-time} & $-$ \\
\cline{2-10}

\rule{0pt}{2ex}
& ~\cite{sgref01} & 2017 & Differential privacy battery supported meter reading & Differential privacy concept is used in conjunction with battery, and multi-armed bandit (MAB) algorithm  & \tabitem Reduces extra cost \newline \tabitem Reduces mutual information sharing & \centering $(\varepsilon, \delta)$-differential privacy & N/A & \comst{Real-time} & $ -$\\
\cline{2-10}
\rule{0pt}{2ex}
& ~\cite{sgref02} & 2017 & Cost friendly differential privacy (CDP) scheme & Differential privacy with battery load balancing and MAB algorithm  & \tabitem Optimized prices in both static and dynamic metering environment \newline \tabitem Reduces mutual information sharing & \centering $(\varepsilon, \delta)$-differential privacy & N/A & \comst{Real-time} & $O(log \; n)$  \\
\cline{2-10}
\rule{0pt}{2ex}
&~\cite{sgref08} & 2017 & Differential privacy for real smart metering data & Point-wise differential privacy with Laplacian noise is used  & Enhanced privacy and smoothing of signal & \centering $\varepsilon$-differential privacy & N/A & \comst{Real-time} & $-$\\
\cline{2-9}
\rule{0pt}{2ex}
&~\cite{addpaper03} & 2018 & \addp{Differential privacy based distributed load balancing for smart grid }& 3$\varepsilon$ based differential privacy using Laplacian noise  & Optimized efficiency, and fine grained reporting without trusted third party & \centering $\varepsilon$-differential privacy & Arduino micro-controller & \comst{Real-time} & $O(kn)$\\
\cline{2-10}

\rule{0pt}{2ex}
&\comst{~\cite{comstref61}} & \comst{2018} & \comst{Differentially private crypto-system based smart metering}& \comst{Perturbation and encryption based aggregation algorithm integrated with task assigning algorithm is used to protect user privacy } & \comst{\tabitem Blocked filtering attack
\newline \tabitem Prevented true value attack} & \centering \comst{$(\varepsilon, \delta)$-differential privacy} & \comst{ N/A } & \comst{Real-time} & \comst{$-$}\\
\cline{2-10}

\rule{0pt}{2ex}
&~\cite{comstref19} & 2018 & \addp{HIDE (Differential privacy for smart \mubcom{micro-grid} architecture)}& Privacy preserving data publishing differential privacy using greedy algorithm and Markov assumptions  & \tabitem Privacy-utility trade-off is minimized \newline \tabitem Max query, count query, and average query is enhanced & \centering $(\varepsilon, \delta)$-differential privacy & N/A & \comst{Real-time} & $-$\\
\cline{2-10}
\rule{0pt}{2ex}
&~\cite{addpaper05} & 2018 & \addp{Data clustering using differential privacy for intelligent electrical IoT}& K-means data clustering algorithm is combined with traditional differential privacy  & \tabitem Enhanced F-score
\newline \tabitem Enhanced clustering privacy & \centering $\varepsilon$-differential privacy & Java & \comst{Real-time} & $O(n)$\\
\cline{2-10}

\hline

\rule{0pt}{2ex}
\multirow{2}{*}{\parbox{2cm}{\textbf{}}}
\rule{0pt}{2ex}
\centering \textbf{\comst{Smart Buildings}}& \comst{\cite{comstref47}} & \comst{2017} & \comst{Differential privacy based monitoring of real-time sensors' data} & \comst{Perturbing, grouping, \& smoothing based differential privacy applied over sensors' streaming} & \comst{\tabitem Protected stream query privacy \newline \tabitem Protected sensors' event monitoring} &\centering { \comst{ $\varepsilon$-differential privacy}} & \comst{N/A} & \comst{Real-time} & \comst{-} \\
\cline{2-10}

\rule{0pt}{2ex}
& \comst{ \cite{comstref21}} & \comst{2018} & \comst{Differentially privacy private traffic obfuscation framework for smart community} & \comst{Utility-aware \& exponential differential privacy mechanism for gateway selection }& \comst{\tabitem Ensure unlinkability in Internet traffic \newline \tabitem Reduction in network resource consumption} & \centering \comst{$(\varepsilon, \delta)$-differential privacy} & \comst{N/A} & \comst{Real-time} & \comst{$O(n^2M)$ }\\
\cline{2-10}
\hline

\centering \textbf{Grid Demand Response} & ~\cite{sgref09} & 2017 & Differential privacy strategy to protect appliance usage in smart metering & Data masking using random Laplacian noise & \tabitem Enhances real-time data privacy \newline \tabitem Improved utility performance by demand response calculation & \centering $\varepsilon$-differential privacy & Simulators$^2$ in C  & \comst{Real-time} & $O(1)$\\
\hline

\centering \textbf{Grid Data Collection using Fog Computing} & ~\cite{sgref10} & 2018 & Differentially private data disclosure in smart grid & Factorial Hidden Markov Model (FHMM) is used to implement differential privacy & \tabitem Enhanced F-1 score \newline \tabitem Optimized kullback leibler divergence & \centering $\varepsilon$-differential privacy & NILMTK & \comst{Real-time} & $O(km)$ \\
\hline

 \end{tabular}
  \end{center}
\end{table*}
%This is related work Table

% ===================== Smart Grid Table  ===========================
\subsubsection{Sensors Data Stream}
\comst{One important feature of smart building is that they produce real-time environmental data from sensors in order to make effective predictions and calculations. However, leakage of this data can cause severe issues towards privacy of that building~\cite{comstref49}. In order to overcome this issue, integration of differential privacy with data of sensors before transmission came up as a viable solution. The authors in~\cite{comstref47} proposed PeGaSus mechanism that incorporates differential privacy over monitoring of real-time sensors’ data. Proposed technique use the concept of perturbing, grouping, and then smoothing of data to protect sensors streaming. The authors further worked over differential privacy based query evaluation for hierarchical streams. In order to evaluate the performance of their proposed strategy, the authors performed experiment over data from 4000 access points collected over a period of 6 months. Similarly, the authors preserved event monitoring, hierarchical aggregation, and different query responses by using differential privacy perturbation. The proposed strategy was over sensors streaming, however to make it more relevant to smart city and smart buildings, the authors presented a next step of this work in~\cite{comstref46}. In~\cite{comstref46}, the authors evaluated PeGaSus on real-world IoT scenarios for smart buildings and developed a tested and differential privacy engine for data stream. The authors concluded that the presented differential privacy based methodology successfully preserve sensors’ streaming privacy for smart building scenarios.}

\subsubsection{Home Traffic Analysis}
\comst{As discussed earlier in this section, residential smart buildings constitute an important part of network of smart buildings. Residential smart buildings also known as smart homes are capable of monitoring and controlling their energy and data flow and these smart homes \yasir{combine to} form a communication known as smart community~\cite{buildref08}. In smart homes, majority of devices are \yasir{connected to the Internet} for monitoring and controlling purposes, \revtwo{this connectivity helps to take timely decisions} via automating tasks. However, on the other hand this connectivity also raises certain threats that can be exploited by adversaries to carry out cyber-attacks on these homes and their residents. One such issue is highlighted by Liu~\textit{et al.} in~\cite{comstref21}, the authors stated that \yasir{the Internet traffic} from smart homes can cause leakage of private information. Furthermore, authors claimed that even cryptographic tools cannot protect data privacy because of effective machine learning algorithms used by adversaries. To tackle this issue, authors proposed a differentially private traffic obfuscation framework for smart homes in a smart community network. The authors proposed utility-aware and exponential differential privacy mechanism for gateway selection of Internet traffic. From this work, authors ensured that accumulated data from such traffic ensure \mubcom{unlinkability and} enhance privacy along with reduction in network resource consumption. The authors modelled this mechanism design as a linear optimization problem and proposed a differentially private strategy to overcome all mentioned issues. Finally, the authors carried out extensive simulation work to show that their algorithm enhanced privacy and reduced delay in a smart community network. Keeping in view this discussion, it is not hard to claim that differential privacy serves as a viable solution to protect privacy of smart homes.}

\subsection{Smart Meter Load Monitoring}\label{dprefes02}
One of the biggest hurdle in the implementation of SG is the privacy concerns of SG users~\cite{sgref01, comstref32}. The smart meters are connected with each other and the main electricity grid utility via strong communication network and they constitute to make a complex network named as advanced metering infrastructure (AMI). Smart meters send their energy utilization information to SG utility after a specific interval of time (e.g., 10 minutes), and if any adversary gets access to these reported readings, then this \mubcom{data can leak} sensitive information of smart meter user~\cite{comstref38, dpsg07}. This can lead to various threatening consequences; for instance, any burglar can detect the occupancy or un-occupancy schedule of a house before attempting any burglary, similarly, potential targets can be selected by vendors to market their campaigns.~\cite{sgref07}. Therefore, certain standardization bodies~\cite{dpsg08, dpsg09} and smart meter users require a privacy-friendly and secure framework for real-time monitoring of SG data that provides useful data to SG utility by keeping in view the confidentiality and privacy of users. \\
Extensive amount of literature \yasir{is presented on smart meter} data aggregation using encryption technologies, so that only the SG utility knows the exact usage information of users~\cite{dpsg10, dpsg11, dpsg12, dpsg13}. However, various implementations showed that encryption is hard to apply over real-time smart metering data because of the requirement of high computational capacity~\cite{dpsg14}. Another obstacle in application of encryption is the necessity of cooperation between all smart meters, because all smart meters have to exchange distributed keys and in case of failure of even a single smart meter, faults may arise in the whole network~\cite{sgref09}. Another popular privacy approach to protect smart metering data is anonymization~\cite{dpsg15}. One approach discussed the idea of hiding individual privacy by providing two different identities, one for billing and other for monitoring purpose~\cite{dpsg15}.  Similarly, the transmission of data using low-frequency and high-frequency ID is also proposed in the literature. \revtwo{However, this data can further be mined, and identification can be carried out using the anonymized data.} This certain piece of information can further be classified to a specific group of people in order to deduce their average behaviours~\cite{dpsg17}. \\
Therefore, protecting real-time data by adding desired amount of noise is one of the most desirable approach to transmit data without compromising privacy~\cite{dpsg18}. \mubcom{Implementing differential privacy in order} to protect real-time \yasir{data has} been employed by various researchers in the past on the basis of different SG scenarios, and different ways of addition of noise in the data. It can be classified into two major categories. One method is protection of real-time smart meter data by combining the advantages of differential privacy and battery load balancing, while the second one is direct perturbation of data using differential privacy, as shown in Fig. ~\ref{fig:tnfig01}.

\subsubsection{Differential Privacy with Battery Load Balancing}  
An effective way to protect smart meter data privacy is to balance the load by using an external battery, \revtwo{this technique is also known as battery load hiding (BLH).} But there are few major downsides of using direct BLH mechanisms that cannot be disregarded. For example, BLH techniques do lack theoretical discussion, as they are usually evaluated in context of their relative entry, regression, and clustering classification~\cite{sgref03}. However, there is no proved evidence to show their relevance directly to privacy protection. Therefore, in order to measure the exact protection and accuracy of privacy mechanisms, researchers used BLH schemes with differential privacy protection. 
In~\cite{sgref01}, the authors proposed a differential privacy based smart meter reading mechanism by perturbing the data without violating the limitations of battery. The authors worked over parameters of noise distribution and also combined multi-armed bandit (MAB) algorithm to further decrease the battery cost. The proposed techniques in~\cite{sgref02} enhanced the privacy loss in a battery based differential privacy scenario. Furthermore, the authors worked over reduction of cost under both dynamic and static pricing environment, and formulated two cost friendly approaches. The study in~\cite{sgref03} analysed previous BLH techniques by identifying their shortcomings, and then proposed a multitasking-BLH algorithm that successfully enhances the constraints of traditional BLH algorithms by optimizing event detection accuracy. The authors in~\cite{sgref04} first analysed the theoretical and practical challenges of differential privacy in BLH mechanisms, and then proposed stateless and stateful differential privacy BLH mechanisms in order to optimize mutual information sharing in different capacities of battery.
\vspace{3mm}

\subsubsection{Direct Data Perturbation} 
To protect smart meter user privacy, perturbing real-time smart metering data has been adopted by many researchers. Noise dimensioning is the most important factor to consider while perturbing the data directly. The correct amount of noise according to the scenario requirement makes differential privacy schemes more useful~\cite{sgref07}. Choice of noise addition parameter $\varepsilon$ cannot be neglected in this discussion, because it determines the level of privacy. Therefore, a vast amount of literature argues over choosing the most efficient $\varepsilon$ value~\cite{dpsg19}. Another parameter which is important to consider while selecting differential privacy for smart metering data is sensitivity. Generally, differential privacy techniques have been applied over counting data according to time-series~\cite{dpsg20}. In counting data, the \revtwo{sensitivity parameter is generally taken as 1}, while smart metering data cannot be depicted as counting data and the value of global sensitivity is not known, therefore the sensitivity cannot be directly taken same as of counting data~\cite{sgref08}.\\
In~\cite{sgref05}, real-time smart metering data is perturbed using Gamma distribution, and in order to make the aggregation secure, the authors used encrypted aggregation strategy. Furthermore, the proposed strategy reduced cluster errors in a SG scenario, and preserved appliance multiple slot privacy. The simulations of this proposed technique is performed using electricity trace simulator. The authors in~\cite{sgref06} proposed a light weight differential privacy approach that generated a random masking value based upon empirical model and error value. Analysis is carried out in the paper after masking residential and industrial profiles with differential privacy approach. Savi \textit{et al.} in~\cite{sgref07} first analytically derived $\varepsilon$ parameter to satisfy privacy guarantee in data aggregation, and then perturbed the data using Gaussian white and colored noise. They aggregated the data from different smart meters and at the end concluded that Gaussian colored noise provides a desirable level of privacy protection. \comst{Similarly in \cite{comstref61}, authors proposed a differentially private crypto-system based smart metering approach to preserve users privacy. In order to make perturbation more efficient and secure, the authors merged Gaussian noise based perturbation with task assigning algorithm and encryption. Furthermore, the authors analysed two privacy attacks named as filtering and time value attack and claimed that their proposed strategy efficiently protects smart meter users' privacy.} \\
In~\cite{sgref08}, authors analysed the trade-off of privacy and accuracy for real-time smart metering data and proved that differential privacy can be applied over real-time data and suitable advantages can be achieved using this approach. The authors further proposed the notion of point-wise privacy stating that the requirements of differential privacy in real-time data are different from differential privacy in statistical databases. \addp{Taking another step ahead, the authors in~\cite{addpaper03} proposed 3$\varepsilon$-differential privacy approach and analysed the outcomes using Arduino micro-controller. The practical output results showed that differential privacy optimized the efficiency value, and provided fine grained data reporting using Arduino even without need of any trusted third party. In~\cite{comstref19}, HIDE mechanism is proposed to address problem of privacy-utility trade-off in smart micro-grid scenario. The authors enhanced max query, count query, and average query along with using greedy algorithm, Markov assumption model, and Laplace noise for differential privacy. Contrary to traditional differential privacy approaches, authors in~\cite{addpaper05} introduced the concept of secure and private data clustering in intelligent energy systems using differential privacy. The authors proposed a light-weight secure clustering algorithm and tested the algorithm over different $\varepsilon$ values to optimize performance and privacy-utility trade-off accordingly.}

\subsection{Grid Data Collection using Fog Computing} \label{dprefes03}

Data collected from smart meters is usually aggregated and stored in data centres operating on cloud computing. \comst{During the transmission and storage, %the data is prone to any intruders' attack. Therefore, privacy needs to be ensured during dual-way communication of SG~\cite{dpsg25}. In this scenario, fog computing comes up as a viable solution to solve many data handling and privacy problems in SG communication up to some extent~\cite{dpsg26, dpsg27, newmbref11}. But
data may encounter delay and can decrease the response time. In order to overcome these issues, fog computing came up as a practical solution. Fog computing can be defined as a computing paradigm which was introduced to overcome burdens of data centers in traditional cloud technology. With time, fog computing emerged as a most viable solution to provide support to latency sensitive, geographically distributed, QoS aware applications of IoT~\cite{comstref36}. However, recent researches demonstrate privacy and security as the most important challenges for fog-computing based IoT applications. As fog computing nodes are not completely trusted and are vulnerable to certain threats and adversaries~\cite{comstref37}. Therefore, protecting privacy of data being communicated from fog-nodes is important.} For example,  what if a fog node aggregating the smart meter data gets compromised. To answer this question, researchers suggested usage of differential privacy along with fog computing in SG systems, to maintain the efficiency and privacy of data~\cite{sgref10, newmbref20}. One of the differential privacy approach considering fog computing in SG scenario has been implemented by H. Cao~\textit{et al.} in~\cite{sgref10}. The authors proposed factorial hidden markov model (FHMM) based differential privacy approach to aggregate data in fog nodes. Authors claimed that the proposed technique protects fog nodes data from any sort of NILM strategies. Furthermore, the given technique improved F-1 score~\cite{newmbref13} along with optimizing kullback leibler divergence in fog computing scenario. Energy consumption of every appliance is perturbed with a noise generated by FHMM, and the data is transmitted to fog node for storage purpose. Thus, this protected data can further be transmitted to analysts or the control centres to carry out certain DSM operations.

\subsection{Summary and Lessons Learnt}

The integration of ICT technologies in energy systems have paved the path for modern, intelligent, and secure energy, collectively named as SG~\cite{dpcps47}. \mubcom{However, plenty of issues still needs to be} resolved, and the most important of them is securing users' private data to maximum extent. On the other hand, utility also requires smart meter data for certain calculations, such as demand response, load forecasting, etc.  Researchers have proposed various strategies to overcome security and privacy issues of SG, including, encryption, battery load balancing, anonymization, and differential privacy. However, from above discussion, we can say that differential privacy provides a suitable solution to majority of SG scenarios. For example, when differential privacy is used with battery load hiding, it proves to be the backbone of mathematical analysis for BLH strategies. Similarly, the use of differential privacy in direct data perturbation showed that it also provides privacy protection at a specific instant of time by perturbing the instantaneous value of measured reading.~\comst{Correspondingly, differential privacy incorporated with smart buildings can efficiently preserve sensors’ \yasir{and the Internet} traffic data.}\\
Demand response calculation strategies do also require a specific level of privacy preservation to protect personal predicted data of users. Therefore, differential privacy comes up as a viable solution to protect demands response data. Likewise, along with \mubcom{privacy protection, efficiency, and speed is also required.} Therefore, the integration of differential privacy with fog computing paved the way for future secure energy systems. However, there are certain field of smart grid that still needs to be preserved using differential privacy. For instance, fault information and transmitting information needs to be preserved, in order to make it private from any intruder that needs to attack any specific damaged area. Similarly, load profiling information and meteorological data also needs critical attention in context of privacy. Moreover, preserving billing information along with maintaining dynamic pricing policy needs to be addressed in different SG scenarios. Furthermore, preserving the identity of buyer and seller using differential privacy must be considered for buying and auction of renewable energy resources (RERs) applications. Similarly,  firmware updates for smart meters needs to be carefully considered to protect the leakage of any specific software component of smart meters. 

%--------------------- FIGURE ------------------------

\begin{figure*}[t]        
\centering
\includegraphics[scale = 0.6]{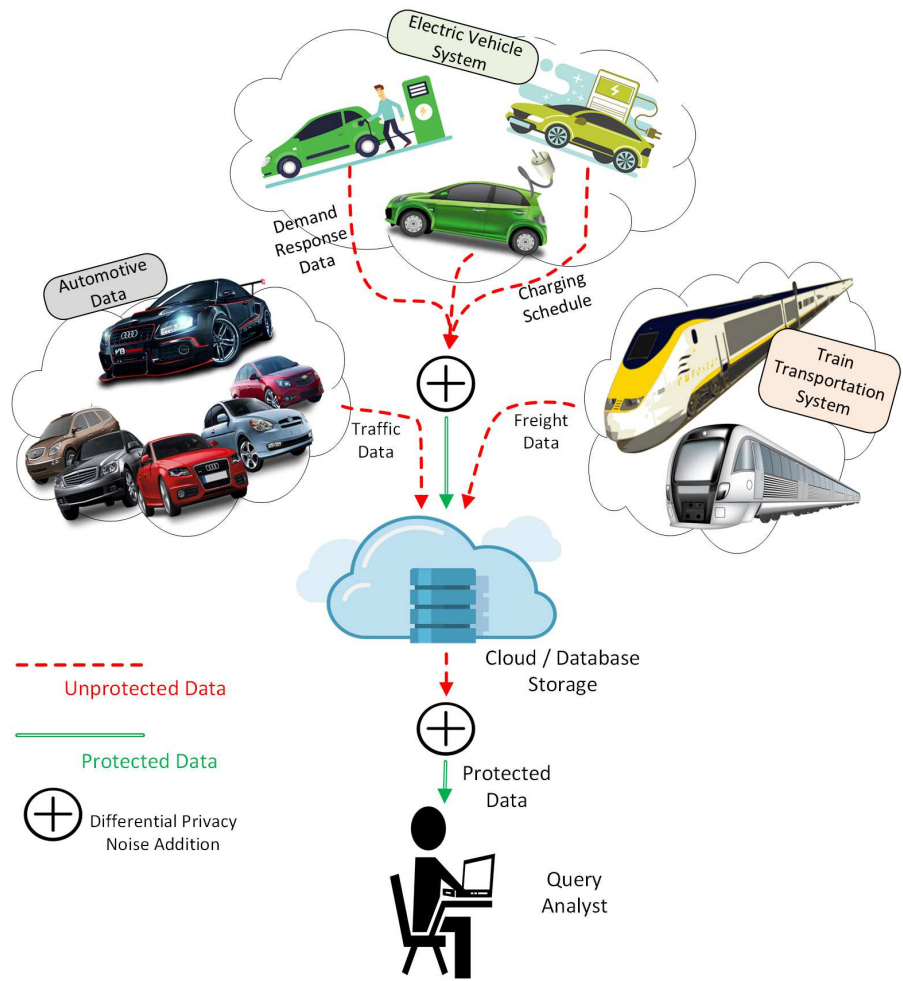}
 \caption{\small{ Illustration of organization of differential privacy (DP) implementation in transportation systems into three scenarios: electric vehicle, automotive data, and train transportation system. When DP is incorporated into these scenarios, automotive, demand response, charging schedule, and train systems data of transportation systems is protected and the data stored by database can be used for query evaluation. }}      
  \label{fig:newtransport}   
\end{figure*}

%--------------------- FIGURE ------------------------

% ################ Flow Chart 2 #################

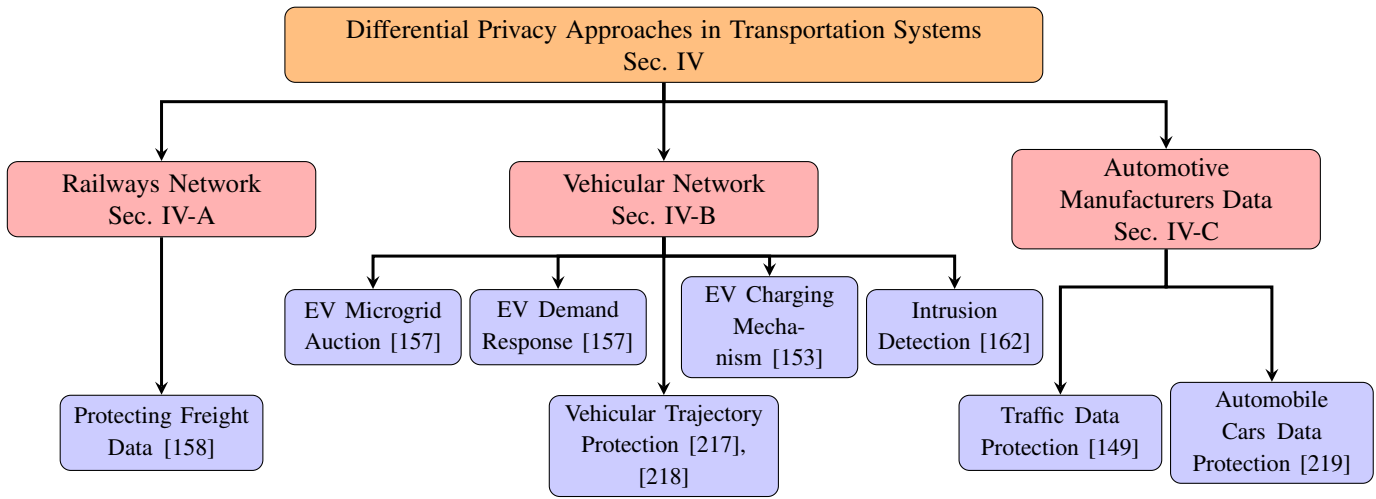
\begin{figure*}[]
     \centering

\begin{tikzpicture}

\node [block,  text centered, fill=orange!50, minimum width = 28em,  text width=28em] (a1) {Differential Privacy Approaches in Transportation Systems\\ Sec.~\ref{dprefts}};
\node[block, below of=a1, yshift=-3em, fill=red!30](b1){Vehicular Network \\ Sec.~\ref{dprefts02}};
\node [block, below of=b1, xshift = -4em, yshift=-2em, text width=6em, minimum width = 2mm] (b1c1) {\small EV Demand Response~\cite{tsref01}};
\node [block, below of=b1, xshift = -11em, yshift=-2em, text width=6em, minimum width = 2mm] (b1c2) {\small EV Microgrid Auction~\cite{tsref01}};
\node [block, below of=b1, xshift = 4em, yshift=-2em, text width=6em,minimum width = 2mm] (b1c3) {\small EV Charging Mechanism~\cite{tsref02}};
\node [block, below of=b1, xshift = 11em, yshift=-2em, text width=6em, minimum width = 2mm] (b1c4) {\small \mubcom{Intrusion} Detection~\cite{tsref03}};
\node [block, below of=b1, xshift = 0em, yshift=-6.5em, text width=8em, minimum width = 8mm] (b1c5) {\small \comst{ \revtwo{Vehicular Trajectory Protection~\cite{comstref58, rev2ref03}}}};

\node[block, below of=a1, yshift=-3em, xshift = -19em, fill=red!30 ](b2){Railways Network \\ Sec.~\ref{dprefts01}};
\node[block, below of=b2, yshift=-6em, text width=7em](b2c1){\small Protecting Freight Data~\cite{tsref04}};

\node[block, below of=a1, yshift=-3em, xshift = 19em, fill=red!30 ](b3){Automotive Manufacturers Data \\ Sec.~\ref{dprefts03}};
\node [block, below of=b3, xshift = -4em, yshift=-6em, minimum width = 2mm,  text width=7em] (b3c1) {\small Traffic Data Protection~\cite{tsref06}};
\node [block, below of=b3, xshift = 4em, yshift=-6em, minimum width = 2mm,  text width=7em] (b3c2) {\small Automobile Cars Data Protection~\cite{tsref05}};

\path [line] (a1)--(b1);
\path [line] (a1)-- ($(a1.south)+(0,-0.25)$) -|(b2);
\path [line] (a1)-- ($(a1.south)+(0,-0.25)$) -|(b3);

\path [line] (b1)-- ($(b1.south)+(0,-0.25)$) -|(b1c1);
\path [line] (b1)-- ($(b1.south)+(0,-0.25)$) -|(b1c2);
\path [line] (b1)-- ($(b1.south)+(0,-0.25)$) -|(b1c3);
\path [line] (b1)-- ($(b1.south)+(0,-0.25)$) -|(b1c4);
\path [line] (b1)-- ($(b1.south)+(0,-0.25)$) -|(b1c5);

\path [line] (b2)--(b2c1);

\path [line] (b3)-- ($(b3.south)+(0,-0.5)$) -|(b3c1);
\path [line] (b3)-- ($(b3.south)+(0,-0.5)$) -|(b3c2);

\end{tikzpicture}

	\small \caption{The differential privacy approaches implemented transportation systems can be divided into railways network, vehicular network and automotive manufacturer’s data privacy techniques.}
     \label{fig:tnfig02}
\end{figure*}

% ################ Flow Chart 2 #################

\section{Differential Privacy in Transportation Systems} \label{dprefts}

Transportation systems are advancing day by day, and the major purpose behind all these advancements is to provide improved services for riders and drivers in the system~\cite{dpits03, dpits04, dpits05}.  Since the beginning of 1970s, ITSs have been developing in various forms and are now considered to be the future of transportation systems~\cite{dpits01}. ITSs incorporate a large number of advanced technologies such as data transmission technologies, intelligent control technologies, and electronic sensing technologies into traditional transportation systems~\cite{dpits02}. In ITSs, every kind of transport (e.g., cars, trains, buses, etc.) is equipped with multiple wireless devices that are generating data for vehicle-to-device (V2D) and vehicle-to-vehicle (V2V) communications~\cite{dpits06}. \revi{ITSs communication is usually carried out using certain modern technologies, such as mobile ad hoc network (MANET), IEEE 1609 or dedicated short-range communication (DSRC), cognitive radio, and heterogeneous vehicular networks (HetVNET)~\cite{addpaper28, addpaper29}.} The participants of ITSs periodically share their vehicular information to communicate with other users in the network. This effective and accurate data is also used by various sources, in order to provide better and reliable services for ITSs~\cite{dpits07, dpits08}. Most of these V2V and V2D communication schemes rely on trustworthiness of sources receiving or aggregating their data~\cite{dpits09}. \\
However, the communication and storage of ITSs data is prone to many attacks and adversaries. For instance, an attacker can record the transmitted messages of any vehicle and can re-use these messages to get access to certain resources like compromising toll services, etc. Similarly, any false accident warning can be transmitted from a compromised vehicular network to block traffic from a certain highway. Moreover, the identity of a train, vehicle, or bus can be impersonated and can be used for unethical causes~\cite{dpits10}. Therefore, the privacy of ITSs data needs to be considered before implementing ITSs in our daily lives~\cite{tsref03}. Many privacy protection strategies have been implemented in the past to consider different scenarios of modern transportation systems. In this section, we divide differential privacy implementation in transportation systems into three subclasses named as railways freight networks, vehicular networks, and automotive manufacturer data, as illustrated in Fig.~\ref{fig:newtransport}.
%we will focus on the implementation of differential privacy in vehicular networks, railways networks, and automotive data protection. 
The taxonomy diagram for differential privacy in transportation systems is given in Fig.~\ref{fig:tnfig02}, and the summary table of literature work of differential privacy in transportation systems is provided in Table~\ref{tab:tstab01}.

\subsection{Railways Freight Network} \label{dprefts01}

Railways are considered as one of the most important mean of freight transportation in the world. The arrival of technology of big data in the railway freight system has brought various opportunities along with some challenges~\cite{dpits23}. Because of involvement of big data analytics technology, customers’ requirements, timeliness, and efficiency can be achieved in railways. However, this evolution also comes up with certain privacy and security risks. One of the major issue in the way of achieving efficiency through big data is data privacy and confidentiality~\cite{dpits24}. Privacy risks can arise in the way of sharing and communication of information, any adversary can try to attack the shared data and can get restricted freight information. This leakage of data can be hazardous to one’s personal privacy. Very few researches are carried out so far in order to protect privacy of railway freight systems~\cite{tsref04}. Previously researchers worked over proposing of game model to protect freight data. However, the proposed models only proved to be a macro solution, and no specific implementation was carried out. \mubcom{\revtwo{Since, railway networks deal with the passenger's data stored in their databases} and this data is further used to calculate the commute rate, hourly, daily, weekly, and monthly travel rates. Therefore, statistical differential privacy is applied to such scenarios.}\\
Similarly, in order to prevent leakage of private information of citizens, Yajuan~\textit{et al.} in~\cite{tsref04} proposed a differential privacy based correlation approach for railway freight systems. In~\cite{tsref04}, the authors first sliced the original service data to an optimal length, and then used apriori and differential privacy algorithm to introduce Laplace noise in the datasets of candidates. By following this method, they ensured that customer’s information is protected even if any adversary is successful in getting access to the background knowledge. Therefore, in context of railway networks, differential privacy strategies prove to be viable and efficient because they provide suitable solution without being computationally complex.

\subsection{Vehicular Networks} \label{dprefts02}
Modern vehicular networks made it possible for drivers or vehicles to communicate with surrounding vehicles or drives. In this way, vehicle is aware of its surroundings environment, which considerably improves on-boards services, and road traffic safety~\cite{dpits11}. For instance, the vehicles in a specific area can detect or expect dangerous situations in their way that may cause severe damages \mubcom{such as collisions or accidents.} As a consequence, the vehicles can take intelligent decisions in order to prevent themselves from such incidents~\cite{dpits12}. Although, this real-time information may also be exploited by any attacker or adversary for unauthorized tracking of vehicles' location~\cite{dpits13}. Generally, wireless medium is used for V2V and V2D communication \revtwo{that can easily be compromised, and broadcasts} can easily be eavesdropped by a passive adversary. This unethical eavesdropping data can be a serious threat to someone’s personal privacy. For instance, getting the information about the frequency of visits to a certain hospital can raise many doubts about the health of driver. Moreover, the life of driver can be put at a risky situation if the adversary eavesdropping the broadcast is a criminal~\cite{dpits06, dpits14}. Similarly, information about charging/discharging of EV and, auction information for EV in microgrids also needs to be protected, in order to prevent adversaries from monitoring the daily routine of EV users~\cite{tsref01}. We have divided the privacy protection in vehicular networks into two subcategories; protection of EVs auction and charging information, and preserving privacy of intrusion detection systems of vehicular networks. 

% ===================== Transportation System Table  ===========================

%This is related work Table
\begin{table*}[htbp]
\begin{center}
 \centering
 \small
  \captionsetup{labelsep=space}
 \captionsetup{justification=centering}
 \caption{\textsc{\\Comparative View of Differential Privacy Techniques in Transportation Systems with their Specific Technique, Optimized Parameters, Privacy Criterion, Scenario, and Experimental Platform.}}
  \label{tab:tstab01}
  \begin{tabular}{|P{1.6cm}|P{0.62cm}|P{0.55cm}|P{2.05cm}|P{2.15cm}|P{2.5cm}|P{1.25cm}|P{1.2cm}|P{1.1cm} |P{0.8cm}|}
  	\hline
\rule{0pt}{2ex}
\bfseries Main Category & \bfseries Ref No. & \bfseries Year & \bfseries Privacy Mechanism & \bfseries Technique of DP Used & \bfseries Enhancement due to Differential Privacy & \bfseries \centering Privacy \newline Criterion & \bfseries Platform Used & \bfseries \comst{Scenario} & \bfseries Time \newline Com- \newline plexity \\
\hline
\rule{0pt}{2ex}
\multirow{3}{*}{\parbox{2cm}{}}%\centering \textbf{Vehicular Network}}}

\rule{0pt}{2ex}
& ~\cite{tsref02} & 2015 & EV Charging truthful mechanism via differential privacy & Used exponential differential privacy along with drawing random vector from distribution  & \tabitem Enhanced truthfulness of privacy mechanism & \centering $(\varepsilon, \delta)$-differential privacy & N/A & \comst{Statistical Database} & $-$\\
\cline{2-10}

\rule{0pt}{2ex}
\centering \textbf{Vehicular Network} & ~\cite{tsref01} & 2017 & ExPO: Exponential based privacy preserving online auction & Exponential differential privacy is used  & \tabitem Demand response \newline \tabitem Improved peak load along with privacy preservation &  \centering $\varepsilon$-differential privacy  & MATLAB & \comst{Real-time} & $-$ \\
\cline{2-10}

\rule{0pt}{2ex}
& \comst{ ~\cite{comstref58}} & \comst{2017} & \comst{Differentially private vehicular trajectory protection algorithm} & \comst{Exponential differential privacy in collaboration with trajectory partition \& clustering algorithm is used}  & \comst{\tabitem Protected information loss \newline \tabitem Enhanced data utility \& efficiency}& \centering \comst{ $(\varepsilon, \delta)$-differential privacy}  & \comst{SUMO} & \comst{Real-time} & $-$ \\
\cline{2-10}

\rule{0pt}{2ex}
& ~\cite{tsref03} & 2018 & Machine learning based collaborative intrusion detection (PML - CIDS) to preserve privacy & Machine learning based differential privacy approach with dual variable perturbation  & \tabitem Enhancement of empirical risk & \centering $(\varepsilon, \delta)$-differential privacy  & N/A & \comst{Statistical Database} & $-$ \\
\cline{2-10}

\rule{0pt}{2ex}
& \revtwo{~\cite{rev2ref03}} & \revtwo{2019}  & \revtwo{Private real-time vehicular trajectory data release} & \revtwo{Laplacian perturbation \& Kalman filter based position protection for dynamically sampled data}  & \tabitem \revtwo{Enhanced accuracy} \newline \tabitem \revtwo{Reduced error rate} \newline \tabitem \revtwo{Enhanced data availability} & \revtwo{\centering $\varepsilon$-differential privacy}  & \revtwo{MATLAB} & \revtwo{Real-time} & $\revtwo{O(m^2)}$ \\
\cline{2-10}

\hline

\rule{0pt}{2ex}
\centering \textbf{Railways Network} & ~\cite{tsref04} & 2017 & Railway freight data correlation analysis using differential privacy & Laplacian noise is added using Apriori algorithm & \tabitem Improved privacy in freight data mining & \centering $\varepsilon$-differential privacy  & N/A & \comst{Statistical Database} & $-$ \\
\hline

\multirow{3}{*}{\parbox{2cm}{\centering \textbf{Automobile Data}}}

\rule{0pt}{2ex}
& ~\cite{tsref06} & 2013 & Differential privacy in intelligent transportation system (ITS) & Smooth sensitivity using Laplacian perturbation & \tabitem Preserved floating car data (FCD) in traffic data centres & \centering $\varepsilon$-differential privacy & N/A & \comst{Real-time} & $-$  \\
\cline{2-10}

\rule{0pt}{2ex}
& ~\cite{tsref05} & 2017 & Differential Privacy scheme in automotive Industry & Laplace, exponential, and randomized mechanisms are discussed  & \tabitem Protected personal identifiable information (PII) & \centering $(\varepsilon, \Delta)$-differential privacy & N/A & \comst{Statistical Database} & $-$ \\
\cline{2-10}

\hline

 \end{tabular}
  \end{center}
\end{table*}
%This is related work Table

% ===================== Transportation System Table  ===========================

\subsubsection{EV Auction and Charging Protection} 
As discussed earlier, the privacy protection of communication between \revtwo{EVs is an important issue that is currently being addressed} by many researchers. Similarly, the auction phenomenon of EVs while selling or buying energy also needs specific attention. In context of flexible storage, EVs can benefit demand response functionalities of energy grids. Particularly, EVs are designed in such a way that they can be charged in low electricity and in the time of need they can be discharged or \mubcom{can sell their energy} to other EVs or SG~\cite{dpits15}. Recently, swap stations are being introduced for charging of EVs. By using swap station technology, EVs charging speed is improved greatly~\cite{dpits16}. Various efforts have been put in by researchers on scheduling the charging of EVs from swap stations~\cite{dpits17}. \comst{Similarly, the discharging EVs sale out their surplus electricity to other EVs or smart homes and this trading is carried out using an auction process. This auction is usually a game-theoretic process in which selling price and incentives are determined after collecting bids and asks from buyers and sellers respectively~\cite{comstref41}.} However, the charging, discharging, auction, and demand response data of EVs needs to be preserved even from swap stations, because making this data public sacrifices the individual privacy of EVs~\cite{tsref01}. Wireless sensor networks and cloud computing is the general medium used for communication and storage between EVs and swap stations. Therefore, it is easy for any passive adversary to eavesdrop the information being transmitted~\cite{dpits18}.\\
Researches have been done in order to protect the auction privacy of EVs, few researchers suggested the use of cryptography to protect auction privacy~\cite{dpits19 }. Due to computational complexity and communication overhead, the \revtwo{performance of cryptographic strategies decreases.} However, the privacy protection mechanism of differential privacy emerged as one of the optimal solution to preserve individual privacy of auction based strategies of EVs~\cite{dpits20}. One of the most significant work in implementation of differential privacy in EV auction is carried out by Haitao~\textit{et al.} in~\cite{tsref01}. The authors used exponential based differential privacy to protect private information of EV owner from being compromised by adversary. They used auctioneer to maximize social welfare of market by matching sellers and buyers. The demand response calculated after using the given scheme improved the peak load requirement without compromising the privacy of EV users. Another noteworthy work in order to protect charging schedule of EVs via differential privacy is carried out by Shuo~\textit{et al.} in~\cite{tsref02}. The authors used the idea of joint differential privacy to limit power and involvement of users at the time of reporting their specifications for Online auction. The proposed strategy ensured that even if any EV misreport its specifications to mediator or swap station, it will not get much benefit from it, that in turn will lead to truthfulness. Moreover, differential privacy can provide secure bidding in conjunction with energy auction scenario of EVs by successfully controlling the information and only displaying the minimum required information. Thus, differential privacy easily surpasses other privacy preservation strategies in context of EVs data protection.

\subsubsection{\comst{Vehicle Trajectory Protection}}
\comst {In modern world it is predicted that every vehicle will be connected to Internet and every vehicle will be sharing its real-time information with the network in order to develop a seamless traffic and transportation system. However, this real-time reporting do also raises certain issues and vehicular trajectory leakage is most crucial among them. These trajectories can be used to predict behaviour of passengers that may further lead to leakage of personal life routine of that individual~\cite{comstref03, comstref11, comstref22}. This data can further be supplied to certain corporation and companies that may exploit this location data for their business purposes such as carpooling companies~\cite{comstref09}. Researches are being carried out to overcome this issue in the most efficient way in which the individual will be able to share its real-time location along with preserving its private information. One such effort is carried out in ~\cite{comstref58}, the authors developed a differentially private vehicular trajectory protection algorithm in which they explored and integrated exponential differential privacy protection with trajectory partition and clustering algorithm. Furthermore, the authors worked over preserving information loss along with enhancing efficiency and data utility of differentially private algorithm.} \revtwo{Similarly, another work which protects real-time user location is carried out by Ma~\textit{et al.} in~\cite{rev2ref03}. The authors first adopted dynamic sampling strategy in order to process real-time location data, and further used Kalman filter to ensure data availability. Afterwards, the authors used Laplacian perturbation of differential privacy to protect user data. By doing so, authors ensured that the protected data provides enough utility along preserving it from malicious adversaries. The output results presented by authors ensured that the privacy and data availability increased as compared to similar approaches.}

\subsubsection{Privacy of Intrusion Detection Systems}
Intrusion detection systems play a vital role to mitigate threats of vehicular networks by detection adversarial behaviour using signature based and/or anomaly based approaches~\cite{dpits21}. An advanced architecture of intrusion detection systems is collaborative intrusion detection systems (CIDS), that enable nodes to share the detected knowledge about attacks, and in return increase detection accuracy~\cite{dpits22}. \yasir{CIDS enable} the EVs to utilize labelled dataset of other vehicles; that speeds up the training process for each EV without burdening the EVs storage capacity. Moreover, the workload is also distributed among all EVs by sharing the laborious task of collecting labelling data. Data communication between EVs is not completely secure and can cause serious privacy threats to training data because of distributed environment. If any adversary is able to successfully extract the private information of EVs, then it can maliciously pretend to be some EV in the network which can observe its surrounding vehicles or it can also observe and manipulate the outcomes of learning process~\cite{tsref03}. Therefore, a privacy preserving mechanism for intrusion detection systems in EVs is important. In order to protect this real-time CIDS data, Tao~\textit{et al.} in~\cite{tsref03} proposed a differential privacy based machine learning CIDS approach that enhances empirical risk in the network by making the data private via dual variable perturbation. The authors in the paper first captured the privacy notation and then worked over dynamic differential privacy for data perturbation in machine learning scenario. Furthermore, the detection accuracy, design, and privacy-security trade-offs of CIDS in context of differential privacy are also considered and enhanced by the authors. From above discussion, it can be concluded that sharing data between modern electric vehicles can be made more secure and private using differential privacy strategies.

\subsection{Automotive and Manufacturer Data} \label{dprefts03}
As discussed earlier, the data from connected vehicles is prone to many security and privacy risks. When it comes to EVs, this problem of privacy protection needs to be tackled and solved by the manufacturing companies in order to maintain trust of drivers. One of the major aspect that manufacturers are considering the most is confidentiality of personally identifiable information (PII)~\cite{tsref05}. A common notion while working with privacy protection of PII is to remove the data that is linked with PII, such as names, tracking numbers, etc. However due to lack of any exact technical definition, it is surprisingly difficult to define and identify perfect PII~\cite{dpits25}. Moreover, Gao~\textit{et al.} in~\cite{dpits26} demonstrated that the speed of driving can be combined with road maps in order to trace the exact location of vehicle. Furthermore, Tockar~\cite{dpits27} showed that anonymized cab data of NYC combined with public data, revealed sufficient information to  detect celebrities and passengers that made visits to sensitive places within the city. By keeping in view the above discussion, we can say that it is also the responsibility of smart car manufacturers to keep the point of view of privacy in mind while designing modern cars.\\
One of the most promising privacy technique that can be implemented by manufactures to preserve privacy of individuals without damaging the original data is differential privacy~\cite{tsref05}. Boes~\textit{et al.} proposed the application of differential privacy for real-time automotive data. Furthermore, the authors discussed various types of noises (e.g., Laplace, exponential, and randomized mechanism) that can be used to perturb data according to manufacturer’s requirement. The authors in~\cite{tsref06} integrated differential privacy with policy-enforcement framework in order to protect floating car data storage in traffic data centres.  Moreover, they provided specific guidelines to ensure the specific privacy guarantee along with providing efficient data accuracy. Thus, differential privacy can prove to be a viable solution to solve certain privacy leakage problems according to manufacturer's point of view. As differential privacy efficiently adds noise in desirable PII, so users’ travelling in vehicles have control over the information they are sharing and that is how they can control their privacy according to the need.

%--------------------- FIGURE ------------------------

\begin{figure*}[t]        
\centering
\includegraphics[scale = 0.7]{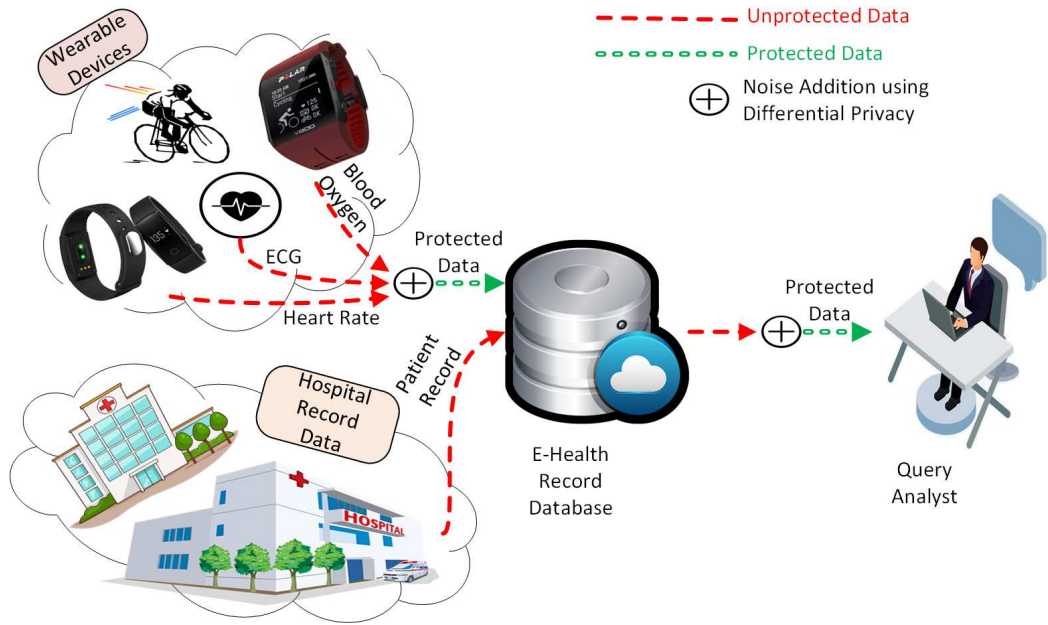}
 \caption{\small{Illustration of organization of differential privacy (DP) implementation in healthcare and medical systems into two scenarios: wearable devices, and hospital record database. When DP is incorporated into these scenarios, heart rate, ECG, blood oxygen level, and patient record data of healthcare and medical systems is protected and the data stored by database can be used for query evaluation.}}      
  \label{fig:newhealth}   
\end{figure*}

%--------------------- FIGURE ------------------------

\subsection{Summary and Lessons Learnt}
Network of connected devices in transportation system ensures reliable service, but it also comes up with certain privacy and security related issues. A major issue being faced in ITSs is privacy leakage of individual identity of EV users. The adversary can compromise a communication channel by the help of passive attacks. Therefore, proper privacy preservation of data being transmitted throughout the network should be maintained. Researchers proposed differential privacy as one of the optimal solution to overcome these real-time and database privacy issues. But still plenty of issues in transportation system needs to be resolved.\\
An important application of transportation systems is modern EVs, which improve reliability, safety, and security in every perspective~\cite{dpits28}. These EVs can be charged at low power and can be used in case of power shortage or failure via discharge process~\cite{dpitsnew01}. However, this modernization comes up with certain privacy issues. For instance, the real-time location, battery status, and charging/discharging schedule reveal a large number of personal information of EV users. Therefore, the most successful approach presented by researchers to prevent this privacy leakage is differential privacy. The above discussion mainly focuses over two major aspects of EVs; charging/discharging protection, and protection of intrusion detection strategies. It is found out that by considering the advantages of differential privacy in EVs, we can publicise the data of EV without worrying about the privacy of individuals.\\
The privacy control according to Manufacturers’ point of view is also discussed in this section along with specifically mentioning the protection of PII in modern vehicles.  Because certain experiments have been conducted by researchers in which they combined two anonymized datasets and successfully achieved the precise information about individual activities. However, if we preserve the data using differential privacy, then the anomaly will not be able to break in to the privacy. Similarly, differential privacy is also implemented in railways freight network to protect customers’ information for data mining. \\
However, many fields of transportation system \revtwo{still need consideration} in order to protect them from anomalies, first and foremost of them is device-to-device (D2D) communication. Whenever, devices in a transportation networks are communicating, they are sharing a considerable amount of personal data that can be a threat to someone’s privacy in case of an attack. Therefore, research efforts to protect D2D via differential privacy needs to be made. Furthermore, securing the storage of big data of ITSs also needs more attention in order to remove any sort of confusion or privacy concern from minds of ITSs users or customers. Similarly, live traffic information also needs to be preserved in order to disrupt any adversary from tracking the lifestyle of any EV user. Moreover, \revtwo{researchers need to focus} about enhancement and protection of privacy in V2V communication as well, because a large number of V2V applications may have crucial significances in case of privacy leakage.

\section{Differential Privacy in Healthcare and Medical Systems} \label{dprefhms}

\begin{figure*}[]
     \centering

\begin{tikzpicture}

\node [block,  text centered, fill=orange!50, minimum width = 25em,  text width=25em] (a1) {Differential Privacy Approaches in Healthcare \\ (Medical Systems) \\ Sec.~\ref{dprefhms}};
\node[block, below of=a1, yshift=-4em, fill=red!30](b1){Electronic Health Record\\ (E - Health)\\ Sec.~\ref{dprefhms02}};
\node [block, below of=b1, xshift = -9.5em, yshift=-3em, minimum width = 2mm, text width=6em] (b1c1) {Improving Processing Speed};

\node [smallblock, below of=b1c1, xshift = -4em, yshift=-3.5em, minimum width = 2.5mm, text width=6em] (b1c1d1) {\comst{Addressing \\ Hierarchical \\ Queries~\cite{hmref02}}};
\node [smallblock, below of=b1c1, xshift = 4em, yshift=-3.5em, minimum width = 2.5mm, text width=6em] (b1c1d2) {\comst{Machine Learning \\~\cite{comstref30, comstref17}}};

\node [block, below of=b1, yshift=-3em, minimum width = 2mm, text width=6em] (b1c2) {Addressing Range Queries~\cite{hmref03}};
\node [block, below of=b1, xshift = 9.5em, yshift=-3em, minimum width = 7mm, text width=7em] (b1c3) {Encryption, Crypto System, \\ and \\ SQL Queries};

\node [smallblock, below of=b1c3, xshift = -7.5em, yshift=-3.5em, minimum width = 2.5mm, text width=6em] (b1c3d1) {Cancer Data\\~\cite{hmref04}};
\node [smallblock, below of=b1c3, xshift = 0em, yshift=-3.5em, minimum width = 2.5mm, text width=6em] (b1c3d2) {Genomic Data\\~\cite{addpaper15}, ~\cite{addpaper14}};
\node [smallblock, below of=b1c3, xshift = 7.5em, yshift=-3.5em, minimum width = 2.5mm, text width=6em] (b1c3d3) {Secure Incentivization \\~\cite{rev2ref04}};

\node[block, below of=a1, yshift=-4em, xshift = -20em, fill=red!30 ](b2){Real-Time Health Data \\ Sec.~\ref{dprefhms01}};
\node[block, below of=b2, yshift=-3em, ](b2c1){From Wearable Devices~\cite{hmref01}};

\node[block, below of=a1, yshift=-4em, xshift = 20em, fill=red!30 ](b3){Health Survey Data Protection \\ Sec.~\ref{dprefhms03}};
\node[block, below of=b3, yshift=-3em, ](b3c1){Users Perspective~\cite{hmref05}};

\path [line] (a1)--(b1);
\path [line] (a1)-- ($(a1.south)+(0,-0.25)$) -|(b2);
\path [line] (a1)-- ($(a1.south)+(0,-0.25)$) -|(b3);

\path [line] (b1)-- ($(b1.south)+(0,-0.25)$) -|(b1c1);
\path [line] (b1)-- ($(b1.south)+(0,-0.25)$) -|(b1c2);
\path [line] (b1)-- ($(b1.south)+(0,-0.25)$) -|(b1c3);

\path [line] (b1c1)-- ($(b1c1.south)+(0,-0.25)$) -|(b1c1d1);
\path [line] (b1c1)-- ($(b1c1.south)+(0,-0.25)$) -|(b1c1d2);

\path [line] (b1c3)-- ($(b1c3.south)+(0,-0.25)$) -|(b1c3d1);
\path [line] (b1c3)-- ($(b1c3.south)+(0,-0.25)$) -|(b1c3d2);
\path [line] (b1c3)-- ($(b1c3.south)+(0,-0.25)$) -|(b1c3d3);

\path [line] (b2)--(b2c1);
\path [line] (b3)--(b3c1);

\end{tikzpicture}

	\small \caption{The differential privacy approaches implemented health care and medical systems can be categorized into real-time health data, electronic health data, and survey data record strategies.}
     \label{fig:tnfig03}
\end{figure*}
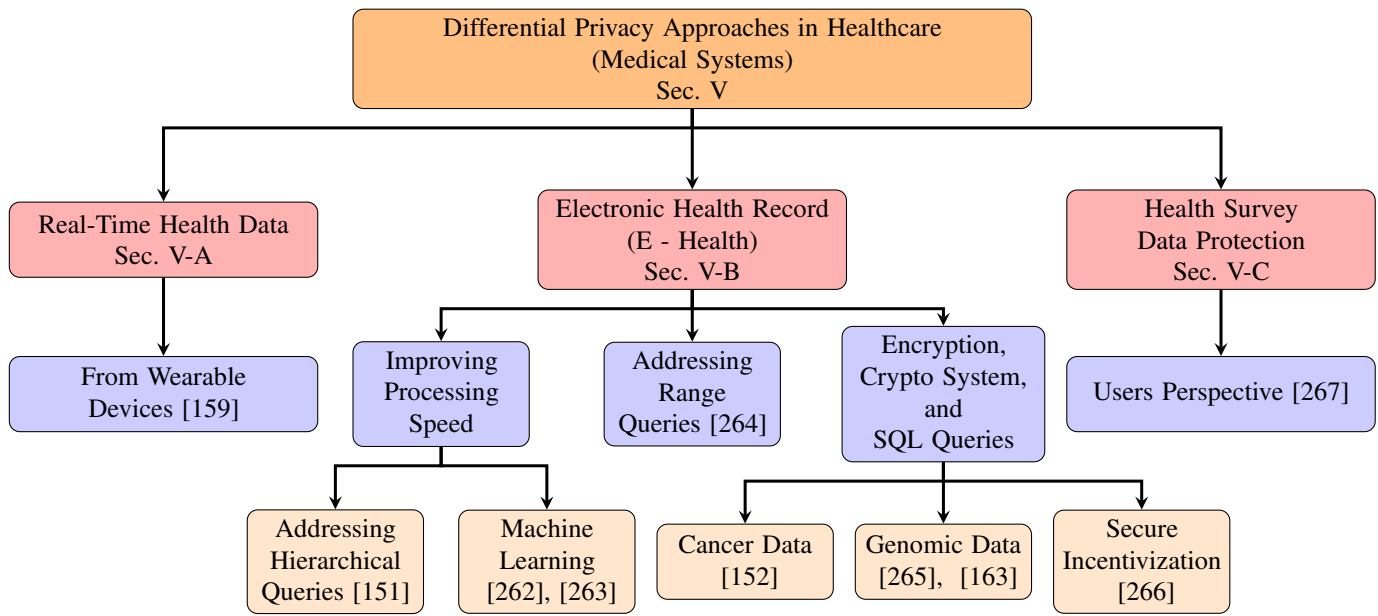

One of the most attractive application of connecting cyber and physical world is healthcare and medical systems~\cite{dphms01}. This connection \revtwo{has a great potential in CPSs, and it gives rise} to many healthcare applications such as real-time health monitoring, fitness programs, remote health monitoring, and elderly care. Medication and treatment from distant places or homes is another potential application of this connection~\cite{dphms02}. Similarly, storage of health records using big data, and performing data analytic surveys for better diagnosing of disease at early stage is also under development phase. Therefore, the healthcare and medical systems are considered to be one of the core part of CPSs. These modern healthcare systems surpass traditional healthcare systems by improving time, cost, and quality of life. Furthermore, the modernization of these systems provides efficient scheduling of finite resources by assuring their most efficient use.\\
One major issue to consider in healthcare systems is the timely measurement and diagnosis of critical factors for treatment of disease. In majority of cases, the late diagnosis leads to dangerous chronic diseases, certain advanced cancer stages, or even death in some cases~\cite{dphms03}. Another important factor to consider in healthcare and medical CPSs practical implementation is its privacy preservation, because even a minor privacy threat can risk someone’s life~\cite{dphms04}. \addp{Majority of these healthcare and medical devices are connected via wireless networks for data reporting and transmission~\cite{addpaper26}.} \revi{This timely measurement and reporting requires seamless communication infrastructure. Generally in healthcare CPSs, 4G long-term evolution (LTE), ultra-narrow band (UNB), ingenu, and low power wide area (LPWA) technologies are used to carry out communication~\cite{addpaper31}. These technologies transmit real-time health data by causing minimum delays.} The medical data contains specific patterns for real-time or e-health monitored data, and these patterns should be protected with certain privacy control because they are directly linked with someone’s personal life. For instance, date of appointment from a specific doctor, health insurance ending date, a specific glucose level in the body, \yasir{diagnosis of any} specific disease, etc. If any intruder gets access to this real-time data, then it can directly or indirectly have an effect on the life of the patient.\\
Researchers have proposed many privacy protection techniques in the past for various applications of healthcare and medical systems. For example, encryption, and data perturbation for real-time data, key-agreement, and anonymization for e-health data sets, etc. However, in this section we divide differential privacy implementation in healthcare and medical systems into three subcategories named as real-time health data, electronic health record, and health survey data protection, as illustrated in Fig.~\ref{fig:newhealth}. 
%we discuss the implementation of differential privacy in real-time health data, e-health records, and health survey data. 
The taxonomy diagram for differential privacy in healthcare and medical systems is given in Fig.~\ref{fig:tnfig03}, and the summary table of literature work of differential privacy in healthcare and medical systems is provided in Table~\ref{tab:hmtab01}.

% ################ Flow Chart 3 #################

% ################ Flow Chart 3 #################

% ===================== Health and Medical Table  ===========================

%This is related work Table
\begin{table*}[htbp]
\begin{center}
 \centering
 \footnotesize
  \captionsetup{labelsep=space}
 \captionsetup{justification=centering}
 \caption{\textsc{\\Comparative View of Differential Privacy Techniques in Health and Medical Systems with their Specific Technique, Optimized Parameters, Privacy Criterion, Scenario, and Experimental Platform.}}
  \label{tab:hmtab01}
  \begin{tabular}{|P{1.05cm}|P{0.6cm}|P{0.5cm}|P{2.05cm}|P{2.3cm}|P{2.6cm}|P{1.2cm}|P{1.1cm}|P{1.1cm}|P{1.33cm}|}
  	\hline
\rule{0pt}{2ex}
\bfseries Main Category & \bfseries Ref No. & \bfseries Year & \bfseries Privacy Mechanism & \bfseries Technique of DP Used & \bfseries Enhancement due to Differential Privacy & \bfseries \centering Privacy Criterion &\bfseries Platform Used & \bfseries Scenario& \bfseries Time \newline Compl- \newline exity \\
\hline

\rule{0pt}{2ex}
\centering \textbf{Real-time Health Data} & ~\cite{hmref01} & 2017 & Real-time health data releasing scheme (Re-Dpoctor) & Data perturbation is used along with adaptive sampling and filtering & \tabitem Mean absolute error and mean relative errors are enhanced & \centering $(\varepsilon, \delta)$-differential privacy & PIP Controller & \comst{Real-time} & $O(n)$ \\
\hline

\multirow{8}{*}{\parbox{2cm}{}}
\rule{0pt}{2ex}
& ~\cite{hmref02} & 2015 & Efficient E-health data release & Heuristic hierarchical query method and private partition algorithm proposed for DP  & \tabitem Enhanced time, overhead, and query error & \centering $(\varepsilon, \Delta)$-differential privacy & N/A & \comst{Statistical Database} & $O(n)$ \\
\cline{2-10}

\rule{0pt}{2ex}
&~\cite{hmref04} & 2015 & Private and Secure management of databases of health care database & Used Laplace mechanism for data privacy & \tabitem Reduced computational overhead & \centering $(\varepsilon, \Delta)$-differential privacy & Prototype in Java16 using Big Integer & \comst{Statistical Database} & $O(N log^2 N)$ \\
\cline{2-10}

\rule{0pt}{2ex}
\centering \textbf{Electronic Health \newline Record \newline Privacy} & ~\cite{hmref03} & 2017 & Health data differential privacy algorithm for range queries & Partitioning by data and work load are implemented with use of Laplacian noise  & \tabitem Optimized error rate of queries & \centering $\varepsilon$-differential privacy & N/A & \comst{Statistical Database} & $-$\\
\cline{2-10}

\rule{0pt}{2ex}
& ~\cite{addpaper14} & 2018 & \addp{ MedCo (Privacy preservation of genomic and distributed clinical data)} & Encryption in combination with differential privacy is used to secure and preserve sensitive data  & \tabitem Enhanced i2b2 database privacy \tabitem Optimized runtime, and network over head & \centering $\varepsilon$-differential privacy & Postgre- \newline SQL & \comst{Statistical Database} & $-$\\
\cline{2-10}

\rule{0pt}{2ex}
& ~\cite{addpaper15} & 2018 & \addp{ Genomic data privacy protection } & Protecting encrypted data using differential privacy and two step decryption  & \tabitem Enhanced execution time \tabitem preserved secret keys leakage in dual decryption & \centering $\varepsilon$-differential privacy & NFLib & \comst{Statistical Database} & $-$\\
\cline{2-10}

\rule{0pt}{2ex}
& \comst{~\cite{comstref17}} & \comst{2018} & \comst{End-to-end differentially private deep learning health record protection} & \comst{Differentially private stochastic gradient descent based deep learning method} & \comst{\tabitem Enhanced training accuracy} \newline \comst{\tabitem Improved computational cost} & \centering \comst{$(\varepsilon, \delta)$-differential privacy }&  \comst{N/A} & \comst{Statistical Database} & \comst{$-$}\\
\cline{2-10}

\rule{0pt}{2ex}
& \comst{~\cite{comstref30}} & \comst{2019} & \comst{Differentially private data clustering (EDPDCS) framework for medical data} & \comst{K-means clustering based differentially private machine learning over MapReduce} & \comst{\tabitem Optimized privacy allocation budget \newline \tabitem Improved learning accuracy} & \centering \comst{$(\varepsilon, \delta)$-differential privacy }&  \comst{Hadoop} & \comst{Statistical Database} & \comst{$-$}\\
\cline{2-10}

\rule{0pt}{2ex}
& \revtwo{~\cite{rev2ref04}} & \revtwo{2019} & \revtwo{Secure E-Health data aggregation with fair incentives} & \revtwo{Combined local differential privacy with Boneh-Goh-Nissim crypto system \& Shamir's secret sharing}  & \tabitem \revtwo{Improved key generation overhead} \newline \tabitem \revtwo{Aggregation privacy} & \revtwo{ \centering $\varepsilon$-differential privacy }& \revtwo{Java (JPBC library)} & \revtwo{Real-time} & $\revtwo{O(\sqrt{t})}$ \\
\cline{2-10}

\hline

\rule{0pt}{2ex}
\centering \textbf{Health Survey Data Protection} & ~\cite{hmref05} & 2018 & Privacy-Utility trade-off in health record systems & K-Anonymity and random data perturbation discussed & \tabitem Discussed and improved survey data according to users perspective & \centering - & N/A & \comst{Statistical Database} & $-$ \\
\hline

 \end{tabular}
  \end{center}
\end{table*}
%This is related work Table
% ===================== Health and Medical Table  ===========================

\subsection{Real-Time Health Data} \label{dprefhms01}
With the rapid expansion of wireless devices in our daily lives, the way we deal with our health is also changing. Real-time health data is being reported to doctors or databases to keep track of user behaviour and activities~\cite{hmref01}. \mubcom{For example, data of heart rate, sleep conditions, blood pressure, walk steps can be shared with doctor, hospital, or with insurance companies.} However, the disclosure of unnecessary data can lead to severe privacy concerns~\cite{dphms05}. While sharing the health data two things are generally considered as first priority,~\textit{(i)} utility (usefulness of data) and ~\textit{(ii)} privacy \yasir{(disclosure of less} than a certain privacy budget). One of the major source of real-time health data are wearable medical devices; that can be defined as non-invasive, and autonomous devices designed to perform any specific medial function such as health data monitoring~\cite{dphms06}. Vital signs of patients such as blood pressure, heart rate, body fat, blood oxygen level, and respiration are constantly being measured and monitored to be aware of any upcoming undesirable situations. Similarly, athletes also use such wearable medical devices in order to measure their calorie burn, pace, heart rate, and speed during exercise and report it to their coaches. This data contains specific patterns, which may provide critical information about any individuals’ health life. However, if this personal data of any patient or athlete gets stolen by any adversary, then the respective individual may face severe health circumstances.\\
Several methods including encryption~\cite{dphms07}, limitation and participation restrictions (privacy by design)~\cite{dphms08} have been proposed by researchers to preserve this crucial data. But none of the proposed method ensures complete privacy protection, because encryption protocols are computationally complex, and the restriction \mubcom{methods have loophole regarding \yasir{definition}} of exact PII. However, differential privacy emerged out to be one of the possible and most viable solution to protect real-time wearable medical devices data. Zhang~\textit{et al.} in ~\cite{hmref01} proposed \textit{Re-DPoctor} scheme to provide budget allocation and adaptive sampling using differential privacy. The proposed strategy satisfies all conditions of differential privacy and reduces mean relative error and mean absolute error of the transmitted data. Moreover, the authors used \yasir{proportional-integral-plus (PIP)} controller and compared utility and privacy trade-off by applying differential privacy over real-time health data. Keeping in view the above discussion, it can be seen that differential privacy can provide a healthy trade-off between privacy and accuracy for real-time health data. As the mathematical models of differential privacy can efficiently be used for data protection by adding desired value of noise, therefore including differential privacy approach in wearable devices can preserve PII to maximum extent. Thus, we can conclude that data perturbation using differential privacy is the most suitable solution, if someone wants to preserve their personal privacy for real-time healthcare and medical systems.

\subsection{E-Health Records} \label{dprefhms02}
Over the past decade, the trend of hospitals adopting electronic way of storing patient records has increased dramatically~\cite{dphms09}. This specific mechanism named as e-health mechanism~\cite{dphms10} integrates advanced ICT features, such as electronic storage, and data outsourcing. This health data contains PII, such as date of birth, presence of any specific disease, medical symptoms, weekly or monthly heart rate, blood pressure level, etc. Data stored in PII datasets is extremely sensitive and should not be disclosed to anyone else except the patient and doctor. Typically, these datasets are protected using obscuring or anonymizing methods during data preparation and cleaning. In obscuring, identifiers such as the quasi identifiers, key identifiers, and certain other types of sensitive and personal data identifiers are masked, after that a separate and protected dataset is prepared for mining~\cite{dphms11, dphms12}. However, these cleaned and protected datasets can easily expose certain PII when they are analysed and mixed with different other feature sets~\cite{dphms13}. Another valuable scheme used to preserve the confidentiality of e-health data is encryption. In this scheme, the data is protected using public and private generated keys. But the major challenge in encryption is to make sure the confidentiality of encrypted data while allowing query execution over it~\cite{hmref04}. Keeping in view all these points, the most suitable scheme that comes up to protect e-health data is differential privacy. \\
Using differential privacy data perturbation algorithm, one can publicize e-health data for query execution without compromising any sort of privacy~\cite{addpaper60}. Similarly, majority of differential privacy algorithms do not have high computational complexity. Therefore, differential privacy can be implemented in basic level e-health databases. \mubcom{Furthermore, e-health records purely deal with statistical data preservation because these health records are further used by various organizations and hospitals to enquire and predict about status of patient. Therefore, statistical differential privacy is usually applied to such electronic health records.} Researches have been carried out till now in order to implement differential privacy in various e-health databases for query execution. Li~\textit{et al.} in~\cite{hmref02} first developed a heuristic hierarchical query method, and then proposed a private partition algorithm for differential privacy in order to enhance time, overhead and query error. \comst{Similarly, the authors in~\cite{comstref17} worked over end-to-end differential privacy based deep learning approach to enhance training accuracy and efficiency. The authors proposed a private stochastic gradient descent based deep learning approach that preserves privacy \yasir{via efficiently} perturbing the clinical data. To secure it further, the authors integrated this differential privacy strategy with cryptographic encryption. Authors claimed that their developed mechanism is private, secure, and efficient as it efficiently protects privacy and security along with reducing computational cost.
Another work over preserving privacy while enhancing learning accuracy is carried by authors in~\cite{comstref30}, the authors proposed a differential privacy based data clustering algorithm that works over k-means clustering and protects private data by integrating differential privacy with machine learning. Furthermore, the presented framework operated over Hadoop and efficiently optimize privacy allocation budget along with improving learning accuracy.} Similarly, the authors in~\cite{hmref03} preserved privacy using Laplacian noise and worked over data partitioning and work load for optimization of error rate of queries. Furthermore, Mohammed~\textit{et al.} in~\cite{hmref04} used the Laplace noise of differential privacy to enhance data privacy by performing experiments over \revtwo{cancer patient's data.} The authors reduced the computational overhead by developing a lightweight framework that supports complex data mining tasks and a variety of SQL queries. \addp{The field of genomic data record protection is explored by researchers in~\cite{addpaper14, addpaper15}. In~\cite{addpaper14}, authors preserved the privacy of genomic and distributed clinical data by \yasir{first encrypting the data and then perturbing it using} differential noise mechanism. Furthermore, they worked over informatics for integrating biology and bedside (i2b2) framework, and enhanced its privacy along with reducing the network overhead. Similarly, the authors in~\cite{addpaper15} also preserved genomic data privacy by using traditional differential privacy approach and two way decryption method to save it from any attacker. The authors enhanced privacy and execution time of i2b2 framework in electronic genomic data records.} \revtwo{Moreover, the authors in~\cite{rev2ref04} developed a differentially private aggregation strategy which aggregated health devices data and do also provides timely incentives to its users. The proposed strategy combined differential privacy, Boneh-Goh-Nissim crypto system, and Shamir’s secret sharing to enhance both the security and privacy of users. The model is developed using JPBC library of java and it ensures the reduction of computational overhead. Therefore, the proposed strategy is more suitable for health IoT devices that have limited computational capacity.} As differential privacy was first designed for statistical databases, therefore the mathematical models of differential privacy perfectly fits healthcare and medical system databases, and this data can easily be secured from intruders using differential privacy perturbation. Keeping in view all the above discussion, we can conclude that differential privacy applied in e-health databases provides a desirable solution to protect privacy during query execution.

\subsection{User Perspective for Health Survey Data Protection} \label{dprefhms03}
As discussed in the above section, sometime the access of databases is given to certain companies and media cells to conduct surveys or query executions in order to learn more about a particular disease or to solve any specific problem. For instance, medical data and patient symptoms data can be used by mobile recommender systems to suggest a medication having less side effects~\cite{dphms14}. Certain therapies can be suggested by recommender system, that matches best with the dispositions of patient~\cite{dphms15, dphms16}. However, these benefits come with a trade-off of privacy. For example, if the query conducting media cell becomes an adversary or get compromised. It can then try to infer in to the personal details of patients, in such cases it is responsibility of data providers to protect users’ data before publicising it for any survey. The major question here arises, how to choose correct and desirable level of privacy without compromising over the benefits. The purpose of this section is to provide users’ perspective over applying privacy in health domain for their personal data. Especially the perspective of a common person towards privacy-precision trade-off of differential privacy. \\
In~\cite{hmref05}, authors analysed the users’ perspective over two different privacy preservation schemes; k-anonymity~\cite{dphms17}, and differential privacy~\cite{intref13}. The authors presented the point of view of common people to give their data towards future health care and for commercial purposes. The experimental results showed that users’ perspective towards their confidentially of data was quite strict if the data was about \yasir{to be given for a commercial use.} While the patients showed reluctance while providing the same data for scientific use. Furthermore, people showed various reservations for anonymization strategies because of the examples that privacy can be breached even after anonymization. However, differential privacy seemed suitable for users upon its idea of preserving privacy by data perturbation and by providing a privacy-precision trade-off. But the most confusing question for users was; how exactly data perturbation protects the privacy? Furthermore, users felt comfortable being the part of large crowd while applying differential privacy to their dataset. By keeping in view all these points, we can conclude that differential privacy used with large datasets is fairly optimal strategy to protect data according to users’ perspective.

%--------------------- FIGURE ------------------------

\begin{figure*}[t]        
\centering
\includegraphics[scale = 0.6]{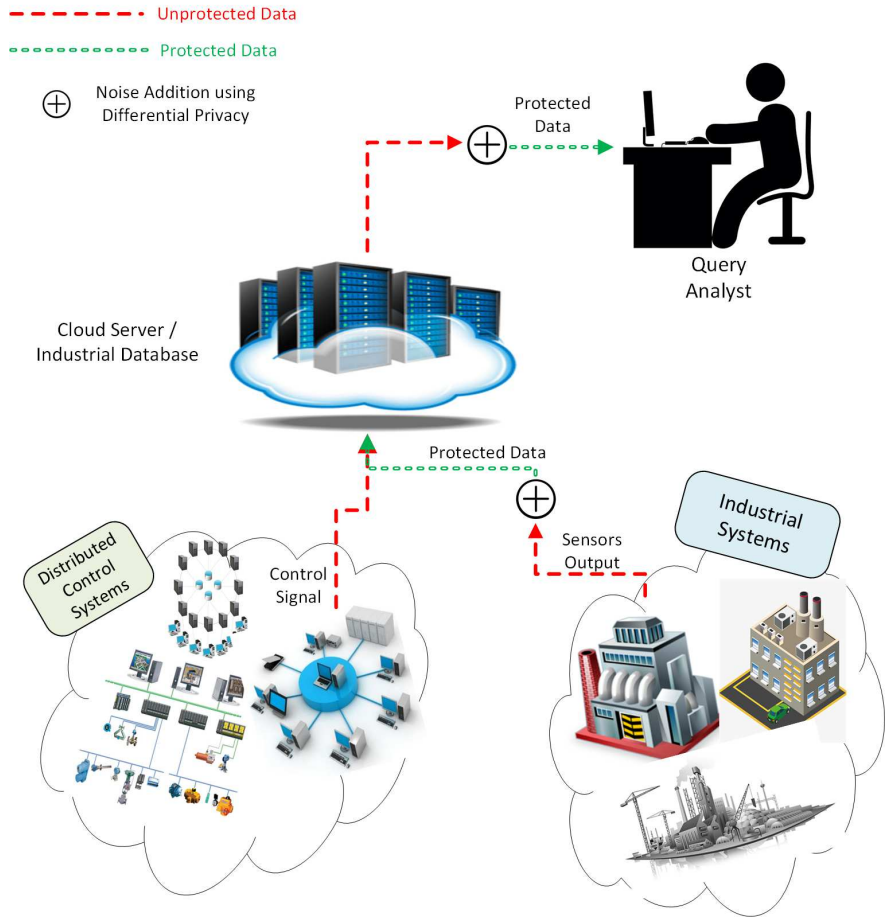}
 \caption{\small{Illustration of organization of differential privacy (DP) implementation in industrial Internet of thing (IIoT) systems into two scenarios: distributed control, and industrial systems. When DP is incorporated into these scenarios, industrial sensors data and distributed control signals of IIoT systems is protected and the data stored by database can be used for query evaluation.}}      
  \label{fig:newindustry}   
\end{figure*}

%--------------------- FIGURE ------------------------

\subsection{Summary and Lessons Learnt}
The new wave of modernizing and digitalizing medical devices and records has seen an exponential growth over the past decade and medical devices are being connected with each other and other databases via wireless networks. This replacement of old system to digitalized medical systems has paved the path to several privacy and security issues~\cite{dphms18}. The actual concern is that medical devices data contains certain PII, such as name, address, heartrate, blood pressure level, symptoms of any disease, certain medical test outcome, etc. This data can be used by malicious attackers to target a specific person, blackmail them, steal their money, and so on~\cite{dphms19}. However, we cannot deny the importance of this digitalized data as well, because doctors and hospitals require this data to overcome any severe consequences within time~\cite{dphms18}. Therefore, a certain level of privacy is required by the medical systems to utilize the data efficiently without risking it.\\
One important application of healthcare and medical systems is real-time data monitoring from certain medical equipment that are used as wearable devices such as smart watches, heart rate sensors, etc. This data contains specific patterns that can be used to judge personal information about any individual. However, this data cannot directly be perturbed because it should be useful for the required observer, e.g., physician, coach, and hospital administration. Certain techniques to protect this data is proposed by researchers, but the most suitable technique to provide efficient results is differential privacy. Differential privacy perturbs the data in such a way that even if any intruders compromises the real-time data, still it will not be able to get the useful information. However, this field is just under consideration and researchers are working to provide efficient ways to protect this real-time data without reducing data efficiency. \\
Another important application of health care data is data mining for early detection of diseases by viewing symptoms. Query execution is carried out by medical companies or hospitals to know better about early symptoms regarding any disease, or to perform statistical analysis of data. However, protecting the privacy during this query execution is a challenging task. But after the introduction of differential privacy in 2006 for statistical databases, the healthcare and medical data is also being protected using the applications of differential privacy. Still, a lot of work needs to be carried out in future. For instance, the artificially intelligent algorithms are being introduced to provide useful results in health databases without compromising the privacy.\\
The protection of healthcare and medical systems using differential privacy has been carried out by many scientists and researchers, however a large number of applications of healthcare systems still needs a considerable attention. For instance, introduction of machine learning in healthcare system is the new trend. Similarly, differential privacy can be incorporated with machine learning algorithms in order to ensure complete privacy of health data. Furthermore, the body sensors or wearable devices are becoming smaller in size with passage of time. Therefore, light-weight and less complex differential privacy algorithms are required to fit in to such devices. To sum up, differential privacy is a vital solution for healthcare and medical systems, but still a lot of efforts are required to address all applications of healthcare and medical system.

% ################ Flow Chart 4 #################

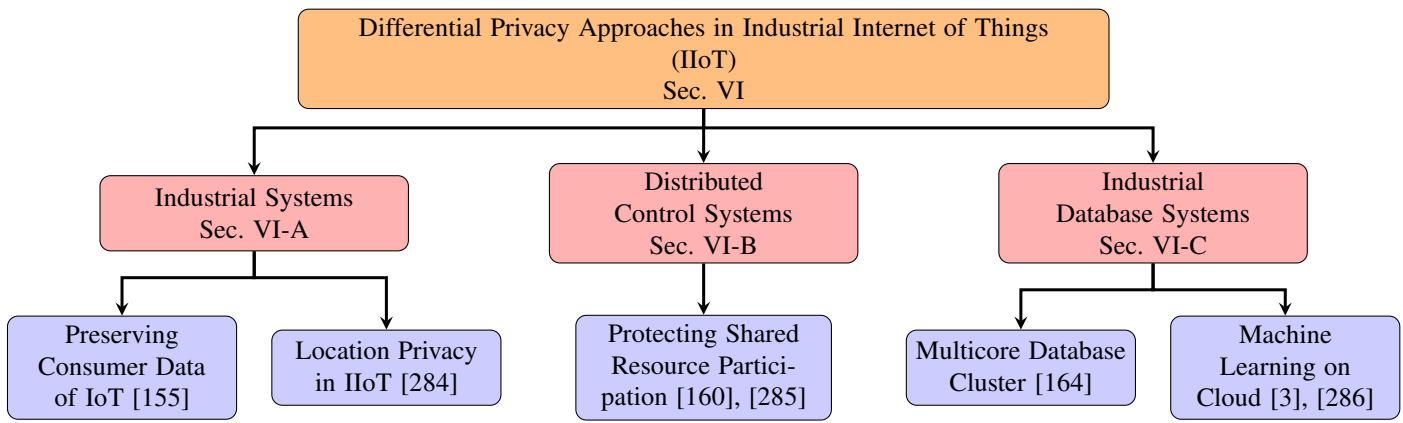
\begin{figure*}[]
     \centering

\begin{tikzpicture}

\node [block,  text centered, fill=orange!50, minimum width = 30em,  text width=30em] (a1) {Differential Privacy Approaches in Industrial Internet of Things \\(IIoT) \\ Sec.~\ref{dprefiiot}};
\node[block, below of=a1, yshift=-3em, xshift = -17em, fill=red!30](b1){Industrial Systems \\ Sec.~\ref{dprefiiot01}};
\node [block, below of=b1, xshift = -5em, yshift=-3em, minimum width = 2mm, text width=8em] (b1c1) {Preserving Consumer Data of IoT~\cite{iotref01}};
\node [block, below of=b1, xshift = 5em, yshift=-3em, minimum width = 2mm, text width=8em] (b1c2) {Location Privacy in IIoT~\cite{iotref02}};

\node[block, below of=a1, yshift=-3em, fill=red!30 ](b2){Distributed Control Systems \\ Sec.~\ref{dprefiiot02}};
\node[block, below of=b2, yshift=-3em, text width=7em, text width=9em](b2c1){Protecting Shared Resource Participation~\cite{iotref03},~\cite{iotref06}};

\node[block, below of=a1, yshift=-3em, xshift = 17em,  fill=red!30](b3){Industrial Database Systems \\ Sec.~\ref{dprefiiot03}};
\node [block, below of=b3, xshift = -5em, yshift=-3em, minimum width = 2mm, text width=8em] (b3c1) {Multicore Database Cluster~\cite{iotref04}};
\node [block, below of=b3, xshift = 5em, yshift=-3em, minimum width = 2mm, text width=8em] (b3c2) {Machine Learning on Cloud~\cite{iotref07},~\cite{iotref05}};

\path [line] (a1)--(b2);
\path [line] (a1)-- ($(a1.south)+(0,-0.25)$) -|(b1);
\path [line] (a1)-- ($(a1.south)+(0,-0.25)$) -|(b3);

\path [line] (b2)--(b2c1);

\path [line] (b1)--($(b1.south)+(0,-0.35)$)-|(b1c1);
\path [line] (b1)--($(b1.south)+(0,-0.35)$)-|(b1c2);

\path [line] (b3)--($(b3.south)+(0,-0.35)$)-|(b3c1);
\path [line] (b3)--($(b3.south)+(0,-0.35)$)-|(b3c2);

\end{tikzpicture}

	\small \caption{The differential privacy approaches implemented industrial Internet of things (IIoT) can be classified into industrial systems, distributed control systems, and industrial database systems.}
     \label{fig:tnfig04}
\end{figure*}
% ################ Flow Chart 4 #################

\section{Differential Privacy in Industrial\\ Internet of Things} \label{dprefiiot}
The term IoT was first introduced to address unique identifiable interoperable objects connected with a wireless technology named as radio-frequency identification (RFID)~\cite{dpiiot02}. This concept of IoT shifted gradually from RFID to Internet. However, with the rapid advancement of IoT technologies, modern Internet has taken over the world in every aspect. Physical layer of IoT devices are connected with each other using Internet protocol (IP) to form an IoT system ~\cite{dpiot01}. The trend of using IoT technologies in industries is exponentially increasing because of its effectiveness~\cite{dpiiot03, comstref26}. \revi{However, standard CPSs communication systems does not cope with IIoT systems because modern IIoT systems have certain extra requirements such as hostile environment operation, predictable throughput, maintenance by some other than communication specialists, and extremely low downtown~\cite{addpaper32}. The two most common IIoT communication systems include Fieldbus, and supervisory control and data acquisition (SCADA)~\cite{addpaper33}.} Large number of projects related to food industry, agriculture, security surveillance, and other similar fields have been conducted using IIoT technologies. As IIoT technologies are being used extensively in a large number of industrial projects, there are various commercial and \mubcom{political interests in it~\cite{dpiiot04}.} Because of those interests, the intruders always try to launch targeted attacks to obtain maximum possible data from IIoT systems and databases in order to damage that specific industry. \\
Keeping in view all these points, it can be said that it is very important to protect the privacy of IIoT systems. A large number of privacy preservation technologies for IoT have been recently discussed in~\cite{dpiiot05, dpiiot06, dpiiot07, dpiiot08}, but most of them focuses \mubcom{over basic IoT systems} privacy preservation. However, the privacy of industrial systems differs from that of general IoT systems because critical decisions without time delays needs to be made in IIoT devices. Few researches proposed limit release~\cite{dpiiot09}, data distortion~\cite{dpiiot10}, and data encryption~\cite{dpiiot11, dpiiot12} as a solution to preserve IIoT privacy, but with constant observation it can be seen that their advantages are limited and they cannot be implied broadly over every IIoT system.\\

Differential privacy is a new standard to preserve the privacy if IIoT systems. \addp{Differential privacy defines a detailed attack model, reduces privacy risks for data disclosure, and ensure data availability at same time of query or decision~\cite{iotref04, addpaper11}.} On the basis of privacy preservation using differential privacy, IIoT systems can be further divided into three subcategories; industrial systems, distributed control systems, and industrial database systems, as illustrated in Fig.~\ref{fig:newindustry}. In this section, we discuss the implementation of differential privacy in these IIoT systems. The detailed taxonomy of differential privacy implementation in IIoT systems is presented in Table~\ref{tab:iottab01} and Fig.~\ref{fig:tnfig04}.

\subsection{Industrial Systems} \label{dprefiiot01}
The rapid development of ICT technologies also changed the perspective of controlling traditional industrial devices. Similarly, merger of IoT, and ICT with traditional industry have revolutionized these systems and a new era of “Fourth Industrial Revolution” is on its way~\cite{dpiiot13}. Modern trends of IoT, and ICT are greatly influencing automation of industrial devices and higher degree of inter-connection among devices is being achieved using these technologies~\cite{dpiiot14}. In industry 4.0, almost every sort of communication will happen via wireless medium, therefore researchers are working over implementation of modern ways of communications (e.g., 5G) in industrial devices and sensors. The communication between these sensors and devices needs to be secured because these sensors and devices generate a large amount of safety-critical and privacy-sensitive data. Safe and secure operation of this data needs to be ensured for smooth running of industry~\cite{dpiiot15}.
Generally, the privacy threats of IIoT systems~\cite{dpiiot16} can be classified into two subcategories; based on preserving data and over preserving of location. Traditional approaches used to preserve data and location privacy are anonymity~\cite{dpiiot19} fuzzification technology~\cite{iotref02}.\\ However, because of multiple data fusion, and techniques of re-identification of anonymized data, these techniques do not show very considerable outcomes in \yasir{IIoT system.} Therefore, differential privacy appeared as one of the most suitable solution to communicate this data from sensors to required devices without compromising the integrity and privacy. Researches are being carried out over implementation of differential privacy in different scenarios of industrial automation. Similarly, the authors in~\cite{iotref01} improved data privacy by implementing differential privacy with k-anonymity.  The authors used the traditional concept of differential privacy, integrated it with k-anonymity and enhances anonymization of data, their preserved data can be aggregated and transmitted without risking the privacy component. However, the authors in~\cite{iotref02} considered the factor of preserving location of industrial sensors using differential privacy. The authors first demonstrated that extracting location of IIoT sensors can prove to be a serious threat to industry and later on authors provided the solution by merging differential privacy with these sensors. In~\cite{iotref02}, MATLAB and PyCharm are used to implement tree node accessing frequency model with Laplacian noise perturbation to enhance data utility timeliness. The authors firmly believe, in order to protect location and data privacy of industrial systems, differential privacy is an optimal solution in context of accuracy and timeliness. By keeping in view the above discussion regarding industrial automated systems privacy, and considering the effectiveness of dynamic nature of differential privacy strategies, it can be said that differential privacy can efficiently preserve privacy when applied with industrial systems.

% ===================== Industrial IoT Table  ===========================

%This is related work Table
\begin{table*}[ht]
\begin{center}
 \centering
 \footnotesize
  \captionsetup{labelsep=space}
 \captionsetup{justification=centering}
 \caption{\textsc{\\Comparative View of Differential Privacy Techniques in Industrial Internet of Things with their Specific Technique, Optimized Parameters, Privacy Criterion, Scenario, and Experimental Platform.}}
  \label{tab:iottab01}
  \begin{tabular}{|P{1.55cm}|P{0.55cm}|P{0.55cm}|P{2cm}|P{2.3cm}|P{2.45cm}|P{1.2cm}|P{1.1cm}|P{1.1cm}|P{1cm}|}
  	\hline
\rule{0pt}{2ex}
\bfseries Main Category & \bfseries Ref No. & \bfseries Year & \bfseries Privacy Mechanism & \bfseries Technique of DP Used & \bfseries Enhancement due to Differential Privacy & \bfseries \centering Privacy Criterion &\bfseries Platform Used & \bfseries \comst{Scenario} & \bfseries Time \newline Compl- \newline exity \\
\hline

\multirow{2}{*}{\parbox{2cm}{\centering \textbf{Industrial Systems}}}
\rule{0pt}{2ex}
& ~\cite{iotref01} & 2016 & Differential privacy for IoT & k-anonymity with traditional differential privacy is used  & \tabitem Enhanced anonymization & \centering $\varepsilon$-differential privacy & N/A & \comst{Real-time} & $-$\\
\cline{2-10}

\rule{0pt}{2ex}
& ~\cite{iotref02} & 2017 & Location privacy for IIoT using Differential privacy & Tree node accessing frequency model is used with Laplacian noise  & \tabitem Maximize data utility and timeliness & \centering $(\varepsilon, \delta)$-differential privacy  & MATLAB and PyCharm & \comst{Real-time} & $-$ \\
\cline{2-9}
\hline

\rule{0pt}{2ex}
\multirow{2}{*}{\parbox{2cm}{\centering \textbf{Distributed Control Systems}}}
\rule{0pt}{2ex}
& ~\cite{iotref03} & 2017 & Differential privacy in linear distributed control systems & Randomized Laplacian noise is used in distributed systems data  & \tabitem Entropy minimization & \centering $(\varepsilon, \Delta)$-differential privacy & N/A & \comst{Real-time} & $O(T^3 / \newline N_{\varepsilon^2})$ \\
\cline{2-10}

\rule{0pt}{2ex}
& ~\cite{iotref06} & 2017 & Differential privacy mechanism for feedback control systems & Minimum required Gaussian noise is calculated  & \tabitem Improved privacy, performance, and attack resiliency & \centering $(\varepsilon, \delta)$-differential privacy & N/A & \comst{Real-time} & $-$\\
\cline{2-10}
\hline

\multirow{3}{*}{\parbox{2cm}{}}%\centering \textbf{Database Systems}}}
\rule{0pt}{2ex}
& ~\cite{iotref04} & 2018 & Differential privacy for multicore DB scan clustering (DP-MCDBScan) & Random Laplacian noise is used with data clustering  & \tabitem Improved efficiency, accuracy and privacy of network data & \centering $\varepsilon$-differential privacy & MATLAB &\comst{ Statistical Database} & $-$ \\
\cline{2-10}

\rule{0pt}{2ex}
\centering \textbf{Industrial Database Systems} & ~\cite{iotref05} & 2018 & Preserving Multiple data-providers privacy  via differential privacy mechanism & Perturbation is added to cipher text after encrypting Laplacian noise  & \tabitem Multiple-providers data is preserved & \centering $\varepsilon$-differential privacy  & MAGMA and Java Simulator & \comst{Real-time} & $-$ \\
\cline{2-10}

\rule{0pt}{2ex}
& ~\cite{iotref07} & 2018 & Differential privacy model machine learning mechanism in CPS using prediction model & Machine learning model publishing is used with traditional differential privacy  & \tabitem Mean absolute error is enhanced & \centering $\varepsilon$-differential privacy  & N/A & \comst{Real-time} & $-$ \\
\cline{2-10}
\hline
 \end{tabular}
  \end{center}
\end{table*}

% ===================== Industrial IoT Table  ===========================

\subsection{Distributed Control Systems} \label{dprefiiot02}

In recent years, the interest in advancement of distributed control systems has increased exponentially in industrial domain~\cite{dpiiot20}. These types of systems involve cooperation of all connected devices in order to take intelligent decisions on the basis of input data. This control is generally achieved using emergent behaviour of various autonomous, simple and cooperative agents/devices~\cite{dpiiot21}. On one hand, the real-time sensing and sharing of this information provides large number of benefits, and on other hand, if any attacker gets access to this information then it can cause serious issues. \revi{For instance, halting or even destroying of industrial machines~\cite{dpiiot34}.} Therefore, preserving crucial information of modern distributed control systems is very important for their complete implementation in industrial sector. \revi{Different approaches to overcome privacy issues of these autonomous distributed control systems} have been proposed by researchers such as encryption, and k-anonymity. But the most promising approach to protect the privacy without losing the originality of data of these systems is differential privacy.\\
In case of linear distributed systems, differential privacy can easily preserve real-time continuously varying data from distributed linear devices by using the metric method presented in~\cite{dpiiot22}. The authors in~\cite{iotref03} used this metric based method of differential privacy and perturbed the data using random Laplacian noise. By following this pattern, the authors minimized entropy of system along with increasing the privacy of data being communicated between devices. Similarly, the authors in~\cite{iotref06} defined inherent differential privacy for systems based on feedback-control. The authors calculated the minimum amount of Gaussian noise that is required to ensure privacy of system. Furthermore, the provided mechanism improves performance, privacy, and attack resiliency of system. Both of the proposed methodologies showed that data perturbation technique of differential privacy can preserve this real-time floating data between control devices. Thus, by viewing the nature and privacy requirements of distributed control systems data, we can say that differential privacy provides is the best light-weight approach to preserve real-time data privacy in distributed control systems.

\subsection{Industrial Database Systems} \label{dprefiiot03}

Advances in IoT systems coupled with social networks are providing more intelligent and comprehensive services in our daily life~\cite{dpiiot23, dpiiot24, dpiiot25, dpiiot26}. These systems generate large proportion of data that is stored usually in cloud servers or databases. Certain functions in these social IoT systems are carried out via predefined interfaces using the stored big data~\cite{dpiiot27, jinref02, dpiiot28}. The data is continuously being shared between clients and servers, and during this process, any information can be leaked if the data is not preserved properly. This in return can generate huge security  and privacy threats to the databases of all social systems, because the individual privacy can easily be compromised during communication and query evaluation~\cite{intref03}. Large number of current researches proposed cryptographic encryption to preserve the privacy of social IoT systems~\cite{intref04, intref05}. However, \revtwo{certain number of keys} needs to be maintained in cryptographic schemes that makes it impossible to implement in databases where data needs to be shared with public or query evaluation. To overcome these issues, Dwork proposed the idea of differential privacy for statistical database~\cite{intref13}. Nevertheless, because of advancements in machine learning algorithms and multiple query releases in datasets, the traditional differential privacy cannot be used in social IoT systems~\cite{dpiiot32}. \\
In order to tackle the problems of traditional differential privacy approaches in social IoT systems, \revtwo{many researchers proposed} different ways to overcome it. In~\cite{iotref04}, Ni~\textit{et al.} proposed a powerful differential privacy approach using the combination of \mubcom{random Laplacian noise} and data clustering technology using MATLAB simulator. Furthermore, as compared to traditional privacy protection approaches, the authors improved efficiency, accuracy, and overall network data privacy. A step ahead of the pack, the authors in~\cite{iotref05} proposed a scheme to preserve multiple data-providers privacy using differential privacy. Instead of data providers, the noise is added via cloud server in this scheme. To further preserve confidentiality, the authors also encrypted the data using double decryption algorithm in combination with differential privacy. Zhu~\textit{et al.} in~\cite{iotref07} worked over implementation of machine learning along with differential privacy for efficient query evaluation. The authors reduced mean absolute error and preserved privacy of data using prediction model. To sum up, databases of social IoT systems needs to be preserved from cyber-attacks. Therefore, modern differential privacy approaches provide efficient privacy protection from all vulnerabilities along with enhanced query evaluation and providing support for machine learning algorithms.

\subsection{Summary and Lessons Learnt}

The development of IoT technology has paved paths for many future applications, one of the most important among them is the involvement of IoT in industry. Modern industrial devices are equipped with sensors that are communicating with each other in real-time to take crucial control decisions. However, leakage of this real-time information \revtwo{can cause severe privacy threats} to the machinery or individuals associated with it~\cite{dpiiot33}. For instance, an unmanned steel mill was halted and destroyed using cyber-attacks in Germany by disrupting and manipulating the control mechanism~\cite{dpiiot34}. Therefore, a certain privacy level for IIoT systems is required in order to operate smoothly. 
\revtwo{Industrial systems are one of the major aspects of IIoT}, these systems include large industrial machines that are communicating with each other using sensors in real-time. With the introduction of fourth generation of industry, \revtwo{these systems are developing very rapidly, and a large amount of data is being transmitted} every second from one device to another. However, protecting the privacy of these real-time systems by focusing on data and location privacy of these sensors and devices is the actual challenge at the moment~\cite{dpiiot35}. Differential privacy came out as an optimal solution to protect the privacy up to a certain extent without compromising the usefulness of data. Till now, differential privacy preserved the data integrity during data aggregation and data transmission of industrial systems. Nevertheless, a large of fields of industrial systems still needs to be preserved. For example, extensive efforts are required to secure the industrial and offices automation control. \\
Another important application similar to industrial systems is distributed control systems. In these systems, intelligent \revtwo{decision is taken on the basis of feedback} or input data. Therefore, this input data needs to be protected before transmitting it to system, because if any intruders gets access to the input or feedback then disastrous results can be seen. For example, in Iran in 2008, a centrifuge is sabotages at a uranium enrichment plant~\cite{dpiiot34}. Therefore, preserving input privacy is very crucial for such systems. Several techniques related to differential privacy have been presented by researchers to protect privacy without losing the originality of data. However, certain fields such as handling of big data of such systems still needs to be explored. Similarly, finding the most efficient trade-off between accuracy and privacy is also a challenge for industrial researchers.\\
Furthermore, the crucial data of IIoT systems is usually stored in cloud or large databases. These \mubcom{databases holders usually} allow companies to perform query execution over them in order to conduct surveys and other statistical tasks. However, if the company or utility becomes a hacker then the sensitive data of users can be leaked. For instance, the database of any social network contains a large amount of personal information whose privacy needs to be ensured before allowing companies to perform query execution. To cope up with this privacy challenge, data scientists apply data perturbation using differential privacy in these databases before allowing any company to perform query evaluation. \revtwo{This data perturbation allows different companies} to execute query evaluation without risking users’ data. Although, \revtwo{there are many fields} that still needs to be explored by researchers. One important application is the implementation of privacy protection in data mining and machine learning systems to make the artificially intelligent systems more secure. Another future domain to protect is buying/auction of industrial products. For example, buyers do not want to disclose their identity to sellers and similarly sellers do not want to disclose their identity to buyers. To sum up, differential privacy actively played the role to protect privacy of industrial systems. However, certain number of domains still needs to be protected and efforts needs to be made over these domains to make secure and reliable further generation IIoT systems.

\section{Open Issues, Challenges, and Future Research Directions}
Currently, differential privacy implementation in cyber physical systems is facing a large number of challenges, because of dynamic nature of CPSs. In this section, we discuss few challenges, open issues, and future research directions for implementation of differential privacy in CPSs.

\subsection{Energy Systems Issues and Research Directions}
Smart grid is the future of energy systems, because it incorporates capabilities of both; traditional energy systems and modern information and communication technologies. However, there are certain applications of smart grid \yasir{that still need considerable} attention in context of user privacy. In this section, we discuss such applications of smart grid in which differential privacy can improve the privacy preservation in an exceptional way.
\subsubsection{Billing with Dynamic Pricing} 

One of the biggest benefit of smart metering is accurate calculation of bills within a dynamic pricing environment~\cite{addpaper30}. This pricing strategy requires detailed energy consumption information, which on other hand may leak private information of smart meter users. Therefore, implementing differential privacy along with accurate dynamic pricing billing is a challenge for researchers. \comst{In recent year, many researches focused over dynamic private billing using differential privacy protection~\cite{fcsref09}.} The trade-off between accuracy and privacy of real-time data reporting is the biggest hurdle in implementation of differential privacy in smart meters. Many researchers are working and developing efficient algorithms to overcome this trade-off to a maximum level. Still, there is a large room that requires to be filled in order to preserve privacy along with dynamic consumption reporting for dynamic billing.

\subsubsection{Auction of Micro-Grid Energy Resources}
The demand of renewable energy is increasing due to rising costs of traditional fossil fuel based energy. Most of these energy resources such as small wind turbines, and solar panels will be deployed in smart homes~\cite{dpcps47}. \revtwo{Energy consumption of each house is different, therefore, some smart homes may use all their produced energy} while others may still have excessive energy left that is not of their use. Smart homes can then auction this excessive energy to other buyers, they can auction this energy and \yasir{buyers purchase it according} to their need~\cite{fcsref10}. However, during this process, buyer and seller usually do not want to disclose their identity to each other. \comst{Therefore, preservation of this information is very crucial for smooth running of auction mechanism in smart grid~\cite{comstref19}.} Few proposed researches considered micro-grid in EV scenario, but the specific application of auction in RERs based smart home is still not addressed in the literature. In order to secure this mechanism, techniques of \revtwo{differential privacy need to be proposed.}

\subsubsection{Firmware Updates}
Smart meters usually operate over an installed firmware that determines every functionality of them. Generally, this firmware is developed by smart meter vendors who usually update them to improve functionality or to remove any detected bug. Similarly, utility companies sometimes do also require firmware updates in case of any change among pricing or laws~\cite{fcsref10}. Since the firmware update file is proprietary, therefore it needs to be communicated to smart meters in a secure and private manner. Furthermore, sometimes update is required for only certain group of smart meters instead of all, in which \yasir{utility requires} case access control. However, to protect this firmware file, certain security and privacy based mechanisms are required. Till now, researchers only proposed security based approaches to overcome this problem. However, privacy requirement cannot be neglected in this application. We believe that incorporation of differential privacy scheme with other security schemes such as encryption can provide optimal results in this scenario.

\subsubsection{Resource Constrained Micro-Grids}
\addp{In order to get a cost-effective supply and power management alternative in remote areas, micro-grid resource constrained architectures are the optimal solution~\cite{addpaper01}. To minimize operational cost, certain lossy networks are used to carry out communication between these resource constrained architectures. However, these lossy networks are prone to many adversaries and privacy attacks because of their unreliable nature~\cite{addpaper02}. This loss of privacy can provoke various crimes, such as energy theft etc. Existing data preservation strategies cannot directly be applied to these lossy networks because \revtwo{these algorithms do not cope-up with cost requirements} in rural areas. As per our point of view, light-weight differential privacy can mitigate these privacy risks and provide a larger control over power management, distribution and anonymization~\cite{addpaper02}. Therefore, researches need to carry out in privacy preservation of resource constrained micro-grids using differential privacy techniques.}

\subsection{Transportation Systems Issues and Research Directions}
Smart transportation system is the need of a smart city. Therefore, governments are trying to improve the quality of transportation systems day by day. Along with the enhancement in quality and rapid communication between vehicles, certain security and privacy issues arises \yasir{that need to be tackled} along with quality enhancement. In this section, we discuss two major applications that require considerable attention in context of privacy preservation.
\subsubsection{Live Traffic Information}
In order to overcome delays due to congestion, live traffic information is generally used by route planning applications~\cite{fcsref15}. These applications consider live feed from connected cars, connected mobile devices, smart signals, and public transport to plan the shortest and less congested path for drivers. This is an advantageous feature that saves time of drivers, but on the other side of coin, it can cause serious threats to location privacy of connected devices and cars. If the network is unprotected, then any intruder can hack the system of these applications and may have access to the live location tracking of connected cars. Therefore, it is important to protect location privacy before reporting it to route planning applications. Differential privacy can provide real-time location privacy by perturbing location or identity in order to preserve drivers' privacy. Therefore, this field of ITSs has a lot potential and it needs to be explored in future.

\subsubsection{Vehicle-to-Vehicle (V2V) Communication}
With the exponential development of wireless communication technologies, vehicular ad-hoc networks, and V2V communication have become progressively popular. Certain services such as VoIP, web browsing, and video conferencing have been carried out with help of these networks and V2V communication~\cite{fcsref16}. In recent years, certain privacy and security issues of V2V communication has arisen. \mubcom{These issues have attracted} attention from both academia and industries. Encryption is considered to be the most famous strategy to secure communication between two vehicles, but it also comes with certain faults and loopholes. Therefore, in order to preserve privacy in communication among two vehicles, differential privacy can be an optimal solution. Future researches should consider integrating differential privacy with different V2V communication scenarios.

\subsection{Healthcare and Medical Systems Issues and Research Directions}
The trend of connecting cyber and \yasir{physical worlds in healthcare} and medical system has increased tremendously. However, this connection comes up with certain issues that needs to be resolved before its successful implementation in daily lives. In this section, we discuss few applications of healthcare systems in which differential privacy can be applied to get advantageous results.
\subsubsection{Body Sensors Data}
With the advancement in wireless technologies, the trend of using body sensors for medical purposes is also increasing dramatically. These sensors (i.e., heart rate sensor, and body temperature sensors) monitor your real-time readings and report them to your physician or trainer~\cite{fcsref12}. Although, this information cannot directly be transmitted to required person without protecting it from adversaries. Encryption is one of the solution for this type of application, but it is computationally complex. On the other hand, differential privacy based real-time reporting of data can be a light weight solution to solve this particular problem. The technique presented by Zhang~\textit{et al.} in~\cite{hmref01} is a great step towards implementing differential privacy in real-time health data sensing and reporting. Still this field has a lot of room and needs to be explored further. For instance, the real challenge is implementing differential privacy noise addition mechanism in low memory devices such as small micro-controllers. We believe that modern differential privacy algorithms can enhance this field and can produce optimal results in real-time health data reporting.
\subsubsection{Elderly Home Sensor Network}
Retirement homes or elderly homes \revtwo{do also need considerable attention because the people living in there require full time care and attention. Therefore, many healthcare devices are placed} in these homes for monitoring and diagnostic purposes~\cite{newmbref21}. However, along with monitoring, these devices \yasir{also need considerable} privacy protection, because even a small loophole in privacy can cause severe circumstances~\cite{fcsref13}. The whole elderly home network can be protected using differential privacy techniques in the devices. The potential applications of differential privacy in elderly homes can be protecting electronic patient records~\cite{fcsref14}, that contains all useful information, identity, and medical records of people living in that home.

\subsection{Industrial Internet of Things Systems Issues and Research Directions}
The advancements in industry is highly influenced by modern IoT technologies. As IoT is taking over industry by providing autonomous control, efficient data storage, and reliable communication, although it also comes up with risk of attacks related to security and privacy of industry. Few past events occurred in industries showed that these advanced technologies can be targeted, hackers can get access to private industrial data. These malicious adversaries \revtwo{can also control machinery or can even destroy} industrial systems. Therefore, certain IIoT \yasir{fields need to be secured} first for smooth running of IoT systems in industry.

\subsubsection{Industrial Big Data Trading (Auction)}
One major issue in applications of big data in industry is to handle auction of products or services. Trusted third-party platforms are generally used to carry out auction between buyer and seller. However, fully trusting third party platform is difficult because of many reasons, including insider adversaries, cyber threats, and platform insecurities~\cite{fcsref17}. We believe that, in order to address this issue of privacy protection in third-party platforms, differential privacy is a suitable solution. Because of strong mathematical modelling background, differential privacy can provide a desirable level of privacy in auction scenario of big data in industries.

\subsection{Other Issues and Research Directions}

\subsubsection{Big Data}
The horizon of big data is not only limited to CPSs, this covers almost every aspect of human life ranging from schools to offices and from farming to industry. In this section, we will be discussing particularly about some future directions and challenges regarding implementation of differential privacy in certain big data applications. We believe that as the data is increasing, differential privacy based algorithms are also becoming advanced and more responsive. However, there are certain challenges that still \yasir{need to be addressed} for big data. \\
%\vspace{0.2mm}
\paragraph{\textbf{Intuitive Privacy Definition}}

In context of differential privacy itself, one of the major challenge is defining the exact privacy. Even after establishing of mathematical proofs and strict privacy model, differential privacy lacks in giving an intuitive definition of privacy according to big data. Thus, finding a more intuitive definition of privacy in accordance with big data analytics is still a challenge for data scientists~\cite{surref02}. 
\paragraph{\textbf{Composition Theorem}}
As we discussed earlier \revi{(see Section~\ref{noisesec}),} that composition theorem plays an active role in designing of algorithm, and allocation of privacy budget. However, existing methods of deciding privacy budget using composition theorem are not optimal~\cite{fcsref01}. Therefore, optimal computation of composition of differential privacy in big data analytics is still an unsolved challenge. Similarly, in the domain of big data, maintaining privacy protection along with issue of dimensionality because of large data volume and computation overhead is a big challenge for researchers~\cite{fcsref02, fcsref03}. \\
\vspace{0.1em}
\noindent In context of future research directions, few novel approaches based on the principal of differential privacy such as local privacy~\cite{fcsref04}, concentrated privacy~\cite{fcsref05},~\textit{w-}event privacy~\cite{fcsref06}, and Bayesian differential privacy~\cite{addpaper47} have very large scope in big data applications. For instance, dealing with time-series data publishing,~\textit{w-}event privacy provides an optimal balance between~\textit{event-level} and~\textit{user-level} privacy~\cite{fcsref08}. As for future work, researches should focus on handling the privacy for large \revtwo{data volumes and designing} of optimal privacy budget for different notions of differential privacy.\\

\vspace{0.5mm}
\subsubsection{Machine Learning} 
The actual purpose of any machine learning algorithm is to extract beneficial information from given data. However, preserving individual privacy along with extracting data is one of the most challenging task of future machine learning algorithms~\cite{jinref06}. For example, if one is analysing sensitive medical data, then first it needs to be made sure privacy is preserved properly, and query evaluation can be performed~\cite{fcsref18}. To tackle this issue, researchers have started working over merging differential privacy data perturbation technique with machine learning algorithms~\cite{fcsref19, surref07, comstref06}. Future research directions in this field needs to examine the merger of efficient differential privacy data preservation techniques with complex machine learning algorithms.

\subsubsection{Cloud Computing}

\addp{
Huge amount of data generated through ubiquitous communication among smart devices paved the path towards a reliable and secure storage named as cloud computing~\cite{addpaper16, jinref03, addpaper17}. Furthermore, cloud computing emerged as a new computing paradigm and business model that enables on-demand supply of storage and computational resources. However, outsourcing this data to any third party can cause certain privacy issues~\cite{addpaper18}. These privacy risks are generally caused due to information redundancy in big data from different sources, multi-tenancy, and ubiquitous access features of platforms of cloud computing~\cite{addpaper19}. Traditional method of protecting cloud privacy is to store encrypted data over cloud platform, and data owners \revtwo{must download and decrypt} the data \yasir{locally to be sent} for processing~\cite{jinref05}. However, with the increase in size of data, it is becoming hard for data owners to afford this computationally complex approach~\cite{addpaper20}. Differential privacy is now emerging as a new practical approach to overcome these privacy issues of cloud computing scenario. Researchers have started work towards privacy preservation of cloud computing data using differential privacy. Privacy of certain cloud applications such as big graphs~\cite{addpaper20}, multi-agent programs~\cite{addpaper21},~\comst{blockchain-based cloud~\cite{comstref23}}, and scalable processing platforms~\cite{addpaper19} have been enhanced via modern differential privacy algorithms. We believe that this field has a large potential and light-weight differential privacy algorithms can revolutionize privacy standards of cloud computing.
}

\subsubsection{Wireless Edge Computing}
\addp{
Along with increase in smart devices, \revtwo{edge computing has now become} a mainstream while dealing with wireless communication scenario. \revtwo{Wireless edge computing provides broad benefits} \yasir{according to aspects of mining} and analysing data, and intelligently perceiving the information of location ~\cite{addpaper09}. These wireless edge computing networks contain large amount of private data that cannot be sent directly for data prediction and processing. Therefore, protecting important features of wireless edge computing needs to be made sure before any sort of query evaluation. In order to tackle this situation, researchers are proposing differential privacy based strategies as an optimal solution~\cite{addpaper10, addpaper12}. Keeping in view all this discussion, it can be said that modern differential privacy algorithms can enhance wireless edge computing, and these algorithms should be explored and presented in future.
}

\subsubsection{Blockchain Technology}

In the past few years, blockchain emerged as one of the novel distributed strategy that allows the secure storage of transactions, or any other type of data without need of any predefined centralized data authority~\cite{newmbref15, addpaper59}. \comst{The notion of blockchain was tightly coupled with Bitcoin for some time, but now it has been adopted widely in many applications e.g., healthcare, finances,~\revi{and logistics}~\cite{newmbref16, comstref59, comstref07}.} The feature of public accessibility without any centralized authority made it famous among its users, but on the other hand it also raised certain security and privacy issues in it. Because of inadequacy of existing blockchain protocols~\cite{newmbref17}, most of the blockchain users are worried about their transaction privacy. To overcome this issue, researchers are proposing certain privacy schemes on the basis of identity, anonymity, and perturbation~\cite{newmbref18, comstref44, newmbref19}.~\revi{\mubcom{Researchers are enhancing data perturbation} strategies by making them artificially intelligent using machine learning algorithm. We believe that, modern differential privacy algorithms in conjunction with blockchain can eradicate the issue of privacy loss even in case of public query evaluation.} Because of mathematical background and light-weight privacy model, differential privacy can preserve transactions and other data storage in blockchain technology. Therefore, researchers should focus on integration of these two modern world technologies to achieve efficient results.

\subsubsection{Game Theory}
\addp{
Game theory addresses issues in which multiple participants compete with each other having contradictory goals or incentives~\cite{addpaper22, addpaper30}. \comst{Similarly, in order to enhance administrators’ decision making, game theory can be used to analyse \revtwo{large number of possible scenarios} before taking the most appropriate action such as smart grid energy trading~\cite{comstref60}.} Security and privacy is also an important aspect of game theory algorithms. As we discussed earlier in context of differential privacy, that trade-off between privacy and utility is a critical issue being \yasir{considered at the moment} (see Section~\ref{existmethod}). This trade-off between privacy and utility is being evolved \yasir{further into a game problem.} Researchers have now started developing modern differential privacy approaches by efficiently handling utility-privacy trade-off with help of game theory based algorithms~\cite{addpaper24, addpaper25}. We believe that game theory based differential privacy techniques can be used to handle privacy of certain differential privacy and CPSs applications. Therefore, researches in this \yasir{field need to be carried out in future.}
}

\section{Conclusion}

With the advancement in information and communication technologies (ICT), cyber physical systems (CPSs) have become an essential part of our lives, ranging from our homes to industries and from offices to hospitals. However, this advancement comes up with certain security and privacy risks attached to it. Various privacy attacks are carried out in CPSs to access critical data or information from private or public datasets.  One of the most optimal solution to overcome these privacy hazards is \mubcom{preserving data by noise addition using differential privacy perturbation mechanisms}. In this article, we have presented a detailed and up-to-date survey of implementation of differential privacy techniques in various CPSs applications. We have comprehensively covered all dimensions and aspects of differential privacy implementation in major CPSs domains. Integration of differential privacy in four application scenarios of CPSs, named as energy systems, transportation systems, healthcare and medical systems, and industrial systems is presented in the paper. Within energy systems, we surveyed privacy protection of demand response data, real-time data, and fog computing communication systems using differential privacy. Similarly, in transportation systems, we covered the aspect of privacy preservation with help of differential privacy in railway networks, vehicular networks, and automotive manufactures databases. Moreover, in healthcare and medical systems, we surveyed differential privacy approaches in real-time health data, and e-health databases. Furthermore, according to industrial point of view, we presented implementation of differential privacy techniques in industrial, \mubcom{distributed control systems,} and industrial database systems. We then concluded the survey article by highlighting challenges, open issues, and future research directions in differential privacy techniques for CPSs.

\bibliographystyle{IEEEtran}


\begin{IEEEbiography}[]{Muneeb Ul Hassan}

received his Bachelor degree in Electrical Engineering from COMSATS Institute of Information Technology, Wah Cantt, Pakistan, in 2017. He received Gold Medal in Bachelor degree for being topper of Electrical Engineering Department. Currently, he is pursuing the Ph.D. degree from Swinburne University of Technology, Hawthorn VIC 3122, Australia. His research interests include privacy preservation, blockchain, game theory, and smart grid. He served in the TPC for the IEEE International Conference on Cloud Computing Technology and Science (CloudCom 2019). He is a reviewer of various journals, such as the IEEE Communications Surveys \& Tutorials, IEEE Journal on Selected Areas in Communications, Elsevier Future Generation Computing Systems, Journal of Network and Computer Applications, Computers \& Electrical Engineering, IEEE ACCESS, Wiley Transactions on Emerging Telecommunications Technologies, IEEE Journal of Communications and Networks, Springer Wireless Networks, Human-centric Computing and Information Sciences, and KSII Transactions on Internet and Information Systems. He also has been a Reviewer for various conferences, such as IEEE Vehicular Technology Conference (VTC)-Spring 2019, Vehicular Technology Conference (VTC)-Fall 2018, IEEE International Conference on Communications (ICC) - 2019, International workshop on e-Health Pervasive Wireless Applications and Services e-HPWAS’18, IEEE Globecom 2018 workshop: Security in Health Informatics (SHInfo2018), Frontiers of Information Technology 2019, Frontiers of Information Technology 2018.

\end{IEEEbiography}
\vspace{-2cm}

\begin{IEEEbiography}[]{Mubashir Husain Rehmani (M’14-SM’15)}

received the B.Eng. degree in computer systems engineering from Mehran University of Engineering and Technology, Jamshoro, Pakistan, in 2004, the M.S. degree from the University of Paris XI, Paris, France, in 2008, and the Ph.D. degree from the University Pierre and Marie Curie, Paris, in 2011. He is currently working as Assistant Lecturer at Cork Institute of Technology (CIT), Ireland. He worked at Telecommunications Software and Systems Group (TSSG), Waterford Institute of Technology (WIT), Waterford, Ireland as Post-Doctoral researcher from Sep 2017 to Oct 2018. He served for five years as an Assistant Professor at COMSATS Institute of Information Technology, Wah Cantt., Pakistan.  He is currently an Area Editor of the IEEE Communications Surveys and Tutorials. He served for three years (from 2015 to 2017) as an Associate Editor of the IEEE Communications Surveys and Tutorials. Currently, he serves as Associate Editor of  IEEE Communications Magazine, Elsevier Journal of Network and Computer Applications (JNCA), and the Journal of Communications and Networks (JCN). He is also serving as a Guest Editor of Elsevier Ad Hoc Networks journal, Elsevier Future Generation Computer Systems journal, the IEEE Transactions on Industrial Informatics, and Elsevier Pervasive and Mobile Computing journal. He has authored/ edited two books published by IGI Global, USA, one book published by CRC Press, USA, and one book with Wiley, U.K. He received “Best Researcher of the Year 2015 of COMSATS Wah” award in 2015. He received the certificate of appreciation, “Exemplary Editor of the IEEE Communications Surveys and Tutorials for the year 2015” from the IEEE Communications Society. He received Best Paper Award from IEEE ComSoc Technical Committee on Communications Systems Integration and Modeling (CSIM), in IEEE ICC 2017. He consecutively received research productivity award in 2016-17 and also ranked \# 1 in all Engineering disciplines from Pakistan Council for Science and Technology (PCST), Government of Pakistan. He also received Best Paper Award in 2017 from Higher Education Commission (HEC), Government of Pakistan.

\end{IEEEbiography}
\vspace{-2cm}

\begin{IEEEbiography}[]{Dr. Jinjun Chen}

is a Professor from Swinburne University of Technology, Australia. He is Deputy Director of Swinburne Data Science Research Institute. He holds a PhD in Information Technology from Swinburne University of Technology, Australia. His research interests include scalability, big data, data science, data systems, cloud computing, data privacy and security, health data analytics and related various research topics. His research results have been published in more than 160 papers in international journals and conferences, including various IEEE/ACM Transactions. 

\end{IEEEbiography}
\vspace{-2cm}

\end{document}